\documentclass{article}
\usepackage{arxiv}
\usepackage[utf8]{inputenc}
\usepackage[T1]{fontenc}
\usepackage{xcolor}
\usepackage{pdflscape}
\usepackage{todonotes}
\usepackage{graphicx, booktabs}
\usepackage{subcaption}
\usepackage[para,online,flushleft]{threeparttable}
\usepackage{float}
\graphicspath{{figures/}}
\usepackage{doi}
\usepackage{subcaption}

\definecolor{blendedblue}{rgb}{0.2, 0.2, 0.6}
\definecolor{forestgreen(web)}{rgb}{0.13, 0.55, 0.13}
\definecolor{darkorange}{rgb}{1.0, 0.55, 0.0}

\usepackage{hyperref}
\hypersetup{
            colorlinks=false,
            citecolor   = blendedblue,
            colorlinks  = true,
            linkcolor   = black,
            urlcolor    = blendedblue}
            
\usepackage{amsfonts, amsmath, amssymb}
\usepackage{theorem}
\usepackage{upgreek}
\usepackage{xcolor}
\usepackage{etoolbox}

\usepackage{orcidlink}
\usepackage[english]{babel}
\theoremstyle{change}
\theorembodyfont{\upshape}
\newtheorem{satz}{Theorem}[section]

\theorembodyfont{\upshape\small}
\newtheorem{bsp}[satz]{Example}
\newtheorem{bem}[satz]{Remark}

\theorembodyfont{\ttfamily}

\newcommand{\ba}{\begin{equation}}
\newcommand{\ea}{\end{equation}}

\usepackage{dsfont}

\newcommand{\0}{\mbox{\boldmath $0$}}
\newcommand{\zeroi}{\mbox{\boldmath\scriptsize $0$}}
\newcommand{\1}{\mbox{\boldmath $1$}}
\newcommand{\onei}{\mbox{\boldmath\scriptsize $1$}}
\newcommand{\2}{\mbox{\boldmath $2$}}
\newcommand{\twoi}{\mbox{\boldmath\scriptsize $2$}}
\newcommand{\3}{\mbox{\boldmath $3$}}

\newcommand{\fd}{\mbox{\boldmath $d$}}
\newcommand{\fdi}{\mbox{\boldmath\scriptsize $d$}}

\newcommand{\fh}{\mbox{\boldmath $h$}}
\newcommand{\fhi}{\mbox{\boldmath\scriptsize $h$}}

\newcommand{\fp}{\mbox{\boldmath $p$}}

\newcommand{\fri}{\mbox{\boldmath\scriptsize $r$}}
\newcommand{\fs}{\mbox{\boldmath $s$}}
\newcommand{\fsi}{\mbox{\boldmath\scriptsize $s$}}

\newcommand{\fti}{\mbox{\boldmath\scriptsize $t$}}

\newcommand{\mY}{\mbox{\textup{\textbf{Y}}}}
\newcommand{\mpi}{\mbox{\boldmath $\uppi$}}

\newcommand{\bin}{\textup{Bin}}

\newcommand{\zip}{\textup{ZIP}}
\newcommand{\poi}{\textup{Poi}}
\newcommand{\norm}{\textup{N}}
\newcommand{\unif}{\textup{U}}

\newcommand{\bbn}{\mathbb{N}}

\newcommand{\bbz}{\mathbb{Z}}

\newcommand{\iid}{i.\,i.\,d.}
\newcommand{\ie}{i.\,e., }
\newcommand{\eg}{e.\,g., }

\newcommand{\cor}{\textup{Corr}}

\usepackage[round]{natbib}

\renewcommand{\headeright}{}

\makeatletter
\def\@seccntformat#1{\@ifundefined{#1@cntformat}%
   {\csname the#1\endcsname\quad}  
   {\csname #1@cntformat\endcsname}
}
\let\oldappendix\appendix 
\renewcommand\appendix{%
    \oldappendix
    \newcommand{\section@cntformat}{\appendixname~\thesection\quad}
}
\makeatother

\title{Nonparametric Monitoring of Spatial Dependence}
\renewcommand{\shorttitle}{Nonparametric Monitoring of Spatial Dependence}
\date{}

\author{
        Philipp Adämmer\orcidlink{0000-0003-3770-0097}\\
        Institute of Data Science\\
        University of Greifswald\\ 
        Greifswald, Germany\\
	\href{mailto:philipp.adaemmer@uni-greifswald.de} 
        {\nolinkurl{philipp.adaemmer@uni-greifswald.de}}\\
	\And
        Philipp Wittenberg\orcidlink{0000-0001-7151-8243}\\
 	Department of Mathematics and Statistics\\
        Helmut Schmidt University\\
	Hamburg, Germany\\
	\href{mailto:pwitten@hsu-hh.de}{\nolinkurl{pwitten@hsu-hh.de}}\\
        \And
        Christian H.\ Wei{\ss}\orcidlink{0000-0002-7492-6194}\footnote{corresponding author}\\
        Department of Mathematics and Statistics\\
        Helmut Schmidt University\\
	Hamburg, Germany\\
	\href{mailto:weissc@hsu-hh.de}{\nolinkurl{weissc@hsu-hh.de}}\\
	\And 
        Murat Caner Testik\orcidlink{0000-0003-2389-4759}\\
        Hacettepe University\\
        Department of Industrial Engineering\\
        Beytepe-Ankara, T\"urkiye\\	\href{mailto:mtestik@hacettepe.edu.tr}    
        {\nolinkurl{mtestik@hacettepe.edu.tr}}\\
}

\begin{document}	
\maketitle

\begin{abstract}
In process monitoring, it is common for measurements to be taken regularly or randomly from different spatial locations in two or three dimensions. While there are nonparametric methods for process monitoring with such spatial data to detect changes in the mean, there is a gap in the literature for nonparametric control charting methods developed to monitor spatial dependence. This study considers streams of regular, rectangular data sets using spatial ordinal patterns (SOPs) as a nonparametric method to detect spatial dependencies. We propose novel SOP control charts, which are distribution-free and do not require prior Phase-I analysis. To uncover higher-order dependencies, we develop a new class of statistics that combines SOPs with the Box-Pierce approach. An extensive simulation study demonstrates the superiority and effectiveness of our proposed charts over traditional parametric approaches, particularly when the spatial dependence is nonlinear or bilateral or when the spatial data contains outliers. The proposed SOP control charts are illustrated using real-world datasets to detect (i) heavy rainfall in Germany, (ii) war-related fires in (eastern) Ukraine, and (iii) manufacturing defects in textile production. This wide range of applications and findings demonstrates the broad utility of the proposed nonparametric control charts. In addition, all methods in this study are provided as a publicly available \texttt{Julia} package on \href{https://github.com/AdaemmerP/OrdinalPatterns.jl}{GitHub} for further implementations.
\end{abstract}

\bigskip
\noindent%
\vfill

{\small* corresponding author}

\section{Introduction}\label{Introduction}
In 2024, we celebrated the centenary of the control chart, which was first proposed by Walter A.\ Shewhart as part of his famous memorandum from May 16, 1924, at the Bell Telephone Laboratories \citep[see][]{Olmstead_1967}. Over the years, control charts have proven to be very effective tools for monitoring processes and many control charts have been developed depending on the different characteristics of measurements. 

During the last two decades, the use of ordinal patterns (OPs), which were originally introduced by \citet{Bandt.Pompe_2002}, has become increasingly popular in time series analysis (see \citet{Bandt_2019, Bandt_2023} for recent surveys). Among others, OPs were utilized to construct nonparametric tests for serial dependence in univariate time series data \citep[see][]{Weiss_2022, Weiss.Schnurr_2024}. Related to this, recently \citet{Weiss.Testik_2023} utilized OPs in statistical process monitoring (SPM), where nonparametric control charts based on OPs were used to detect the occurrence of serial dependence in time-series data. 
 
In the present study, we propose several novel types of OP-based control charts for SPM with data that are collected at the points in a regular two-dimensional grid (regular lattice data):
\ba
\label{data_rectangle}
\begin{array}{cccc}
y_{0,0}, & y_{0,1}, & \cdots & y_{0,n},\\
\vdots & \vdots & \ddots & \vdots \\
y_{m,0}, & y_{m,1}, & \cdots & y_{m,n}.\\
\end{array}
\ea
Here, we use the short-hand notation $(y_{\fsi})$ for the outcomes of such a rectangular data set, where the spatial position $\fs=(s_1,s_2)$ takes values in the regular grid $\{0,\ldots, m\}\times \{0,\ldots, n\}$ with $m,n\in\bbn=\{1,2,\ldots\}$.

The monitoring of spatial data is of utmost importance in various areas, such as health surveillance \citep[see][]{knox64,kulldorff97,yang20}, environmental applications \citep[see][]{qiu23,yang24}, and SPM of industrial data \citep[see][]{yan18}. For a comprehensive discussion, also see the book by \citet{qiu24}.
For the last few years, increasing attention has been paid to monitoring streams of rectangular data \eqref{data_rectangle} in the SPM literature (see \eg \citet{Megahed.etal_2011, Colosimo_2018} for surveys). Especially, applications related to (video-)image data have been reported, covering quality-related areas (\eg monitoring of phone displays or textile images) as well as health surveillance (see \citet{Wang.Tsung_2005, Jiang.etal_2011, Megahed.etal_2012, Bui.Apley_2018, Tsiamyrtzis.etal_2022} for details and further references). 
Some earlier research also considered vector representation of streams of spatial positions. For instance, \citet{Runger.Fowler_1998} considered the wafer fabrication step of an integrated circuits manufacturing process, where silicon dioxide thickness measurements were taken from nine sites on each of four wafers in a furnace at each run. 
Another manufacturing example for monitoring rectangular data, the multilayer chip capacitors process described by \citet{Barton.Gonzalez-Barreto_1996}, is briefly discussed later in Remark~\ref{bemEWMAm1n1}. 
A survey of automated visual-based defect detection approaches in industrial applications for detecting defects on surfaces through images can be found in \citet{Czimmermann.etal_2020}. 
However, monitoring streams of rectangular data sets is not limited to traditional SPM applications. In Section~\ref{Real-Data Application}, in addition to a manufacturing example, we also consider geographic data on weather events and war-related fires. 

\smallskip
In the following, we develop novel nonparametric control charts for monitoring streams of rectangular data. Existing nonparametric SPM procedures for spatial data mainly focus on changes in the mean (see \citet{yang20,qiu23,yang24} as examples), while our approaches are designed to discover spatial dependence, thus complementing existing approaches. To enable a nonparametric monitoring of spatial dependence, we consider Spatial OPs (SOPs), an extension of OPs to the plane, which were first proposed by \citet{Ribeiro.etal_2012} and studied in detail by \citet{Bandt.Wittfeld_2023}. A comprehensive discussion is provided in Section~\ref{Spatial Ordinal Patterns and Types}. A first group of novel control charts are proposed in Section~\ref{Control Charts for Spatial Dependence}, where we also consider control charts based on the spatial autocorrelation function (ACF) as a competing (but parametric) monitoring approach. 
Section~\ref{Simulation Study} and Appendix \ref{sec:appendix_ooc_results} present the results of a comprehensive simulation study that examines the average run length (ARL) performance of the proposed SOP charts with respect to various out-of-control scenarios. 
While the aforementioned investigations refer to the case of uncovering first-order dependencies, which are usually most pronounced in applications even if additional dependencies exist at higher spatial lags, we demonstrate in Section~\ref{Detecting Higher-order Spatial Dependence} that one can also design SOP charts focusing on higher-order spatial dependencies. Among others, we develop a novel SOP statistic for SPM, which is inspired by the classic Box--Pierce (BP) statistic known from time series analysis. The performances of our newly proposed nonparametric control charts for higher-order dependence are analyzed by simulations (also see Appendix \ref{sec:appendix_ooc_results}), again considering appropriate parametric competitors. 
The great relevance of our novel SOP charts and their application in practice is demonstrated in Section~\ref{Real-Data Application}, where three real-world examples (precipitation data, war-related fires, and textile images) are discussed. Finally, Section~\ref{Conclusions} concludes and outlines directions for future research. 
To facilitate further implementations, we also provide a publicly available \texttt{Julia} package on \href{https://github.com/AdaemmerP/OrdinalPatterns.jl}{GitHub}, which contains all the methods used in this study.

\section{Spatial Ordinal Patterns and Types}
\label{Spatial Ordinal Patterns and Types}
Let $(y_{\fsi})$ be a rectangular data set as given in \eqref{data_rectangle}, which is assumed to originate from some stationary spatial process in the plane (random field), say $(Y_{\fsi})_{\bbz^2} = (Y_{\fsi})_{\fsi\in\bbz^2}$ with $\bbz=\{\ldots, -1, 0, 1, \ldots\}$, where the~$Y_{\fsi}$ are real-valued and continuously distributed random variables (RVs). According to \citet{Ribeiro.etal_2012}, a SOP is computed by first picking out a $w_1\times w_2$-rectangle from the data $(y_{\fsi})$, where $w_1,w_2\geq 2$. Then, the original real numbers are replaced by integer ranks from $\{1,2,\ldots,w_1\cdot w_2\}$, leading to one of the $(w_1\cdot w_2)!$ different SOPs of order~$w_1\cdot w_2$. 
As rank statistics, SOPs are robust against outliers, and their distribution does not depend on the actual (continuous) marginal distribution of $(Y_{\fsi})_{\bbz^2}$. 
The importance of SOPs for developing nonparametric (distribution-free) approaches for dependence analysis arises from the fact that under the additional assumption of spatial independence, \ie if $(Y_{\fsi})_{\bbz^2}$ is independent and identically distributed (\iid), the SOPs follow a discrete uniform distribution with probability $1/(w_1\cdot w_2)!$ for each SOP, see \citet{Bandt.Wittfeld_2023, Weiss.Kim_2024} for details. 
Therefore, if testing the null hypothesis of the spatial independence of $(Y_{\fsi})_{\bbz^2}$ against spatial dependence, one uses test statistics that compare the SOPs' actual distribution to the discrete uniform one. 
However, the uniform probability $1/(w_1\cdot w_2)!$ gets rather small for $w_1\cdot w_2>4$ such that the different SOPs are hardly observed with reasonable frequency in practice. For this reason, like in \citet{Bandt.Wittfeld_2023, Weiss.Kim_2024}, we restrict our discussion to $2\times 2$-SOPs (\ie $w_1=w_2=2$), where always one of the 24~possible squares of ranks from $\mathcal{S}=\big\{\mpi = \left(\begin{smallmatrix} r_1 & r_2 \\ r_3 & r_4 \end{smallmatrix}\right)\ \big|\ \{r_1,r_2,r_3,r_4\}=\{1,2,3,4\}\big\}$ is observed. Denoting the extracted $2\times 2$-rectangle by $\mY = \left(\begin{smallmatrix} y_{1} & y_{2} \\ y_{3} & y_{4} \end{smallmatrix}\right)$, \ie the entries are read row-by-row, the corresponding SOP $\mpi = \left(\begin{smallmatrix} r_1 & r_2 \\ r_3 & r_4 \end{smallmatrix}\right)$ is defined by
\ba
\label{ordpatternrank}
r_k<r_l\qquad\Leftrightarrow\qquad
y_{k}< y_{l} \quad\text{or}\quad
(y_k=y_l \text{ and } k<l)
\ea
for all $\{k,l\}\subset \{1,2,3,4\}$. Note that this definition covers the case of ties within~$\mY$, although ties are observed with probability~0 for continuously distributed RVs. In practice, however, due to the limited numerical precision of measurement devices, ties in~$\mY$ might occasionally be observed. 

\smallskip
If analyzing $(y_{\fsi})$ for spatial dependence, we do not only extract a single $2\times 2$-square for SOP computation, but all $m\cdot n$ possible squares $\mY_{\fsi} = \left(\begin{smallmatrix} y_{s_1-1, s_2-1} & y_{s_1-1,s_2} \\ y_{s_1, s_2-1} & y_{s_1,s_2} \end{smallmatrix}\right)$ for $\fs\in \{1,\ldots,m\}\times \{1,\ldots,n\}$. 
It is also possible to consider further integer ``delay parameters'' $d_1,d_2\in\bbn$ (spatial lags) and to extract squares of the form $\mY_{\fsi}^{(\fdi)} = \left(\begin{smallmatrix} y_{s_1-d_1, s_2-d_2} & y_{s_1-d_1,s_2} \\ y_{s_1, s_2-d_2} & y_{s_1,s_2} \end{smallmatrix}\right)$ with $\fd=(d_1,d_2)$. This approach allows for detecting higher-order spatial dependencies in the data, see Section~\ref{Detecting Higher-order Spatial Dependence} for details. However, for now, we concentrate on $\fd=\1=(1,1)^\top$, with $\mY_{\fsi} = \mY_{\fsi}^{(\onei)}$, as first-order dependence is often most pronounced in applications even if additional higher-order dependencies exist. After having computed all SOPs, one determines the vector of relative frequencies of all SOPs to estimate the true SOP probabilities, where deviations from $(\frac{1}{24},\ldots,\frac{1}{24})^\top$ indicate the presence of spatial dependence. Note that for $(Y_{\fsi})_{\bbz^2}$ being \iid, \citet{Weiss.Kim_2024} proved asymptotic normality for the SOP frequencies and derived a closed-form expression for their covariance matrix.

\smallskip
Although we already focus on the smallest possible SOPs, their frequencies will commonly be rather low unless the sample size $m\cdot n$ is very large. For this reason, \citet{Bandt.Wittfeld_2023} proposed a partition of~$\mathcal S$ into larger subsets and to determine the frequencies with respect to these subsets only. More precisely, each SOP~$\mpi\in\mathcal S$ is assigned to one out of three possible ``types'', where~$\mpi$ is said to have type~$k\in\{1,2,3\}$ iff~$\mpi\in\mathcal S_k$, with the subsets~$\mathcal S_k$ of size~8 being defined as
\ba
\label{types}
\begin{array}{rl}
\mathcal S_1\ =& \Big\{\left(\begin{smallmatrix} \textbf{1} & 2 \\ 3 & 4 \end{smallmatrix}\right),
\left(\begin{smallmatrix} \textbf{1} & 3 \\ 2 & 4 \end{smallmatrix}\right),
\left(\begin{smallmatrix} 2 & \textbf{1} \\ 4 & 3 \end{smallmatrix}\right),
\left(\begin{smallmatrix} 2 & 4 \\ \textbf{1} & 3 \end{smallmatrix}\right),
\left(\begin{smallmatrix} 3 & \textbf{1} \\ 4 & 2 \end{smallmatrix}\right),
\left(\begin{smallmatrix} 3 & 4 \\ \textbf{1} & 2 \end{smallmatrix}\right),
\left(\begin{smallmatrix} 4 & 2 \\ 3 & \textbf{1} \end{smallmatrix}\right),
\left(\begin{smallmatrix} 4 & 3 \\ 2 & \textbf{1} \end{smallmatrix}\right)\Big\}, 
\\[.5ex]
& \text{(monotonic behaviour along both rows and columns)}
\\[2ex]
\mathcal S_2\ =& \Big\{\left(\begin{smallmatrix} 1 & \textbf{2} \\ 4 & 3 \end{smallmatrix}\right),
\left(\begin{smallmatrix} 1 & 4 \\ \textbf{2} & 3 \end{smallmatrix}\right),
\left(\begin{smallmatrix} \textbf{2} & 1 \\ 3 & 4 \end{smallmatrix}\right),
\left(\begin{smallmatrix} \textbf{2} & 3 \\ 1 & 4 \end{smallmatrix}\right),
\left(\begin{smallmatrix} 3 & \textbf{2} \\ 4 & 1 \end{smallmatrix}\right),
\left(\begin{smallmatrix} 3 & 4 \\ \textbf{2} & 1 \end{smallmatrix}\right),
\left(\begin{smallmatrix} 4 & 1 \\ 3 & \textbf{2} \end{smallmatrix}\right),
\left(\begin{smallmatrix} 4 & 3 \\ 1 & \textbf{2} \end{smallmatrix}\right)\Big\},  
\\[.5ex]
& \text{(increase/decrease along one of the dimensions, either rows or columns)}
\\[2ex]
\mathcal S_3\ =& \Big\{\left(\begin{smallmatrix} 1 & \textbf{3} \\ 4 & 2 \end{smallmatrix}\right),
\left(\begin{smallmatrix} 1 & 4 \\ \textbf{3} & 2 \end{smallmatrix}\right),
\left(\begin{smallmatrix} 2 & \textbf{3} \\ 4 & 1 \end{smallmatrix}\right),
\left(\begin{smallmatrix} 2 & 4 \\ \textbf{3} & 1 \end{smallmatrix}\right),
\left(\begin{smallmatrix} \textbf{3} & 1 \\ 2 & 4 \end{smallmatrix}\right),
\left(\begin{smallmatrix} \textbf{3} & 2 \\ 1 & 4 \end{smallmatrix}\right),
\left(\begin{smallmatrix} 4 & 1 \\ 2 & \textbf{3} \end{smallmatrix}\right),
\left(\begin{smallmatrix} 4 & 2 \\ 1 & \textbf{3} \end{smallmatrix}\right)\Big\}.
\\[.5ex]
& \text{(both lowest and both highest ranks on a diagonal)}
\end{array}
\ea
It is easily verified from \eqref{types}, see the ranks printed in bold font, that the type of the SOP $\mpi = \left(\begin{smallmatrix} r_1 & r_2 \\ r_3 & r_4 \end{smallmatrix}\right)$ is recognized as that rank number which shares a diagonal with the rank~4. As indicated in parentheses in \eqref{types}, the different types are well interpretable; see \citet{Bandt.Wittfeld_2023}. Under the \iid\ assumptions, the types are again uniformly distributed, namely with probability vector $\fp=(\tfrac{1}{3},\tfrac{1}{3},\tfrac{1}{3})^\top$. For their relative frequencies~$\widehat{\fp}$, closed-form normal asymptotics have been derived by \citet{Weiss.Kim_2024}. 

\smallskip
To analyze for spatial dependence, the following pairs of statistics based on~$\widehat{\fp}$ have been proposed by \citet{Bandt.Wittfeld_2023}:
\ba
\label{StatType}
\begin{array}{l}
\widehat{\tau}\ =\ \widehat{p}_1 - 1/3
\quad\text{and}\quad
\widehat{\kappa}\ =\ \widehat{p}_2 - \widehat{p}_3,
\\[2ex]
\widetilde{\tau}\ =\ \widehat{p}_3 - 1/3
\quad\text{and}\quad
\widetilde{\kappa}\ =\ \widehat{p}_1 - \widehat{p}_2.
\end{array}
\ea
Again, asymptotic normality holds, and deviations from zero are indicators of spatial dependence. While \citet{Bandt.Wittfeld_2023} recommend~$\widehat{\tau}$ and~$\widehat{\kappa}$ for applications in image analysis, \citet{Weiss.Kim_2024} found the $\widetilde{\tau}$-test to show the best power with respect to various spatial data generating processes (DGPs). For their simulation study, they also considered the spatial ACF $\rho(\fh)=\cor[Y_{\fsi}, Y_{\fsi-\fhi}]$ as a competitor, where they focused on the spatial lag $\fh=\1$ in accordance to the choice $\fd=\1$ for the delay parameters. More precisely, they applied the corresponding sample version $\hat\rho(\1) = \big(\sum_{\fsi} (Y_{\fsi}-\overline{Y})(Y_{\fsi-\onei}-\overline{Y})\big) \big/ \big(\sum_{\fsi} (Y_{\fsi}-\overline{Y})^2\big)$ with~$\overline{Y}$ denoting the sample mean, see \citet{Meyer.etal_2017} for the asymptotics. While the $\hat\rho(\1)$-test is well-suited for uncovering linear dependence in $(Y_{\fsi})_{\bbz^2}$, as it is caused by unilateral spatial autoregressive (SAR) processes, SOP-based tests from \eqref{StatType} turned out {\color{red}to} be advantageous for nonlinear or bilateral DGPs, and for unilateral linear processes in the presence of outliers or zero inflation \citep[see][]{Weiss.Kim_2024}. 

\begin{figure}[t]
\centering\footnotesize
\begin{tabular}{l@{\qquad}l}
(a)\hspace{-3ex}$\begin{array}{r|ccccc}
\toprule
s_1\setminus s_2 & 0 & 1 & 2 & 3 & 4 \\
\midrule
0 & 0.0598 & 0.0591 & 0.0587 & 0.0582 & 0.0576 \\
1 & 0.0600 & 0.0597 & 0.0590 & 0.0583 & 0.0581 \\
\cline{3-4}
2 & 0.0602 & \multicolumn{1}{|c}{0.0596} & \multicolumn{1}{c|}{0.0594} & 0.0581 & 0.0570 \\
3 & 0.0598 & \multicolumn{1}{|c}{0.0596} & \multicolumn{1}{c|}{0.0589} & 0.0585 & 0.0571 \\
\cline{3-4}
4 & 0.0600 & 0.0593 & 0.0587 & 0.0584 & 0.0569 \\
\bottomrule
\end{array}$
&
(b)\begin{tabular}{@{}l}
\includegraphics[viewport=30 65 265 295, clip=, scale=0.45]{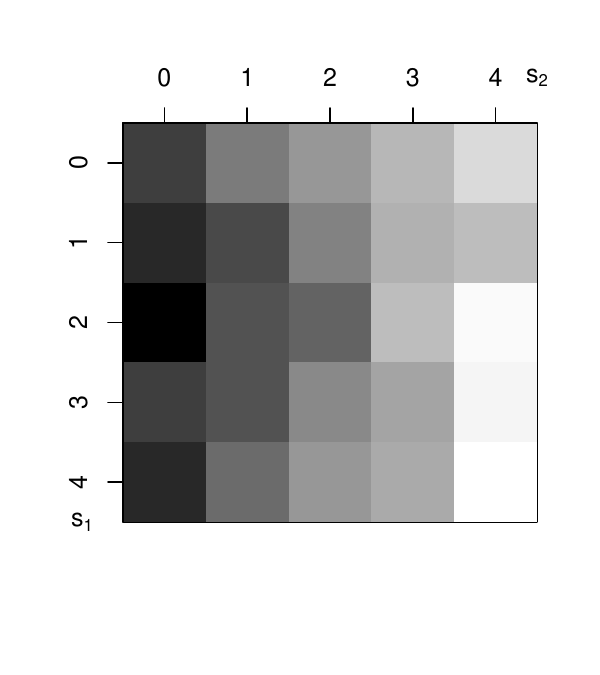} \\
\end{tabular}
\\
\\[-2ex]
(c)$\begin{array}{r|cccc}
\toprule
s_1\setminus s_2 & 1 & 2 & 3 & 4 \\
\midrule
1 & \left(\begin{smallmatrix} 3 & 1 \\ 4 & 2 \end{smallmatrix}\right) & \left(\begin{smallmatrix} 3 & 1 \\ 4 & 2 \end{smallmatrix}\right) & \left(\begin{smallmatrix} 3 & 1 \\ 4 & 2 \end{smallmatrix}\right) & \left(\begin{smallmatrix} 3 & 1 \\ 4 & 2 \end{smallmatrix}\right) \\[1ex]
2 & \left(\begin{smallmatrix} 3 & 2 \\ 4 & 1 \end{smallmatrix}\right) & \left(\begin{smallmatrix} 4 & 1 \\ 3 & 2 \end{smallmatrix}\right) & \left(\begin{smallmatrix} 3 & 2 \\ 4 & 1 \end{smallmatrix}\right) & \left(\begin{smallmatrix} 4 & 2 \\ 3 & 1 \end{smallmatrix}\right) \\[1ex]
3 & \left(\begin{smallmatrix} 4 & 1 \\ 3 & 2 \end{smallmatrix}\right) & \fbox{$\left(\begin{smallmatrix} 3 & 2 \\ 4 & 1 \end{smallmatrix}\right)$} & \left(\begin{smallmatrix} 4 & 1 \\ 3 & 2 \end{smallmatrix}\right) & \left(\begin{smallmatrix} 3 & 1 \\ 4 & 2 \end{smallmatrix}\right) \\[1ex]
4 & \left(\begin{smallmatrix} 3 & 2 \\ 4 & 1 \end{smallmatrix}\right) & \left(\begin{smallmatrix} 4 & 2 \\ 3 & 1 \end{smallmatrix}\right) & \left(\begin{smallmatrix} 4 & 2 \\ 3 & 1 \end{smallmatrix}\right) & \left(\begin{smallmatrix} 4 & 2 \\ 3 & 1 \end{smallmatrix}\right) \\
\bottomrule
\end{array}$
&
(d)$\begin{array}{r|cccc}
\toprule
s_1\setminus s_2 & 1 & 2 & 3 & 4 \\
\midrule
1 & 1 & 1 & 1 & 1 \\[1ex]
2 & 2 & 2 & 2 & 1 \\[1ex]
3 & 2 & \fbox{2} & 2 & 1 \\[1ex]
4 & 2 & 1 & 1 & 1 \\
\bottomrule
\end{array}$
\end{tabular}
\caption{Bottle thickness data from Example~\ref{bspBottles} in (a), the corresponding plot in (b), computed SOPs in (c), and types in (d). Highlighted $2\times 2$-square in (a) leads to highlighted SOP and type in (c) and (d), respectively.}
\label{figBottles}
\end{figure}

\begin{bsp}
\label{bspBottles}
To illustrate the computation of SOPs and their types, let us analyze one of the data examples discussed by \citet[p.~135]{Grimshaw.etal_2013}, shown in Figure~\ref{figBottles}\,(a). For evaluating the quality of a manufactured bottle, its thickness (in inches) is measured at regular spatial locations~$\fs$ around the cylindrical surface ($5\times 5$ grid, so $m=n=4$), and the rectangular data set in~(a) results from unwrapping the surface. A plot of the data, where increasing darkness expresses increasing thickness, is shown in Figure~\ref{figBottles}\,(b). 
The SOPs corresponding to the $2\times 2$-squares at locations $\fs\in\{1,\ldots,4\}^2$ (bottom-right corner of extracted squares) are shown in~(c), and the resulting types in~(d). The highlighted $2\times 2$-square at $\fs=(3,2)$ is one out of three cases where a tie is observed (twice the value ``0.0596''), which is caused by the limited measurement precision (units of 0.0001\, inch). Definition {\color{red}in} \eqref{ordpatternrank} accounts for the tie by assigning the lower rank to the first occurrence of ``0.0596''.

For an ideal bottle, we expect a unique thickness in the mean with only random disturbances for different measurement points, such that it would satisfy the null hypothesis of \iid\ measurements. However, the type frequencies implied by Figure~\ref{figBottles}\,(d), $\widehat{\fp} = (0.5625, 0.4375, 0.0000)^\top$, deviate considerably from the probability vector $\fp=(\tfrac{1}{3},\tfrac{1}{3},\tfrac{1}{3})^\top$ as expected under the null hypothesis of spatial independence. In particular, type~3 is never observed. The statistics in \eqref{StatType} take the values $\widehat{\tau}\approx 0.229$ ($\pm 0.239$), $\widehat{\kappa}\approx 0.438$ ($\pm 0.424$), $\widetilde{\tau}\approx -0.333$ ($\pm 0.248$), and $\widetilde{\kappa}\approx 0.125$ ($\pm 0.410$), respectively, where the critical values ({\color{red}at} 5\% level) shown in parentheses are computed according to \citet{Weiss.Kim_2024}. So~$\widehat{\kappa}$ and~$\widetilde{\tau}$ indicate significant spatial dependence, which seems plausible as these two statistics depend on~$\widehat{p}_3$. By contrast, $\hat\rho(\1)\approx 0.301$ does not violate its critical values $\pm 0.392$, with the latter being computed according to \citet{Meyer.etal_2017}. Certainly, any of these test decisions should be interpreted cautiously because the sample size is very small, so the validity of asymptotic approximations is questionable. However, spatial dependence is also supported by Figure~\ref{figBottles}\,(b), where a kind of decreasing trend from left to right is visible. Note that such visual representations are especially useful as they provide fault diagnostic information in practice.
\end{bsp}
We conclude this section with a discussion of the problem of ties. Up to now, we assumed that the rectangular data sets $(y_{\fsi})$ originate from continuously distributed RVs $(Y_{\fsi})_{\bbz^2}$ such that ties have probability zero. 
Nevertheless, ties might be observed in practice due to limited measurement precision, as demonstrated by Example~\ref{bspBottles}, yet usually with negligible frequency. However, things change if the rectangular data originates from discrete RVs, such as count RVs with the range being contained in the set of nonnegative integers, $\bbn_0=\{0,1,\ldots\}$. 
To distinguish the discrete integer case from the continuous real-valued one, let us use the letters $X,x$ instead of $Y,y$ for denoting the RVs and data, respectively (although we restrict to $X,x$ being integer-valued, the case of ordinal RVs and data is covered as well because the order information required for SOP-computation is already contained in the integer rank counts corresponding to the ordinal categories, see \citet{Weiss.Schnurr_2024} for details). For such discrete RVs $(X_{\fsi})_{\bbz^2}$, ties occur with positive probability, and the frequency of ties in given data $(x_{\fsi})$ might be quite large (also see \citet{Weiss.Schnurr_2024, Weiss.Kim_2024} for related discussions). Then, it is not recommended anymore to handle the ties according to definition \eqref{ordpatternrank} since this might cause a notable bias in estimating the true distribution of SOPs and types. 

\smallskip
However, it is still possible to adapt the aforementioned SOP~methodology to the integer case, namely by using a randomization approach \citep[``jittering'', see][]{Machado.SantosSilva_2005} as proposed by \citet{Weiss.Kim_2024}. More precisely, before computing the SOPs and types, we first add \iid\ uniform noise $(U_{\fsi})_{\bbz^2} \sim \unif(0,1)$ to the discrete integer RVs $(X_{\fsi})_{\bbz^2}$, where $\unif(0,1)$ denotes the uniform distribution in the open interval $(0;1)$. Then, the resulting
\ba
\label{noise}
Y_{\fsi}\ :=\ X_{\fsi}+U_{\fsi}
\quad\text{with the noise }
(U_{\fti})_{\bbz^2} \text{ being \iid\ } \unif(0,1)
\ea
are continuously distributed RVs such that ties occur with probability zero. Moreover, if $(X_{\fsi})_{\bbz^2}$ are even \iid, then the SOPs and types computed from $(Y_{\fsi})_{\bbz^2}$ are again discrete uniformly distributed such that the tests based on \eqref{StatType} are still of nonparametric nature. Furthermore, the added noise only affects the ranks within the ties, whereas $X_{\fri} < X_{\fsi}$ necessarily implies that also $Y_{\fri} < Y_{\fsi}$, \ie strict orders in $(X_{\fsi})_{\bbz^2}$ are always preserved. 
An unavoidable drawback of this randomization approach, however, is given by the fact that the actual outcomes of the SOP analysis, \eg computed values of the statistics~$\widetilde{\tau}$, depend on the chosen noise. This issue is discussed in more detail later in Section~\ref{Real-Data Application}, and a possible alternative solution is proposed for future research in Section~\ref{Conclusions}.

\section{Nonparametric Control Charts for Spatial Dependence}
\label{Control Charts for Spatial Dependence}
In what follows, we are concerned with monitoring a \emph{stream} of rectangular data sets, as it arises, for example, in the SPM applications surveyed in Section~\ref{Introduction}. Formally, the DGP is given by $\big[(Y_{\fsi}^{(t)})_{\fsi}\big]_t$, where $t\in\bbn$ is the inspection time, and the spatial locations are $\fs\in \{0,\ldots,m\}\times \{0,\ldots,n\}$ with $m,n\in\bbn$ being constant over time~$t$. In the case of integer-valued RVs~$X_{\fsi}^{(t)}$, we first transform them into real-valued RVs~$Y_{\fsi}^{(t)}$ by adding uniform noise according to \eqref{noise}. 

Let us now discuss the possible in-control (IC) assumptions that would be suitable for the intended SPM procedures. The basic requirement for any IC-model is that the rectangular sets~$(Y_{\fsi}^{(t)})_{\fsi}$ are generated independently of each other for different times~$t$, but by a unique stationary spatial process (random field). This spatial process is specified by the chosen IC-model, and it implies some (unique) joint distribution for the RVs within the rectangular sets~$(Y_{\fsi}^{(t)})_{\fsi}$ at each time~$t$. Given such an IC-assumption, we declare the true DGP as being out-of-control (OOC) if it violates the IC-model.

It is important to note that the considered IC-models may be quite different for different applications. For the simulation study described in Section~\ref{Simulation Study} as well as for the data applications in Sections~\ref{Rainfall data} and~\ref{War related fires in Ukraine}, we consider the following special case:
\begin{description}
	\item[(IC-iid)] For each $t\in\bbn$, $(Y_{\fsi}^{(t)})_{\fsi}$ is \iid\ with continuous distribution.
\end{description}
For this particular IC-model, any rectangular set~$(Y_{\fsi}^{(t)})_{\fsi}$ exhibiting spatial dependence constitutes an OOC-situation and should thus be detected by our SPM procedures. In Section~\ref{Textile images}, by contrast, the chosen IC-model allows for a certain form of spatial dependence while other forms of dependence are classified as OOC.

\smallskip
For process monitoring, it is suggested to use the $t$th data set $(Y_{\fsi}^{(t)})_{\fsi}$ for the computation of SOPs and types, and to derive the corresponding vector~$\widehat{\fp}_t$ of type frequencies. Hence, altogether, the process $\big[(Y_{\fsi}^{(t)})_{\fsi}\big]_t$ is transformed into the process~$(\widehat{\fp}_t)_t$ of frequency vectors, which are then used for constructing control charts according to \eqref{SOPcharts} below. 
Note that under the assumption of ``IC-iid'', as explained in Section~\ref{Spatial Ordinal Patterns and Types}, the SOPs and types computed from $(Y_{\fsi}^{(t)})_{\fsi}$ are discrete uniformly distributed, independent of the actual distribution of the~$Y_{\fsi}^{(t)}$. In particular, the process~$(\widehat{\fp}_t)_t$ is \iid\ with $E\big[\widehat{\fp}_t\big]=(\tfrac{1}{3},\tfrac{1}{3},\tfrac{1}{3})^\top$. This shall allow us to define nonparametric control charts in this case. 

\smallskip
Inspired by \citet{Weiss.Testik_2023}, we distinguish two general approaches for process monitoring. These approaches differ in how the type frequencies are computed (the monitored statistics computed afterwards, according to \eqref{SOPcharts} are the same for both approaches):
\begin{itemize}
	\item If the $t$th control statistic \eqref{SOPcharts} is computed directly from the raw relative frequencies~$\widehat{\fp}_t$, then we finally obtain a kind of Shewhart control chart \citep[see][]{Montgomery_2009}. It has the property of being memory-less because the previous~$\widehat{\fp}_u$, $u<t$, do not affect~$\widehat{\fp}_t$ and thus the $t$th control statistic from \eqref{SOPcharts}.
	\item To later obtain a control chart with an inherent memory, we use a (multivariate) exponentially weighted moving average (EWMA) approach \citep[see][]{Roberts_1959} for computing ``smoothed'' type frequencies. Having specified the smoothing parameter $\lambda\in (0;1)$ as well as the initial probability vector~$\fp_0$ for the types, we compute the smoothed frequency vectors via
\ba
\label{EWMA}
\hat{\fp}_0^{(\lambda)}\ =\ \fp_0,\qquad
\hat{\fp}_t^{(\lambda)}\ =\ \lambda\,\widehat{\fp}_t + (1-\lambda)\,\hat{\fp}_{t-1}^{(\lambda)} \quad\text{for } t=1,2,\ldots
\ea
Afterwards, $\hat{\fp}_t^{(\lambda)}$ is used for calculating the $t$th control statistic according to \eqref{SOPcharts}. Since the final control chart relies on EWMA-smoothed relative frequencies (instead of the raw frequencies used for the previous Shewhart approach), we simply refer to it as an ``EWMA chart'', although it is not identical to the original EWMA chart of \citet{Roberts_1959}.
\end{itemize}
The EWMA's~$\lambda$ controls the strength of the inherent memory, with increasing memory for decreasing~$\lambda$, see \citet{Roberts_1959}. Note that the boundary case $\lambda\to 1$ leads to the aforementioned Shewhart approach. Hence, it is sufficient to introduce the subsequent control charts with respect to the sequence $(\hat{\fp}_t^{(\lambda)})$ computed via \eqref{EWMA}, where $(\hat{\fp}_t^{(1)})$ represents the Shewhart version of the control chart. The choice of~$\fp_0$ (only relevant if $\lambda<1$) depends on the actual IC-assumption; under the assumption of ``IC-iid'', we choose $\fp_0=\big(\tfrac{1}{3},\tfrac{1}{3},\tfrac{1}{3}\big)^\top$. 

\smallskip
After having described our approaches for computing the type frequencies, we now propose four novel classes of control charts, which are obtained by plugging in $\hat{\fp}_t^{(\lambda)}$ into one of the four spatial-dependence statistics defined in \eqref{StatType}, and plot the resulting sequence of statistics against symmetric two-sided control limits (CLs): 
\ba
\label{SOPcharts}
\begin{array}{lll}
\widehat{\tau}\text{-chart:} &
\text{plot } \widehat{\tau}_t^{(\lambda)} = \widehat{p}_{t,1}^{(\lambda)} - \tfrac{1}{3}, & \text{trigger alarm if } \big|\widehat{\tau}_t^{(\lambda)}\big| > l_{\widehat{\tau}, \lambda};
\\[2ex]
\widehat{\kappa}\text{-chart:} &
\text{plot } \widehat{\kappa}_t^{(\lambda)} = \widehat{p}_{t,2}^{(\lambda)} - \widehat{p}_{t,3}^{(\lambda)}, & \text{trigger alarm if } \big|\widehat{\kappa}_t^{(\lambda)}\big| > l_{\widehat{\kappa}, \lambda};
\\[2ex]
\widetilde{\tau}\text{-chart:} &
\text{plot } \widetilde{\tau}_t^{(\lambda)} = \widehat{p}_{t,3}^{(\lambda)} - \tfrac{1}{3}, & \text{trigger alarm if } \big|\widetilde{\tau}_t^{(\lambda)}\big| > l_{\widetilde{\tau}, \lambda};
\\[2ex]
\widetilde{\kappa}\text{-chart:} &
\text{plot } \widetilde{\kappa}_t^{(\lambda)} = \widehat{p}_{t,1}^{(\lambda)} - \widehat{p}_{t,2}^{(\lambda)}, & \text{trigger alarm if } \big|\widetilde{\kappa}_t^{(\lambda)}\big| > l_{\widetilde{\kappa}, \lambda}.
\end{array}
\ea
Chart design and performance evaluation shall be done based on ARL considerations. Specifically, we focus on the so-called ``zero-state ARL'' \citep[see][]{Knoth_2006}, which expresses the mean number of statistics plotted on a chart, right from the beginning of process monitoring at $t=1$ until the first alarm. Furthermore, if the process deviates from the specified IC-model (\ie if the process has changed into an OOC-state), then the zero-state ARL also assumes that this change already happened at $t=1$. There exist further ARL concepts in the literature where a delayed change point is assumed, see \citet{Knoth_2006}. However, for a Shewhart chart (case $\lambda=1$), they all lead to the same ARL value, and for conventional EWMA charts such as defined in \eqref{EWMA}, the differences between the different ARL values are usually negligible for practice (also see the study by \citet{Weiss.Testik_2023} on related OP-based EWMA charts). Thus, we focus on the zero-state ARL, and use it in a two-fold manner:
\begin{itemize}
	\item If computed under IC-conditions, the resulting IC-ARL is used for chart design, \ie the CL~$l_{\cdot, \lambda}$ is chosen such that the considered chart's IC-ARL is close to a specified target value ARL$_0$ (we use the common choice ARL$_0=370$ for illustration).
	\item If computed under OOC-conditions, the resulting OOC-ARL expresses the performance of the considered chart in detecting the process change.
\end{itemize}
The ARL values can be approximated based on simulations. For a given DGP scenario and chart design, one simulates $R$~times the DGP and determines the time until the first alarm (\ie the run length). The sample mean across these $R$ run lengths provides an estimate of the true ARL. Recall that for the special case of assumption ``IC-iid'', the distribution of the process~$(\widehat{\fp}_t)_t$ does not depend on the distribution of the~$Y_{\fsi}^{(t)}$. Hence, our charts are nonparametric/distribution-free. Therefore, for ARL simulation, one can use any continuous distribution for the~$Y_{\fsi}^{(t)}$, \eg a standard normal distribution. 

\begin{bem}
\label{bemACFchart}
A natural competitor of our SOP-based control charts given in \eqref{SOPcharts} for spatial dependence is an analogous control chart based on the spatial ACF (recall the discussion after \eqref{StatType}). 
Here, the idea is to compute $\hat\rho(\1)$ from each data set $(Y_{\fsi}^{(t)})_{\fsi}$, and to use the obtained sequence of ACF values, say $(\widehat{\rho}_t)$, for process monitoring. In analogy to \eqref{EWMA}, this can be combined with an EWMA approach, as
\ba
\label{ACF-EWMA}
\widehat{\rho}_0^{(\lambda)}\ =\ \rho_0,\qquad
\widehat{\rho}_t^{(\lambda)}\ =\ \lambda\,\widehat{\rho}_t + (1-\lambda)\,\widehat{\rho}_{t-1}^{(\lambda)} \quad\text{for } t=1,2,\ldots,
\ea
and an alarm is triggered at time~$t$ if $\big|\widehat{\rho}_t^{(\lambda)}\big| > l_{\widehat{\rho}, \lambda}$. Under the assumption {\color{red}of} ``IC-iid'', the initial value is chosen as $\rho_0=0$. A fundamental difference is given by the fact that the distribution of $\hat\rho(\1)$ depends on one of {\color{red}the} $(Y_{\fsi}^{(t)})_{\fsi}$. So the $\widehat{\rho}$-chart is of a parametric nature, and chart design requires a full specification of the in-control model. On the other hand, $\hat\rho(\1)$ can also be computed from integer data $(X_{\fsi}^{(t)})_{\fsi}$, \ie the $\widehat{\rho}$-chart can be directly applied to $(X_{\fsi}^{(t)})_{\fsi}$ without prior randomization.
\end{bem}

\numberwithin{satz}{subsection}

\section{Simulation Study}
\label{Simulation Study}
To analyze the ARL performance of our novel control charts proposed in Section~\ref{Control Charts for Spatial Dependence}, we did a comprehensive simulation study covering various IC- and OOC-scenarios (which is later continued in Section~\ref{Performance Analyses}). To ensure a high quality of approximation, we used $R=10^6$ replications per scenario to determine the control limits and $R=10^5$ replications to approximate the IC and OOC-ARLs for performance analyses. The selected DGPs and sample sizes $m,n$ are similar to those used by \citet{Weiss.Kim_2024} in their research on SOP-based hypothesis testing, as this allows us to compare the ARL performance of our ``sequential tests'' with the power performance of their classical hypothesis tests. 
Since ARL~simulations are much more time-consuming than power simulations, and to also make the continuous and discrete cases more balanced, we restrict to count distributions with mean~$5$ (``medium counts''). 

\begin{table}[!htb]
    \centering
    \small
    \tabcolsep3.1pt
    \caption{IC-designs of SOP-EWMA charts in \eqref{SOPcharts} and $\widehat{\rho}$-chart in \eqref{ACF-EWMA} for different $m, n$ combinations, smoothing parameters $\lambda$, and target ARL$_0\approx 370$; simulated with $10^6$ replications. SOP-EWMA charts use $\norm(0,1)$ simulations, $\widehat{\rho}$-charts use either $\norm(0,1)$ or $\poi(5)$ simulations.}
    \label{tab:ICiid_design}
\resizebox{\linewidth}{!}{
    \begin{threeparttable}
    \begin{tabular}{ccccccccccccccccccc}
    \toprule
    &&\multicolumn{2}{c}{$\widehat{\tau}$-chart}
    &&\multicolumn{2}{c}{$\widehat{\kappa}$-chart}
    &&\multicolumn{2}{c}{$\widetilde{\tau}$-chart}
    &&\multicolumn{2}{c}{$\widetilde{\kappa}$-chart}
    &&\multicolumn{2}{c}{$\widehat{\rho}$--$\norm(0,1)$}
    &&\multicolumn{2}{c}{$\widehat{\rho}$--$\poi(5)$}\\
    \cmidrule{3-4}\cmidrule{6-7}\cmidrule{9-10}\cmidrule{12-13}\cmidrule{15-16}\cmidrule{18-19}
    $m,n$ & $\lambda$ 
              &  $l_{\widehat{\tau}, \lambda}$ & ARL & 
              & $l_{\widehat{\kappa}, \lambda}$ & ARL & 
              & $l_{\widetilde{\tau}, \lambda}$ & ARL & 
              & $l_{\widetilde{\kappa}, \lambda}$ & ARL &
              & $l_{\widehat{\rho}, \lambda}$ & ARL &
              & $l_{\widehat{\rho}, \lambda}$ & ARL\\
\midrule
$1,1$ & $0.05$ & $0.18549$ & $369.5$ &  & $0.32399$ & $370.3$ &  & $0.18555$ & $370.8$ &  & $0.32398$ & $370.3$ &  & $0.15106$ & $370.4$ &  & $0.15059$ & $369.7$\\
$\text{}$ & $0.10$ & $0.2809$ & $370.7$ &  & $0.49178$ & $369.2$ &  & $0.28085$ & $369.9$ &  & $0.49177$ & $369.6$ &  & $0.19448$ & $369.2$ &  & $0.19378$ & $370.2$\\
$\text{}$ & $0.25$ & $0.48164$ & $370.4$ &  & $0.78644$ & $370.3$ &  & $0.48162$ & $369.8$ &  & $0.7864$ & $370.8$ &  & $0.28253$ & $370.1$ &  & $0.28117$ & $369.9$\\
    \midrule
$10,10$ & $0.05$ & $0.01962$ & $370.2$ &  & $0.03491$ & $370.2$ &  & $0.02042$ & $370.3$ &  & $0.03351$ & $369.8$ &  & $0.03537$ & $370.5$ &  & $0.03533$ & $370.6$\\
$\text{}$ & $0.10$ & $0.03049$ & $369.6$ &  & $0.05426$ & $369.7$ &  & $0.03174$ & $369.8$ &  & $0.05209$ & $370.1$ &  & $0.05313$ & $369.8$ &  & $0.05305$ & $369.5$\\
$\text{}$ & $0.25$ & $0.05386$ & $370.1$ &  & $0.09587$ & $370.0$ &  & $0.05611$ & $369.9$ &  & $0.09200$ & $370.0$ &  & $0.09137$ & $370.5$ &  & $0.09119$ & $369.8$\\
    \midrule
$15,15$ & $0.05$ & $0.0131$ & $369.2$ &  & $0.02333$ & $370.4$ &  & $0.01365$ & $369.4$ &  & $0.02237$ & $369.9$ &  & $0.02425$ & $370.6$ &  & $0.02423$ & $370.5$\\
$\text{}$ & $0.10$ & $0.02036$ & $369.7$ &  & $0.03626$ & $369.5$ &  & $0.02122$ & $369.7$ &  & $0.03476$ & $370.0$ &  & $0.03701$ & $369.7$ &  & $0.03698$ & $369.5$\\
$\text{}$ & $0.25$ & $0.03597$ & $369.9$ &  & $0.06408$ & $369.7$ &  & $0.03751$ & $370.6$ &  & $0.06142$ & $369.9$ &  & $0.06450$ & $370.4$ &  & $0.06443$ & $370.3$\\
    \midrule
$25,25$ & $0.05$ & $0.00787$ & $369.8$ &  & $0.01402$ & $369.9$ &  & $0.00821$ & $370.4$ &  & $0.01343$ & $370.3$ &  & $0.01497$ & $369.8$ &  & $0.01496$ & $370.5$\\
$\text{}$ & $0.10$ & $0.01223$ & $369.9$ &  & $0.0218$ & $370.0$ &  & $0.01276$ & $370.5$ &  & $0.02087$ & $369.0$ &  & $0.02310$ & $370.4$ &  & $0.02309$ & $369.5$\\
$\text{}$ & $0.25$ & $0.02162$ & $370.6$ &  & $0.03853$ & $370.4$ &  & $0.02255$ & $370.1$ &  & $0.0369$ & $370.7$ &  & $0.04059$ & $370.2$ &  & $0.04058$ & $370.8$\\
    \midrule
$40,25$ & $0.05$ & $0.00622$ & $370.1$ &  & $0.01109$ & $369.9$ &  & $0.00649$ & $369.9$ &  & $0.01062$ & $370.5$ &  & $0.01195$ & $370.5$ &  & $0.01194$ & $370.2$\\
$\text{}$ & $0.10$ & $0.00967$ & $369.3$ &  & $0.01724$ & $370.0$ &  & $0.01009$ & $370.6$ &  & $0.01650$ & $369.2$ &  & $0.01847$ & $369.7$ &  & $0.01847$ & $370.5$\\
$\text{}$ & $0.25$ & $0.01710$ & $369.9$ &  & $0.03048$ & $370.7$ &  & $0.01784$ & $370.5$ &  & $0.02918$ & $370.4$ &  & $0.03254$ & $371.0$ &  & $0.03252$ & $369.8$\\
    \bottomrule
    \end{tabular}
    \begin{tablenotes}
    NOTE: Designs of SOP charts ($\widehat{\tau}, \widehat{\kappa}, \widetilde{\tau}, \widetilde{\kappa}$) apply to any continuously distributed process. The table's largest standard error for the ARL is 0.4. \end{tablenotes} \end{threeparttable}}
\end{table}

\subsection{In-control ARL Performance}
\label{In-control ARL Performance}
Let us start by investigating the chart design and ARL performance under the assumption {\color{red}of} ``IC-iid''. As our novel SOP-EWMA charts in \eqref{SOPcharts} are nonparametric, it suffices to focus on one particular kind of continuous distribution for the~$Y_{\fsi}^{(t)}$ for computing IC-ARLs and for determining the chart design---an obvious choice is the standard normal distribution $\norm(0,1)$. Here, chart design means computing the CL~parameter $l_{\cdot, \lambda}$ for the considered kind of control chart, for the selected value of~$\lambda$, and the chosen target ARL$_0$. Following the common practice in SPM, we focus on the target ARL$_0\approx 370$. For the smoothing parameter~$\lambda$, we choose the levels $\lambda\in\{0.25, 0.10, 0.05\}$ like in \citet{Weiss.Testik_2023}, which corresponds to an increasing memory of the EWMA charts, ranging from a ``medium'' up to a ``very strong'' level. In our OOC-analyses in Section~\ref{Out-of-control ARL Performance}, however, we only focus on the ``compromise choice'' $\lambda=0.10$ (strong memory) to keep the amount of simulations manageable. Furthermore, we did not consider larger values of~$\lambda$ (in particular, we did not consider the Shewhart case $\lambda=1$) because the sample sizes $m,n$ are often rather small in quality applications (such as the bottle thickness in Example~\ref{bspBottles}) such that the frequency vectors $\widehat{\fp}_t$ used for \eqref{EWMA} are highly discrete. In our simulations, we focus on the sample sizes $(m,n)\in\{(10,10), (15,15), (25,25), (40,25)\}$ like in \citet{Weiss.Kim_2024}, but we also briefly discuss the case $m=n=1$ later in Remark~\ref{bemEWMAm1n1}. While the IC-simulations for the SOP-EWMA charts \eqref{SOPcharts} can be restricted to \iid\ $Y_{\fsi}^{(t)}\sim\norm(0,1)$ due to their nonparametric nature, the competing $\widehat{\rho}$-chart relies on a parametric statistic such that an individual chart design is necessary for each specific IC-marginal distribution. Therefore, in view of the OOC-scenarios considered in Section~\ref{Out-of-control ARL Performance} below, the $\widehat{\rho}$-chart was also designed for \iid\ Poisson data with mean~$5$, \ie $Y_{\fsi}^{(t)}\sim\poi(5)$. The obtained results are summarized in Table~\ref{tab:ICiid_design}. 

\begin{table}[!htb]
    \centering
    \caption{IC-ARLs of EWMA $\widehat{\rho}$-charts with $\lambda=0.1$ and target ARL$_0\approx370$, for different $m, n$ combinations and different marginal distributions, but using the $\norm(0,1)$-CLs from Table~\ref{tab:ICiid_design}; simulated with $10^5$ replications.}
    \label{tab:DGP0}
\resizebox{\linewidth}{!}{
    \begin{threeparttable}
    \renewcommand{\arraystretch}{.9}
    \tabcolsep7.6pt
    \begin{tabular}{ccccccccccc}
    \toprule
$m,n$ & t$(2)$ & BPoi & Wei & Exp & Poi & Lap & SkN & U & MixN & Ber \\
\midrule
$10,10$ & $590.76$ & $514.94$ & $465.06$ & $464.82$ & $410.59$ & $392.42$ & $389.60$ & $357.43$ & $352.20$ & $349.14$ \\
$15,15$ & $525.78$ & $429.13$ & $418.39$ & $417.40$ & $391.38$ & $383.36$ & $377.54$ & $364.85$ & $360.10$ & $360.17$ \\
$25,25$ & $457.68$ & $392.11$ & $390.43$ & $389.41$ & $380.83$ & $374.48$ & $374.69$ & $368.64$ & $365.42$ & $366.65$ \\
$40,25$ & $433.12$ & $382.04$ & $380.19$ & $383.35$ & $376.12$ & $373.13$ & $371.13$ & $368.55$ & $367.40$ & $366.57$ \\
    \bottomrule
    \end{tabular}
    \begin{tablenotes}
    NOTE: The table's largest standard error for the ARL is 1.84. Abbreviations used for column headings:
    BPoi = Ber$(0.2)$ $\cdot$ Poi$(5)$; Wei = Weibull$(1,1.5)$, Exp = Exp$(1)$, Poi = Poi$(0.5)$, Lap = Laplace$(0,1)$, SkN = SkewN$(0,1,10)$, U = U$(0,1)$, MixN = 0.5\,N$(-9,1)$+0.5\,N$(9,1)$, Ber = Bernoulli$(0.5)$. 
    \end{tablenotes}
\end{threeparttable}}
\end{table}

The designs of the SOP-EWMA charts in Table~\ref{tab:ICiid_design} can be immediately used together with any continuously distributed DGP $(Y_{\fsi}^{(t)})$ (or randomized integer DGP $(X_{\fsi}^{(t)})$). In particular, no prior Phase-I analysis is necessary, \ie no model fitting to historical IC-data. Quite the contrary, the SOP-EWMA charts could even be applied within Phase~I for a retrospective data analysis. For the parametric $\widehat{\rho}$-chart, tailor-made chart designs after prior Phase-I analysis are necessary. Comparing the CLs for $\norm(0,1)$ and $\poi(5)$ in Table~\ref{tab:ICiid_design}, one might certainly get the impression that their differences are negligible. Moreover, we observed in simulation experiments that the $\norm(0,1)$-CLs are somehow robust: if the shape of the true IC-distribution is close to normal, then the IC-ARL obtained by using the $\norm(0,1)$-CLs is reasonably close to the target ARL$_0\approx 370$. The same holds if the sample sizes $m,n$ are sufficiently large, as the central limit theorem implies an asymptotic normal distribution for $\hat{\rho}(\1)$. 
However, if the shape clearly deviates from normality and the sample sizes are small, using the $\norm(0,1)$-CLs might result in a misleading IC-ARL performance. This is illustrated by Table~\ref{tab:DGP0}, where various marginal distributions are considered for the \iid\ DGP. For the smallest sample size considered, $(m,n)=(10,10)$, the true IC-ARLs vary considerably, from about~349.14 up to~590.76. Analogous variations, though dampened, are observed for larger sample sizes $(m,n)$. In particular, depending on the actual marginal distribution, it might happen that the EWMA $\widehat{\rho}$-chart produces too many false alarms or is very conservative, where the latter causes a deterioration in OOC-performance. More precisely, for skewed distributions or distributions with heavier tails than the normal, we observe moderately to severely increased IC-ARLs, whereas platykurtic distributions (we selected examples from \citet[Table~1]{Westfall_2014}) led to a decrease of the IC-ARL. Altogether, it becomes clear that the pronounced parametric nature of the EWMA $\widehat{\rho}$-chart requires a careful Phase-I analysis in practice, where an appropriate marginal distribution must be identified and fitted to the historical IC data. By contrast, as highlighted before, our novel nonparametric SOP-EWMA charts can be applied immediately from the beginning of process monitoring. 

\begin{bem}
\label{bemEWMAm1n1}
It should be noted that the SOP-EWMA charts are well defined already for $m=n=1$ (provided that~$\lambda$ is clearly smaller than~1 to ensure a reasonable memory), \ie if only one SOP is computed per rectangle. Such an application is described by \citet{Barton.Gonzalez-Barreto_1996}, where a manufacturing process of multilayer chip capacitors is monitored. 
There, for each manufactured clay flat, the so-called registration error is determined for four pads (one from each quadrant), leading to a $2\times 2$ measurement for each clay flat. 
This is illustrated by Table~\ref{tabBarton96}, where the $2\times 2$ registration errors for the first six clay flats from Table~1 in \citet{Barton.Gonzalez-Barreto_1996} are shown in the first row, the corresponding SOPs in the second row, their types in the third row, the raw frequency vectors~$\widehat{\fp}_t$ in the fourth row, and their EWMA-smoothed counterparts $\hat{\fp}_t^{(\lambda)}$ in the last row. While the~$\widehat{\fp}_t$ can take one out of only three possible values and are, thus, not suitable for defining a (Shewhart) control chart, the~$\hat{\fp}_t^{(\lambda)}$ can be used for the control charts in \eqref{SOPcharts} despite the small sample size $m=n=1$---at least theoretically. While we were able to determine the IC-designs, see Table~\ref{tab:ICiid_design}, the OOC-simulations showed that the OOC-ARLs decrease only very slowly for increasing deviation from the ``IC-iid'' assumption, such that one has to expect large detection delays in practice. This is plausible in view of the low information content of a single $2\times 2$-SOP, and we decided not to further analyze the case $m=n=1$.
\end{bem}

\begin{table}[t]
\centering
\caption{Clay flats at times $t=1,\ldots,6$ according to Table~1 in \citet{Barton.Gonzalez-Barreto_1996}: step-wise computation of EWMA-smoothed type frequencies $\hat{\fp}_t^{(\lambda)}$ according to \eqref{EWMA} with $\lambda=0.1$ and  $\fp_0=(\tfrac{1}{3},\tfrac{1}{3},\tfrac{1}{3})^\top$, where $(y_{\fsi}^{(t)})$ comprises the four registration errors of the $t$th clay flat.}
\label{tabBarton96}

\smallskip
\begin{tabular}{l|cccccc}
\toprule
$(y_{\fsi}^{(t)})$ &
$\left(\begin{smallmatrix} 3.30 & 3.95
\\ 5.89 & 3.20 \end{smallmatrix}\right)$
&
$\left(\begin{smallmatrix} 0.27 & 3.71
\\ 0.39 & 4.33 \end{smallmatrix}\right)$
&
$\left(\begin{smallmatrix} 3.06 & 1.66
\\ 2.93 & 2.12 \end{smallmatrix}\right)$
&
$\left(\begin{smallmatrix} 2.74 & 2.86
\\ 1.31 & 2.10 \end{smallmatrix}\right)$
&
$\left(\begin{smallmatrix} 1.36 & 3.42
\\ 2.21 & 1.80 \end{smallmatrix}\right)$
&
$\left(\begin{smallmatrix} 2.00 & 2.44
\\ 3.65 & 1.64 \end{smallmatrix}\right)$
\\
\midrule
$\mpi_t$ &
$\left(\begin{smallmatrix} 2 & 3 \\ 4 & 1 \end{smallmatrix}\right)$
&
$\left(\begin{smallmatrix} 1 & 3 \\ 2 & 4 \end{smallmatrix}\right)$
&
$\left(\begin{smallmatrix} 4 & 1 \\ 3 & 2 \end{smallmatrix}\right)$
&
$\left(\begin{smallmatrix} 3 & 4 \\ 1 & 2 \end{smallmatrix}\right)$
&
$\left(\begin{smallmatrix} 1 & 4 \\ 3 & 2 \end{smallmatrix}\right)$
&
$\left(\begin{smallmatrix} 2 & 3 \\ 4 & 1 \end{smallmatrix}\right)$
\\
\midrule
Type & 3 & 1 & 2 & 1 & 3 & 3 \\
\midrule
$\widehat{\fp}_t$ & $\left(\begin{smallmatrix} 0 \\ 0 \\ 1 \end{smallmatrix}\right)$ & $\left(\begin{smallmatrix} 1 \\ 0 \\ 0 \end{smallmatrix}\right)$ & $\left(\begin{smallmatrix} 0 \\ 1 \\ 0 \end{smallmatrix}\right)$ & $\left(\begin{smallmatrix} 1 \\ 0 \\ 0 \end{smallmatrix}\right)$ & $\left(\begin{smallmatrix} 0 \\ 0 \\ 1 \end{smallmatrix}\right)$ & $\left(\begin{smallmatrix} 0 \\ 0 \\ 1 \end{smallmatrix}\right)$ \\
\midrule
$\hat{\fp}_t^{(\lambda)}$ & 
$\left(\begin{smallmatrix}  0.300 \\  0.300 \\  0.400 \end{smallmatrix}\right)$ & 
$\left(\begin{smallmatrix}  0.370 \\  0.270 \\  0.360 \end{smallmatrix}\right)$ & 
$\left(\begin{smallmatrix}  0.333 \\  0.343 \\  0.324 \end{smallmatrix}\right)$ & 
$\left(\begin{smallmatrix}  0.400 \\  0.309 \\  0.292 \end{smallmatrix}\right)$ & 
$\left(\begin{smallmatrix}  0.360 \\  0.278 \\  0.362 \end{smallmatrix}\right)$ & 
$\left(\begin{smallmatrix}  0.324 \\  0.250 \\  0.426 \end{smallmatrix}\right)$
\\
\bottomrule
\end{tabular}
\end{table}

\subsection{Out-of-control ARL Performance}
\label{Out-of-control ARL Performance}
Let us now turn to an analysis of the OOC-ARL performance. 
As before, we use the $\widehat{\rho}$-chart as the competitor to our novel SOP-based charts. As recognized in Section~\ref{In-control ARL Performance}, the design of the $\widehat{\rho}$-chart is highly sensitive with respect to the actual IC-distribution. In what follows, we always assume to know the correct IC-distribution, which implies a kind of ``starting advantage'' for the $\widehat{\rho}$-chart. 
Various spatially dependent DGPs for the rectangular sets $(Y_{\fsi}^{(t)})_{\fsi}$ are considered, in analogy to the power analyses in \citet{Weiss.Kim_2024}. We begin our analyses with the most well-known case, unilateral SAR DGPs. More precisely, we consider the continuously distributed unilateral SAR$(1,1)$ process \citep[see][]{Pickard_1980} defined by
\begin{equation} \label{SAR11}
Y_{t_{1}, t_{2}}= \alpha_1\cdot Y_{{t_1}-1, t_{2}} + \alpha_2\cdot  Y_{t_{1}, t_{2}-1}+ \alpha_3\cdot  Y_{t_{1}-1, t_{2}-1} + \varepsilon_{t_{1}, t_{2}}
\end{equation}
with \iid\ innovations $\varepsilon_{t_{1}, t_{2}}\sim\norm(0,1)$ on the one hand, and its integer counterpart, the SINAR$(1,1)$ process \citep[see][]{Ghodsi.etal_2012}, on the other hand:
\begin{equation}\label{SINAR11}
X_{t_{1}, t_{2}}= \alpha_1 \circ X_{{t_1}-1, t_{2}} + \alpha_2 \circ X_{t_{1}, t_{2}-1}+ \alpha_3 \circ X_{t_{1}-1, t_{2}-1} + \epsilon_{t_{1}, t_{2}},
\end{equation}
with \iid\ innovations $\epsilon_{t_{1}, t_{2}}\sim\poi(5)$. Here, ``$\circ$'' denotes the binomial thinning operator introduced by \citet{Steutel.Harn_1979}, constituting an integer-valued substitute of the ordinary multiplication ``$\cdot$''. It is defined by requiring that $\alpha\circ X|X\sim\bin(X,\alpha)$, \ie $\alpha\circ X$ is conditionally binomially distributed with $\alpha\in [0;1)$. The obtained OOC-ARLs are summarized in Appendix \ref{sec:appendix_ooc_results} in Tables~\ref{tab:DGP3} and~\ref{tab:DGP4}, respectively, with the minimal OOC-ARLs being highlighted by bold font. Not surprisingly, for such unilateral linear DGPs, the $\widehat{\rho}$-chart performs best without exception. Nevertheless, the SOP-EWMA charts have also clearly reduced ARLs, where usually the $\widetilde{\tau}$-chart performs best (with lower OOC-ARL for increased $\alpha_1,\alpha_2$). Only for the model specification $(\alpha_1, \alpha_2, \alpha_3)=(0.2,0.2,0.5)$, where the ``diagonal'' AR parameter~$\alpha_3$ is particularly large, the $\widetilde{\tau}$-chart has a poor performance  and the $\widehat{\kappa}$-chart does best among the SOP-charts. This might be explained from the definition of~$\widetilde{\tau}$ in \eqref{StatType}, which solely focuses on type~3, where the maximal ranks occur along one of the diagonals. Type~3, however, does not distinguish if the maximal ranks occur along the main diagonal (as caused by large~$\alpha_3$) or the anti-diagonal (which rarely happens for a unilateral DGP with large~$\alpha_3$). Therefore, the $\widetilde{\tau}$-chart fails in detecting a large~$\alpha_3$. 

In practice, however, we are often confronted with more demanding situations where the DGP deviates from the simple, well-behaved first-order autoregressive process. As the first non-textbook scenario, let us consider the same SAR DGPs \eqref{SAR11} and \eqref{SINAR11} as before, but with the resulting data being contaminated by additive outliers (AOs). More precisely, 10\,\% of the generated data were randomly selected and manipulated by a further summand. In the case of the continuously distributed SAR$(1,1)$ DGP \eqref{SAR11}, we either added the fixed contamination~$+10$, or one of~$+10$ or~$-10$ with probability 0.5 each. For the discrete SINAR$(1,1)$ DGP \eqref{SINAR11}, the contamination was randomly chosen from the $\poi(25)$-distribution. The resulting OOC-ARLs are summarized in Appendix \ref{sec:appendix_ooc_results} in Tables~\ref{tab:DGP5b} and~\ref{tab:DGP6_mu5}, respectively. Except for the parametrization $(\alpha_1, \alpha_2, \alpha_3)=(0.2,0.2,0.5)$ already discussed before, the $\widehat{\rho}$-chart performs considerably worse. In addition, for the small grid $(m,n)=(10, 10)$ and $(\alpha_1, \alpha_2, \alpha_3)=(0.1,0.1,0.1)$, the OOC-ARL of the $\widehat{\rho}$-chart is considerably larger than the IC-ARL. The SOP charts, in turn, are robust against the outliers (recall that SOPs consist of ranks) and outperform the $\widehat{\rho}$-chart. More precisely, except for $(\alpha_1, \alpha_2, \alpha_3)=(0.2,0.2,0.5)$, the $\widetilde{\tau}$-chart leads to the lowest OOC-ARLs in nearly any case. 

\smallskip
In the discrete count case, we considered a further variation of the basic unilateral SINAR$(1,1)$ DGP \eqref{SINAR11}: the recursive scheme used for data generation is the same as in \eqref{SINAR11}, but the innovations~$\epsilon_{t_{1}, t_{2}}$ are not Poisson-distributed anymore and follow a zero-inflated Poisson (ZIP) distribution instead. More precisely, $\epsilon_{t_{1}, t_{2}}\sim \zip(0.9,5)$ with mean~$5$ again, but the value~$0.9$ of the zero-inflation parameter implies that at least 90\,\% of the innovations are equal to zero. In contrast, the remaining truly positive counts cause positive shocks to the DGP. Since such positive shocks are usually followed by multiple zero innovations, the binomial thinnings in the SINAR$(1,1)$ recursion lead to decaying count values. Altogether, the generated spatial dependence is combined with many zeros and decaying cascades of counts. The resulting OOC-ARLs are summarized in Table~\ref{tab:DGP7} in Appendix \ref{sec:appendix_ooc_results}. It can be seen that the $\widehat{\rho}$-chart quickly recognizes the apparent spatial dependence. However, the $\widetilde{\tau}$-chart has---with the exception $(\alpha_1, \alpha_2, \alpha_3)=(0.2,0.2,0.5)$---mostly similar or lower OOC-ARLs. Also, the $\widehat{\tau}$-chart performs comparatively well in this scenario.

Up to now, we only considered linear (unilateral) DGPs, although sometimes under ``demanding conditions'', recall Tables~\ref{tab:DGP5b}--\ref{tab:DGP7}. Next, we investigate truly nonlinear (unilateral) DGPs, namely the two kinds of quadratic moving average (QMA) process proposed by \citet{Weiss.Kim_2024}. The continuously distributed version, the unilateral SQMA$(1,1)$ process, is defined by
\begin{equation} \label{SQMA11}
Y_{t_{1}, t_{2}}= \beta_{1} \cdot \varepsilon_{{t_1}-1, t_{2}}^a + \beta_{2} \cdot  \varepsilon_{t_{1}, t_{2}-1}^b+ \beta_{3} \cdot  \varepsilon_{t_{1}-1, t_{2}-1}^c + \varepsilon_{t_{1}, t_{2}}
\end{equation}
with \iid\ innovations $\varepsilon_{t_{1}, t_{2}}\sim \norm(0,1)$. The DGP in \eqref{SQMA11} has powers $a,b,c\in\{1,2\}$, so the MA-terms are either linear or squared. In our simulations, we considered the following four combinations for $(a,b,c)$: the combination $(2,2,2)$ expresses that all MA-terms are squared, $(2,1,2)$ that the $\beta_1$- and $\beta_3$-term are squared, $(1,1,2)$ that only the $\beta_3$-term is squared, and $(2,1,1)$ that only the $\beta_1$-term is squared, also see Table~\ref{tab:DGP8} in Appendix \ref{sec:appendix_ooc_results}. The integer counterpart to \eqref{SQMA11} is the SQINMA$(1,1)$ process with \iid\ $\poi(5)$-innovations~$\epsilon_{t_{1}, t_{2}}$, where
\begin{equation} \label{SQINMA11}
X_{t_{1}, t_{2}}= \beta_{1} \circ \epsilon_{{t_1}-1, t_{2}}^a + \beta_{2} \circ  \epsilon_{t_{1}, t_{2}-1}^b+ \beta_{3} \circ  \epsilon_{t_{1}-1, t_{2}-1}^c + \epsilon_{t_{1}, t_{2}}.
\end{equation}
In Appendix \ref{sec:appendix_ooc_results}, Table~\ref{tab:DGP9}, we consider the same combinations for $(a,b,c)\in\{1,2\}^3$ as before. The OOC-ARLs in Tables~\ref{tab:DGP8} and~\ref{tab:DGP9} show that such nonlinear spatial dependence is often best detected by using the $\widetilde{\tau}$-chart. Especially for the small sample size $(m,n)=(10,10)$, the $\widetilde{\tau}$-chart can be faster than the $\widehat{\rho}$-chart by a factor up to~22. Note again that, similar to the SAR$(1,1)$ model with additive outliers, for the small grid $(m,n)=(10,10)$ and $(a,b,c)=(1^2,2^1,3^1)$, the OOC-ARL of the $\widehat{\rho}$-chart is larger than the IC-ARL. In the few cases where the $\widehat{\rho}$-chart has the lowest OOC-ARLs, the $\widetilde{\tau}$-chart still does reasonably well, so using the $\widetilde{\tau}$-chart for uncovering QMA-like spatial dependence appears to be a good universal solution.

Finally, let us turn to (continuously distributed) bilateral DGPs. For unilateral DGPs, the observation~$Y_{t_1,t_2}$ is generated by observations and innovations with ``time'' indices $\leq t_1,t_2$ only (``past information''). Bilateral DGPs, in turn, also incorporate information from times $\geq t_1,t_2$ (``future information''). Although this might look artificial at first glance, the bilateral approach constitutes a parametrically parsimonious way of generating an intensified spatial dependence structure. According to Whittle's representation theorem \citep{Whittle_1954}, under mild conditions, any bilateral spatial DGP can be represented (at least approximately) by a unilateral DGP having the same spatial ACF. This unilateral counterpart, however, would need much more model parameters than the bilateral formulation. The first-order simultaneous AR (SAR$(1)$) model is defined by
\ba
\label{simultAR1}
Y_{t_{1},t_{2}} = a_{1} \cdot Y_{t_{1}-1,t_{2}} +   a_{2} \cdot Y_{t_{1},t_{2}-1} +  a_{3} \cdot Y_{t_{1},t_{2}+1} +  a_{4}\cdot Y_{t_{1}+1,t_{2}} +  \varepsilon_{t_{1}, t_{2}},
\ea
where $\varepsilon_{t_{1}, t_{2}}$ are \iid\ $\norm(0,1)$. The resulting OOC-ARLs are summarized in Table~\ref{tab:DGP10} with different parameter values for ($a_1$, $a_2$, $a_3$, $a_4$).  Table~\ref{tab:DGP11} considers additional contamination by AOs of $\pm 5$. The results in both tables clearly show the superiority of the $\widetilde{\tau}$-chart in all scenarios. In two cases with the small grid $(m,n)=(10,10)$, the OOC-ARL of the $\widehat{\rho}$-chart is again larger than the IC-ARL. 

The DGP for Table~\ref{tab:DGP12}, in turn, is a bilateral counterpart to the SQMA$(1,1)$ model \eqref{SQMA11}. 
The first-order simultaneous QMA (SQMA$(1)$) model is defined by
\ba
\label{simultQMA1}
Y_{t_{1}, t_{2}}= b_{1} \cdot \varepsilon_{{t_1}-1, t_{2}-1}^{a} + b_{2} \cdot \varepsilon_{t_{1}+1, t_{2}-1}^{b}+ b_{3}\cdot \varepsilon_{t_{1}+1, t_{2}+1}^{c} + b_{4} \cdot \varepsilon_{t_{1}-1, t_{2}+1}^{d} + \varepsilon_{t_{1}, t_{2}},
\ea
with \iid\ $\varepsilon_{t_{1}, t_{2}}\sim\norm(0,1)$, where the powers $a,b,c,d$ are again either chosen as~$1$ (linear term) or~$2$ (quadratic term). The considered combinations are $(2,2,2,2)$, $(2,1,2,1)$, and $(2,2,1,1)$. Table~\ref{tab:DGP12} shows the emprical results. In the first case (all error terms are squared), either the $\widehat{\kappa}$- or $\widetilde{\kappa}$-chart have the lowest OOC-ARLs (but~$\widetilde{\tau}$ does not much worse). In the other two cases, the $\widetilde{\tau}$-chart is again superior. 

In a nutshell, we find that the proposed $\widetilde{\tau}$-EWMA chart is a universally applicable solution to uncover the bilateral and nonlinear spatial dependence. In contrast to the $\widehat{\rho}$-chart, it has the additional advantage of being nonparametric and robust, \ie its performance is neither affected by the actual IC-distribution nor by a possible contamination with outliers. Therefore, it is our preferred choice for the real-data applications presented in Section~\ref{Real-Data Application} as well as for the novel higher-order extensions proposed in the following Section~\ref{Detecting Higher-order Spatial Dependence}.

\section{Detecting Higher-Order Spatial Dependence}
\label{Detecting Higher-order Spatial Dependence}
Our newly proposed SOP-charts in Section~\ref{Control Charts for Spatial Dependence} and the corresponding performance analyses in Section~\ref{Simulation Study} focus on the default case where SOPs are calculated on the basis of directly neighboring observations (\ie with delay $\fd=\1$, recall Section~\ref{Spatial Ordinal Patterns and Types}). This case is most important for applications as we usually observe the strongest dependence between adjacent observations (first-order dependence). Nevertheless, there may be situations where the dependence is more pronounced for larger spatial lags (higher-order dependence), in which case SOPs with a corresponding delay $d_1+d_2>2$ appear to be more suitable. Furthermore, it may happen that the spatial dependencies are weak but slowly decaying with increasing spatial lag. Then, it appears advantageous to consider SOPs with multiple delays at once to uncover the apparent spatial dependence. In Section~\ref{Possible Approaches}, we propose novel SOP-based control charts that are tailor-made to detect such higher-order dependence. Afterwards, in Section~\ref{Performance Analyses}, we analyze their performance with additional simulation experiments, where we also consider possible competitors based on the spatial ACF.

\subsection{Possible Approaches}
\label{Possible Approaches}
If SOPs are to be used for detecting higher-order dependencies, we propose to compute the SOPs and types ``with delays'', \ie to compute the SOPs and types from the squares $\mY_{\fsi}^{(\fdi)} = \left(\begin{smallmatrix} y_{s_1-d_1, s_2-d_2} & y_{s_1-d_1,s_2} \\ y_{s_1, s_2-d_2} & y_{s_1,s_2} \end{smallmatrix}\right)$ with delay $\fd=(d_1,d_2)\in\bbn^2$ and $d_1+d_2>2$. Let us use ``$\fd$-SOPs'' and ``$\fd$-types'' as the corresponding short-hand notations. 
Although the use of delays has been shown to be successful in the case of time series, see \cite{keller14,Weiss_2022} for details, 
the use of delays for spatial data is new and has not been investigated by either \cite{Bandt.Wittfeld_2023} or \cite{Weiss.Kim_2024}. In addition, ordinal patterns with delays have not been considered in process monitoring so far \citep[see][]{Weiss.Testik_2023}. Thus, our novel proposals and the corresponding performance results may also serve as an inspiration for future research in the aforementioned areas. 
For the sake of brevity, we will only focus on extensions of the $\widetilde{\tau}$-chart in the sequel, which showed a superior performance in Section~\ref{Simulation Study}. In addition, we restrict our simulations to continuously distributed DGPs, because we found in Section~\ref{Simulation Study} that the SOP-charts perform similarly well for discrete DGPs if we use the randomization approach \eqref{noise}.

Our EWMA approach \eqref{EWMA} is easily adapted to the use of delay~$\fd$ by defining
\ba
\label{EWMAdelay}
\hat{\fp}_0^{(\lambda,\fdi)}\ =\ \fp_0^{(\fdi)},\qquad
\hat{\fp}_t^{(\lambda,\fdi)}\ =\ \lambda\,\widehat{\fp}_t^{(\fdi)} + (1-\lambda)\,\hat{\fp}_{t-1}^{(\lambda,\fdi)} \quad\text{for } t=1,2,\ldots,
\ea
where~$\widehat{\fp}_t^{(\fdi)}$ is the frequency vector of $\fd$-types in the $t$th data set $(Y_{\fsi}^{(t)})_{\fsi}$, and where~$\fp_0^{(\fdi)}$ is an initial probability vector. Under the assumption of ``IC-iid'', the $\fd$-SOPs and $\fd$-types are again discrete uniformly distributed, so one would choose $\fp_0^{(\fdi)}=(\tfrac{1}{3},\tfrac{1}{3},\tfrac{1}{3})^\top$ for initializing \eqref{EWMAdelay}. The resulting ``delayed $\widetilde{\tau}$-chart'', abbreviated as $\widetilde{\tau}^{(\fdi)}$-chart, is then defined as
\ba
\label{tilde_tau_d}
\widetilde{\tau}^{(\fdi)}\text{-chart:}\quad
\text{plot } \widetilde{\tau}_t^{(\lambda,\fdi)} = \widehat{p}_{t,3}^{(\lambda,\fdi)} - \tfrac{1}{3}, \quad \text{trigger alarm if } \big|\widetilde{\tau}_t^{(\lambda,\fdi)}\big| > l_{\widetilde{\tau}, \lambda,\fdi}.
\ea
The other SOP-charts in \eqref{SOPcharts} could also be adapted analogously. In order to have a reasonable competitor to our novel $\widetilde{\tau}^{(\fdi)}$-chart, we also adapt the $\widehat{\rho}$-chart from Remark~\ref{bemACFchart} to the case of higher-order dependence, namely by computing the spatial ACF with user-specified spatial lag $\fh\in\bbz^2\setminus\{\0\}$ via $\hat\rho(\fh) = \big(\sum_{\fsi} (Y_{\fsi}-\overline{Y})(Y_{\fsi-\fhi}-\overline{Y})\big) \big/ \big(\sum_{\fsi} (Y_{\fsi}-\overline{Y})^2\big)$. The resulting $\widehat{\rho}^{(\fhi)}$-chart is defined by
\ba
\label{ACF-EWMA-h}
\widehat{\rho}_0^{(\lambda, \fhi)}\ =\ \rho_0^{(\fhi)},\qquad
\widehat{\rho}_t^{(\lambda, \fhi)}\ =\ \lambda\,\widehat{\rho}_t^{(\fhi)} + (1-\lambda)\,\widehat{\rho}_{t-1}^{(\lambda, \fhi)} \quad\text{for } t=1,2,\ldots,
\ea
where $\widehat{\rho}_t^{(\fhi)}$ is the $\hat\rho(\fh)$-value computed from the $t$th data set $(Y_{\fsi}^{(t)})_{\fsi}$. The IC-value of $\rho(\fh)$, which is used for the initialization in \eqref{ACF-EWMA-h}, is given by $\rho_0^{(\fhi)}=0$ for $\fh\not=\0$ under the assumption ``IC-iid''. 

If we anticipate that violations of the ``IC-iid'' assumption manifest themselves in, \eg second-order dependencies, then it would be natural to use the $\widetilde{\tau}^{(\twoi)}$-chart and $\widehat{\rho}^{(\twoi)}$-chart for process monitoring. However, if we do not have such a specific OOC-DGP in mind, but only expect possible higher-order dependencies, then it seems reasonable to consider multiple delays~$\fd$ or spatial lags~$\fh$ simultaneously.  To avoid using a large number of control charts in parallel, an appealing solution would be to aggregate the information for all delays or lags into one statistic, which is then plotted on a single control chart. One possible solution is a BP-like construction, as recently proposed by \citet{Bui.Apley_2018} for monitoring image data.  
Based on their valuable insights, we develop a BP-SOP version of the $\widetilde{\tau}^{(\fdi)}$-chart that, in contrast to \citet{Bui.Apley_2018}, does not require prior fitting of a supervised learning model, but instead allows directly analyzing streams of continuously distributed rectangular datasets for higher-order spatial dependencies. 
Having specified the maximal ``window size'' $w\in\bbn$, we propose to define the $\widetilde{\tau}^{\textup{BP}(w)}$-statistic as
\ba
\label{tilde_tau_BP}
\widetilde{\tau}_t^{\textup{BP}(\lambda,w)}\ =\ \sum_{d_1,d_2=1}^w \big(\widetilde{\tau}_t^{(\lambda,\fdi)} - \widetilde{\tau}_0^{(\fdi)}\big)^2,
\ea
where the IC-value of the $\widetilde{\tau}^{(\fdi)}$-statistic, $\widetilde{\tau}_0^{(\fdi)}$, is equal to zero under the assumption ``IC-iid''.
Analogously, the competing $\widehat{\rho}^{\textup{BP}(w)}$-statistic is defined as
\ba
\label{ACF-EWMA-BP}
\widehat{\rho}_t^{\textup{BP}(\lambda,w)}\ =\ \sum_{h_1,h_2=-w,\ \fhi\not=\zeroi}^w \big(\widehat{\rho}_t^{(\lambda,\fhi)} - \rho_0^{(\fhi)}\big)^2,
\ea
where the trivial summand for $h_1=h_2=0$ is omitted during summation. It is also worth noting that the computational effort for \eqref{ACF-EWMA-BP} gets reduced by considering the identities $\rho(-h_1,-h_2) = \rho(h_1,h_2)$ and $\rho(-h_1,h_2) = \rho(h_1,-h_2)$. Recall that $\rho_0^{(\fhi)}=0$ under the assumption ``IC-iid''.

\subsection{Performance Analyses}
\label{Performance Analyses}
Let us extend the simulation-based performance analyses from Section~\ref{Simulation Study} to our novel control charts for higher-order spatial dependence as proposed in Section~\ref{Possible Approaches}. More precisely, we consider the SOP-based $\widetilde{\tau}^{(\fdi)}$- and $\widetilde{\tau}^{\textup{BP}(w)}$-charts as well as the competing ACF-based $\widehat{\rho}^{(\fhi)}$- and $\widehat{\rho}^{\textup{BP}(w)}$-charts for $\fd,\fh\in\{\1,\2,\3\}$ and $w\in\{1,2,3\}$, respectively. 
Like before, we use $\lambda=0.1$ as the EWMA's smoothing parameter and all previous combinations for the sample size $(m,n)$. The control limits are again determined with respect to the target ARL$_0\approx 370$ under ``IC-iid'' (using $10^6$~replications), where all spatial ACF-based charts additionally assume that the data are $\norm(0,1)$-distributed. This additional assumption is crucial for the performance of the $\widehat{\rho}^{(\fhi)}$- and $\widehat{\rho}^{\textup{BP}(w)}$-charts. In analogy to Table~\ref{tab:DGP0} in Section~\ref{In-control ARL Performance}, we determined their IC-ARLs under non-normality, namely if the spatial data are either \iid\ t$(2)$ or Exp$(1)$. The obtained results in Table~\ref{tab:rename11} show strong deviations from the target ARL$_0\approx 370$, especially if using the $\widehat{\rho}^{\textup{BP}(w)}$-chart. Therefore, the subsequent ARL~performances of the ACF-based control charts are expected to be misleading if the IC-model has not been correctly identified.

We now turn to the analysis of OOC-ARL performances (where the ACF-based control charts use the correct IC model). To identify situations where the higher-order charts might be beneficial, we consider the following OOC-DGPs:

\begin{itemize}
\item The unilateral SAR$(1,1)$ process \eqref{SAR11} with $(\alpha_1, \alpha_2, \alpha_3)=(0.4,0.3,0.1)$, without and with AOs (+10) like in Tables~\ref{tab:DGP3} and~\ref{tab:DGP5b}, is juxtaposed with a unilateral SAR$(2,2)$ process having the same parameter values but at lags~2 instead of lags~1. The obtained simulation results are summarized in Tables~\ref{tab:rename12}--\ref{tab:rename13} for the SAR$(1,1)$ DGP, and in Tables~\ref{tab:rename14}--\ref{tab:rename15} for the SAR$(2,2)$ DGP.

\item The unilateral SQMA$(1,1)$ process \eqref{SQMA11} with powers $(2,2,2)$ like in Table~\ref{tab:DGP8} is contrasted with a corresponding unilateral SQMA$(2,2)$ process, which only differs by replacing lag~1 by lag~2. The results are summarized in Table~\ref{tab:rename16} and~\ref{tab:rename17}, respectively.
        
\item The bilateral SAR$(1)$ process \eqref{simultAR1} with parameters $(0.1,\ldots,0.1)$ is considered again, with and without outliers ($\pm 5$) like in Tables~\ref{tab:DGP10} and~\ref{tab:DGP11}, because it exhibits an intensified spatial dependence structure such that higher-order dependencies also exist. The ARL results are shown in Tables~\ref{tab:rename18} and~\ref{tab:rename19}.
\end{itemize}
In Table~\ref{tab:rename12} regarding the SAR$(1,1)$ process, the ``first-order charts'' (\ie with $\fd=\1$ or $\fh=\1$) perform better than their higher-order competitors, which is plausible since first-order spatial dependence dominates here. While the charts in Table~\ref{tab:rename12} focus on a single delay~$\fd$ or lag~$\fh$, the BP-type charts in Table~\ref{tab:rename13} aggregate across all delays or lags within the square determined by~$w$. Here, we do not have a unique preference for one~$w$, and the ARL differences for different~$w$ are often quite small. Again, this is plausible as the first-order dependence is considered by any of the BP-charts. Finally, in analogy to Table~\ref{tab:DGP5b}, all ACF-based charts are rather sensitive to outliers.

Looking at the SAR$(2,2)$-counterparts in Tables~\ref{tab:rename14}--\ref{tab:rename15}, it becomes clear that the second-order charts perform much better than the first-order ones. For the BP-type charts, both $w=2$ and $w=3$ lead to quite similar OOC-ARLs, which can be explained in analogy to the situation of Table~\ref{tab:rename13}. In any case, higher order charts are now superior, which is plausible given the AR dependence only at lag~2. Analogous conclusions apply to the SQMA processes in Tables~\ref{tab:rename16} and~\ref{tab:rename17}. Here, the different performances are even more pronounced due to the MA-dependence: while AR-dependence at some lag~$h$ also implies dependence at higher lags, MA dependence manifests itself only at specific lags. Therefore, the BP-approach is advantageous here, because it monitors several spatial lags at once instead of focusing on one particular lag only. The $\widetilde{\tau}^{(\fdi)}$-charts with suitable~$\fd$ are clearly superior to their $\widehat{\rho}^{(\fhi)}$-competitors in detecting the non-linear dependence, while the $\widehat{\rho}^{\textup{BP}(w)}$-chart constitutes a clear improvement in this respect (which still relies on appropriate distributional assumptions).

Finally, in Tables~\ref{tab:rename18} and~\ref{tab:rename19} regarding the bilateral SAR$(1)$ process, the $\widetilde{\tau}$-charts are superior in most cases, and if not, they perform only slightly worse than their $\widehat{\rho}$-counterparts. It is interesting to note that for both types of BP-charts, the ARL~performance slightly deteriorates with increasing~$w$. While the SAR$(1)$'s memory decays only slowly with increasing spatial lag, it is most pronounced at lag~1. Increasing~$w$, the relative importance of lag~1 for the BP-statistic gets reduced, so it is plausible that the BP-charts with $w=1$ are superior here. In summary, the $\widetilde{\tau}^{(\fdi)}$-charts with $\fd,\fh\geq\1$ are attractive solutions for such DGPs where only higher-order dependencies exist in OOC-situations. If there are also (or solely) first-order dependencies under OOC-conditions, as it is often the case in practice, the default choice $\fd=1$ is recommended. If one is not sure about the likely OOC-scenarios, the $\widetilde{\tau}^{\textup{BP}(w)}$-chart with $w\geq 2$ seems to be a good compromise, as it can detect both first-order and higher-order spatial dependencies reasonably well. While the latter property also holds for the $\widehat{\rho}^{\textup{BP}(w)}$-chart, due to its parametric nature, it sincerely depends on appropriate distributional assumptions concerning the IC-model. As real-world data hardly follow one of the textbook distributions, we advise using the nonparametric (and robust) $\widetilde{\tau}$-charts for applications in practice. 

\section{Real-Data Applications}
\label{Real-Data Application}
To illustrate the application of SOPs and types for monitoring streams of spatial grid data, we present three real-world data examples. Given the findings from the previous Section~\ref{Simulation Study}, we focus on the $\widetilde{\tau}$-EWMA chart for process monitoring. In the first two examples, we are concerned with discrete-valued grid data, namely rainfall data (Section~\ref{Rainfall data}) and counts of war-related fires in eastern Ukraine (Section~\ref{War related fires in Ukraine}). In both cases, the ``IC-iid'' assumption considered in Section~\ref{Simulation Study} is adequate. The third application (Section~\ref{Textile images}) is more demanding in this respect. Here, we monitor (quasi-continuous) textile images, where we use a bootstrap approach to tailor the $\widetilde{\tau}$-chart.

\subsection{Rainfall data}
\label{Rainfall data}
\begin{figure}[!htb]
     \begin{subfigure}{0.28\textwidth}
         \centering
         \includegraphics[width=\linewidth]{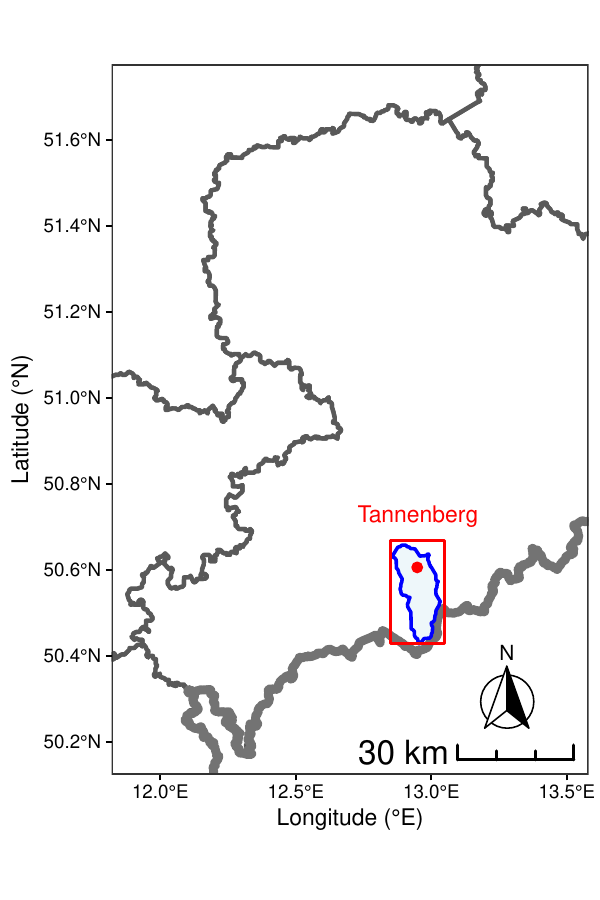}
         \caption{Map of western Saxony with highlighted  Tannenberg catchment.}
         \label{fig:map_tannenberg}
     \end{subfigure}    
     \begin{subfigure}{0.68\textwidth}
         \centering
         \includegraphics[width=\linewidth]{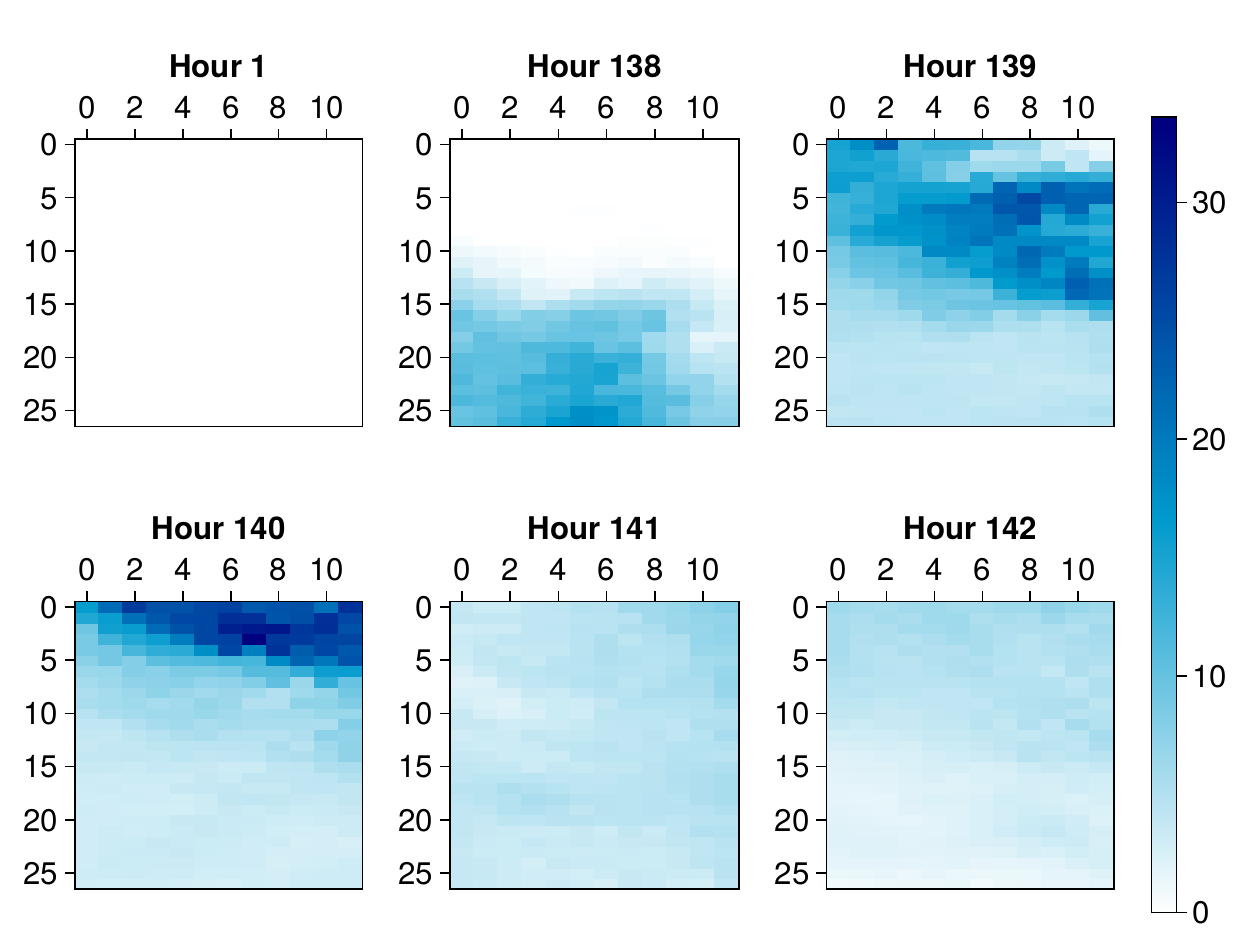}
         \caption{Selected matrices of hourly precipitation sums in August 2007.}
         \label{fig:hourly_rain_tannenberg}
     \end{subfigure}    
        \caption{
        Map of the Tannenberg catchment in Germany with a rectangle grid of interest (left). Heat maps of hourly rainfall data for the Tannenberg catchment in August 2007. Grid sizes are $m = 26$ and $n = 11$.
        }
       \label{fig:rain_map}
\end{figure}

As our first application, we consider the RADOLAN (``RAdar-OnLine-AN\-eichung'') precipitation data set of \citet{Winterrath.etal_2018}, which provides reprocessed gauge-adjusted radar data expressing the one-hour precipitation sums (in $0.1$\,mm steps) in a 1\,km\,$\times$\,1\,km grid. 
We focus on the Tannenberg catchment, located in the federal state of Saxony (Germany). This region is covered by the grid shown in Figure~\ref{fig:map_tannenberg} (with $m=26$ and $n=11$). It was previously investigated by, among others, \cite{Fischer.etal_2024}. 
According to \citet[Figure~9, p.~13]{Kaiser.etal_2021}, the Tannenberg catchment constitutes one of Germany's hot spot regions for heavy rainfall-related flood events. Hence, applying a monitoring system for extraordinary precipitation events is of utmost practical importance. 

To illustrate the application of the $\widetilde{\tau}$-EWMA chart, we consider the week between August 3--9, 2007, leading to $168$ hourly observations. As we shall recognize below, two severe rain events occurred during this week; see also Figure \ref{fig:hourly_rain_tannenberg}. 
Since we aim to detect such severe rain events, the IC-model should refer to the situation of, at most, moderate precipitation. Then, however, ties are frequently observed, especially in times of no precipitation at all (like in ``Hour~1'' of Figure \ref{fig:hourly_rain_tannenberg}). Hence, we apply the randomization approach \eqref{noise}, now with $\unif(0,0.1)$-noise in view of the measurement accuracy $0.1$\,mm, such that no-rain hours are transformed into \iid\ uniform noise satisfying the ``IC-iid'' assumption. Note that during hours with strong precipitation, the $0.1$\,mm steps are sufficiently fine such that ties hardly happen. Therefore, the randomization approach \eqref{noise} has only a little effect on the SOPs referring to such an OOC-case as it does not affect any strict order among the precipitation values. In particular, extraordinary precipitation events manifest themselves in clusters of high precipitation values, so we get a pronounced spatial structure implying spatial dependence under OOC-conditions.

\begin{figure}[!htb]
     \begin{subfigure}{0.49\textwidth}
         \centering
         \includegraphics[width=\linewidth]{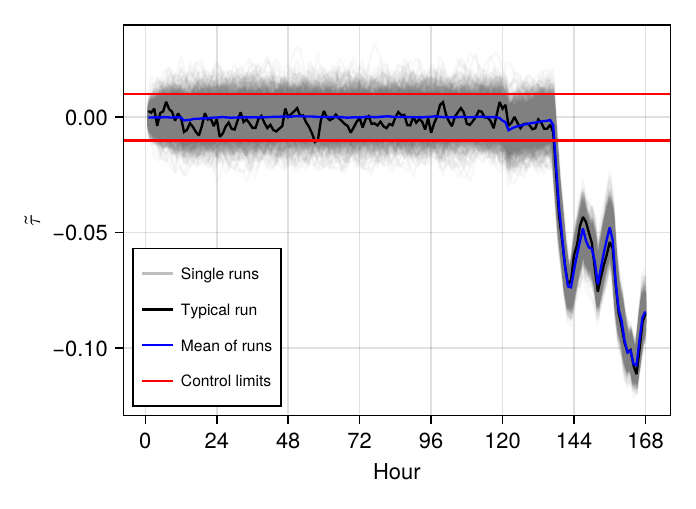}
         \caption{EWMA chart ($\lambda = 0.1$)}
         \label{fig:EWMA_rain}
     \end{subfigure}    
     \begin{subfigure}{0.49\textwidth}
         \centering
         \includegraphics[width=\linewidth]{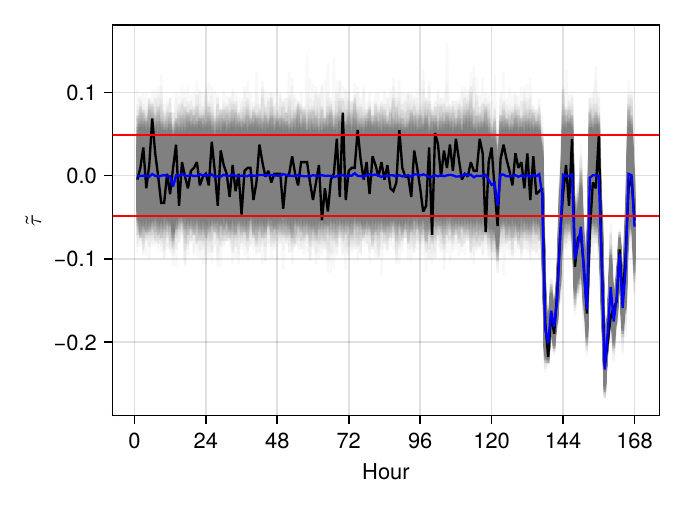}
         \caption{Shewhart chart ($\lambda = 1.0$)}
         \label{fig:Shewhart_rain}
     \end{subfigure}    
    \centering
    \caption{$\widetilde{\tau}$-charts with $\lambda=0.1$ (left) and $\lambda=1$ (right) of hourly rainfall data for the Tannenberg catchment in a single week interval 03/08/2007 - 09/08/2007. The light-gray curves show the control charts computed for $1000$ single runs. Grid sizes are $m = 26$ and $n = 11$.
    }
    \label{fig:SOP_charts_rain}
\end{figure}

Figure~\ref{fig:EWMA_rain} shows the resulting $\widetilde{\tau}$-EWMA chart with $\lambda = 0.1$, where $l_{\widetilde{\tau}, \lambda}=0.018819$ is calibrated for ARL$_0=370$ based on $10^6$ IC-iid replications. The black sample path serves as our illustrative example since, in practice, one concentrates on one sequence of EWMA statistics. But to illustrate the possible consequences of the randomization approach \eqref{noise}, we added $999$ additional runs (which are computed from the same data but using newly generated noise) as light-gray curves. 
The mean of the altogether $1000$ runs is highlighted in blue, so it becomes clear that the black line represents a ``typical run''. Recall that the ARL computations already account for the randomization (as this is covered by the ``IC-iid'' assumption), \ie any of the grey curves satisfies the IC-ARL target. This also gets clear from the first part of EWMA statistics (say, hours $\leq 100$), where the process is IC and only a few gray curves lead to a (false) alarm (ARL$_0=370$ allows for a false alarm after 370 EWMA statistics in the mean). 

The $\widetilde{\tau}$-EWMA chart in Figure~\ref{fig:EWMA_rain} triggers alarms of increasing severity at times $\geq 138$, clearly indicating a process change. At this point, it is useful to look at the $\widetilde{\tau}$-Shewhart chart (obtained by setting $\lambda=1$, \ie by removing any memory), which we added for the sake of interpretation in Figure \ref{fig:Shewhart_rain}. Recall that we did not consider Shewhart charts in Section~\ref{Simulation Study} as they are known to be insensitive to small process changes (but may react faster to sudden extreme events), and as they are difficult to design due to discreteness. In fact, it was not possible to closely narrow ARL$_0=370$, but only ARL$_0=362.29$ for the control limit $l_{\widetilde{\tau}, \lambda}=0.092075$.  Thus, Figure \ref{fig:Shewhart_rain} mainly serves as a time series plot of the non-smoothed $\widetilde{\tau}$-statistics for explaining the EWMA's alarms. It becomes clear that there seems to be a sudden severe rain event starting at $t=138$, which essentially continues until $t=165$, with two short phases of restraint between $t=143$ and $156$. 
Figure~\ref{fig:EWMA_rain} shows that this event is quickly detected by the $\widetilde{\tau}$-EWMA chart, and the second main rain event (starting around $t=157$) leads to a further drop-down of the EWMA-statistics (irrespective of the actual noise). Looking back to Figure~\ref{fig:hourly_rain_tannenberg}, we recognize that the abnormal precipitation behavior goes along with a pronounced spatial dependence structure, which explains why it is successfully uncovered by the $\widetilde{\tau}$-EWMA chart.

\begin{figure}[!htb]
     \begin{subfigure}{0.49\textwidth}
         \centering
         \includegraphics[width=\linewidth]{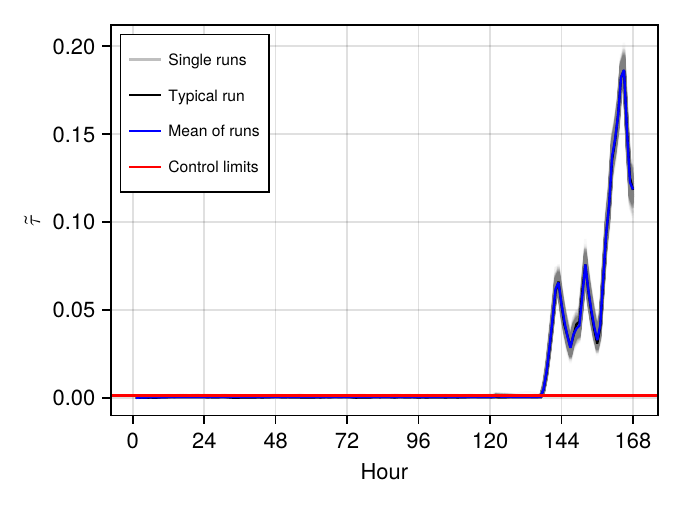}
         \caption{BP-EWMA chart ($\lambda=0.1$)}
         \label{fig:EWMA_rain_bp_full}
     \end{subfigure}    
     \begin{subfigure}{0.49\textwidth}
         \centering
         \includegraphics[width=\linewidth]{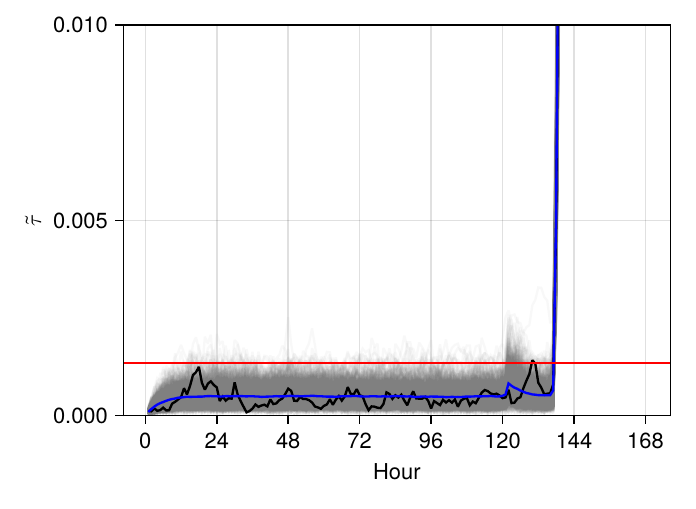}
         \caption{BP-EWMA chart ($\lambda=0.1$, zoomed)}
         \label{fig:EWMA_rain_bp_zoom}
     \end{subfigure}    
    \caption{$\widetilde{\tau}^{\textup{BP}(3)}$-charts with $\lambda=0.1$ (left) and zoomed in (right) of hourly rainfall data for the Tannenberg catchment in a single week interval 03/08/2007 - 09/08/2007. The light-gray curves show the control charts computed for $1000$ single runs. Grid sizes are $m = 26$ and $n = 11$.
    }
    \label{fig:SOP_charts_rain_BP}
\end{figure}

Finally, we applied the novel BP-extension of the $\widetilde{\tau}$-EWMA chart to the rainfall data, as proposed in Section~\ref{Detecting Higher-order Spatial Dependence}. The resulting control chart based on the $\widetilde{\tau}^{\textup{BP}(w)}$-statistics \eqref{tilde_tau_BP} with $w=3$ is shown in Figure~\ref{fig:SOP_charts_rain_BP}, where the control limit equals $l_{\widetilde{\tau}^{\textup{BP}(3)}, \lambda}=0.0013443$. While the absolute value of the statistics increases severely for times $t\geq 138$ (for this reason, we show an additional zoomed version of the $\widetilde{\tau}^{\textup{BP}(3)}$-EWMA chart in part~(b)), both chart types ``$\widetilde{\tau}$ with $\fd=\1$'' and ``$\widetilde{\tau}^{\textup{BP}(3)}$'' signal at the same time $t=138$, \ie the additional BP-feature does not lead to an advantage here. This is reasonable as already the first-order dependence is quite strong for the rainfall data, recall Figure~\ref{fig:rain_map}, so it suffices to use the more simple $\widetilde{\tau}$-EWMA chart.

\subsection{War-related fires in Ukraine}
\label{War related fires in Ukraine}
The Ukrainian-Russian war, which culminated in Russia's full-scale invasion of Ukraine on the morning of 24 February 2022, is one of the most extensively documented wars in history \citep{npr_ukraine_war}.  Especially at the beginning of the invasion, social media platforms were flooded with footage of airstrikes, first-hand accounts from bloggers embedded with military units, and updates on the ever-changing front lines. Despite the wealth of material available, however, this documentation offers only a fragmented view as events that go unrecorded or unpublished remain obscured. By contrast, cameras positioned high above the battlefield offer a distinct advantage, providing a broader and alternative perspective on the ongoing war. 

The British newspaper ``The Economist'' has monitored the war's progression by utilizing the ``Fire Information for Resource Management System'' (FIRMS), a NASA program engineered to identify fires globally \citep{Economist.Solstad_2023}. In a nutshell, the authors use gradient-boosted trees from 10 non-war years in Ukraine to predict fires since the Russian invasion. If the true number of fires exceeds a certain predicted quantile, the fires are classified as war-related. The disadvantage of their approach is that FIRMS cannot detect fires through clouds. In addition, as the model is probabilistic, fires may be misclassified as war-related and vice versa. However, their data track the front lines exceptionally well, so we use their data as given and do not apply any post-processing. The data are publicly available on \href{https://github.com/TheEconomist/the-economist-war-fire-model}{GitHub} (accessed on 2025-01-25) and updated twice a day. 

\begin{figure}[!htb]
     \begin{subfigure}{0.44\textwidth}
         \centering
         \includegraphics[width=\linewidth]{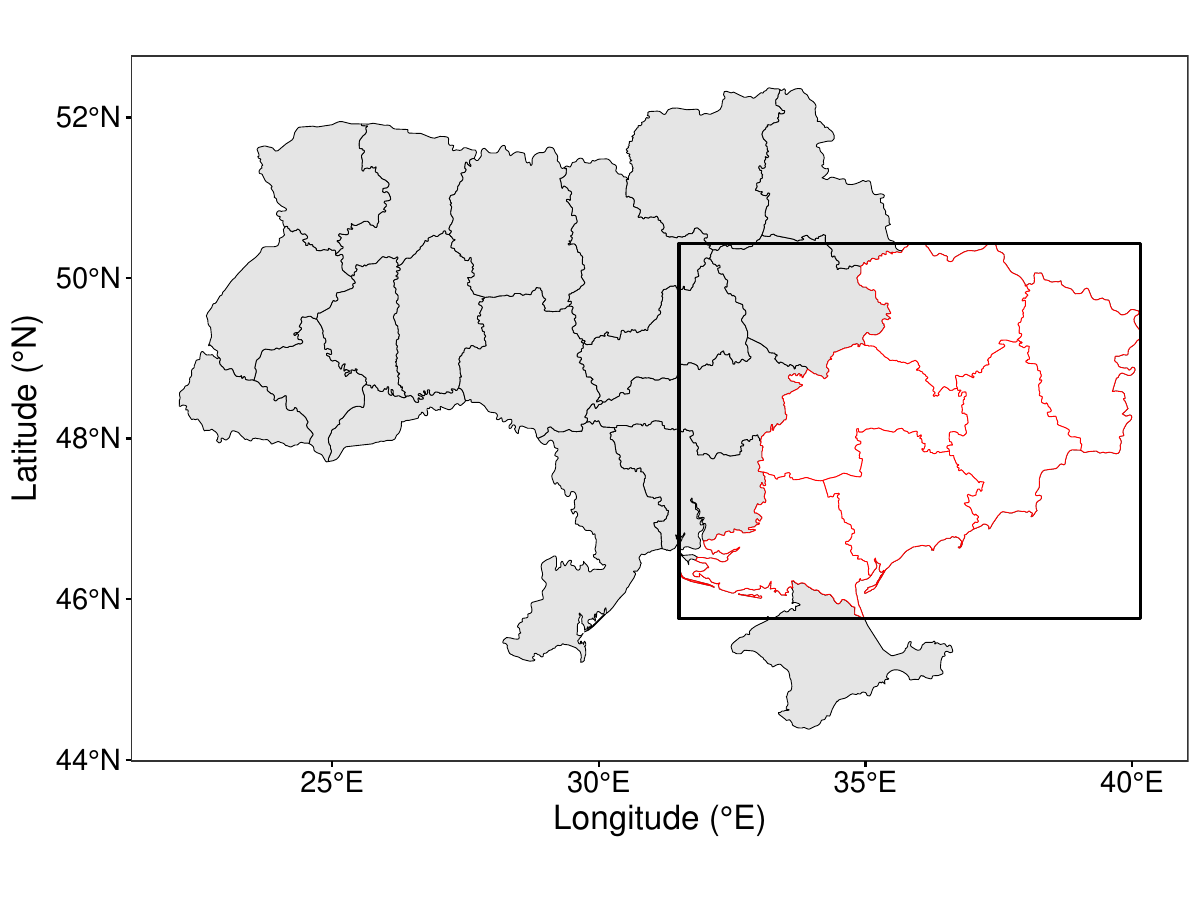}
         \caption{Map of Ukraine with highlighted eastern regions.}
         \label{fig:Map_Ukraine}
     \end{subfigure}    
     \begin{subfigure}{0.55\textwidth}
         \centering
         \includegraphics[width=\linewidth]{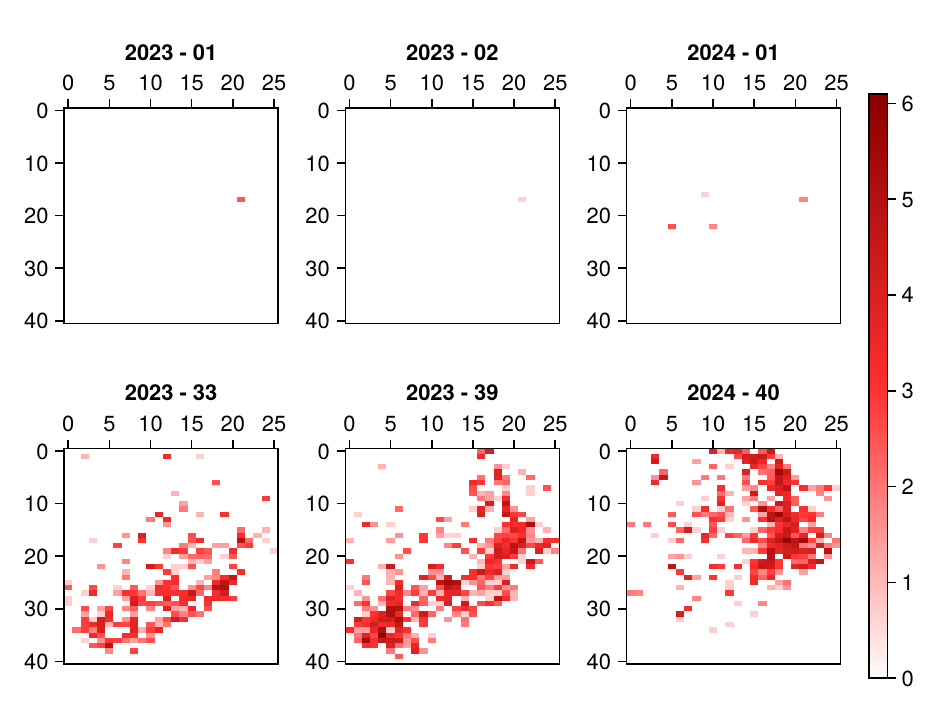}
         \caption{Selected count matrices of war-realted fires in 2023 {\color{red}and 2024}.}
         \label{fig:WeeksFire_Ukraine}
     \end{subfigure}    
        \caption{
    Map of the Ukraine (left) highlighting the eastern Ukraine region by red outlines. The rectangle is our grid of interest. The right panel shows chosen weekly heatmaps of fires (log-scale) with grid sizes $m = 40$ and $n = 25$ (right).
    }
    \label{fig:Ukraine_maps}
\end{figure}

We focus on the weekly numbers of war-related fires in 2023 and 2024 in the eastern provinces of Ukraine, where most of the fighting has been taking place, namely Luhanska, Dnipropetrovska, Donetska, Zaporizka, Kharkivska, and Khersonska. The provinces are highlighted in red in the left panel of Figure \ref{fig:Ukraine_maps}. We divide our chosen rectangle into bins (41 for latitude and 26 for longitude) and count the number of fires that occurred during the corresponding weeks in 2023 and 2024 for each bin.\footnote{In total, the data contains in 160,966  war-related fires for eastern Ukraine in 2023 and 2024} The count matrix is used to compute the SOPs, resulting in a matrix with $m = 40$ rows and $n = 25$ columns, corresponding to one of the simulation scenarios in Section~\ref{Simulation Study}. The right panel of Figure \ref{fig:Ukraine_maps} shows the calculated count matrix for different weeks in the beginning and middle of 2023 and 2024, respectively. It can be seen that almost no war-related fires were detected in the early weeks of 2023 and 2024, which is most likely because FIRMS cannot detect fires through clouds.
As the monitored data are discrete counts, we observe a vast number of ties in phases of low fire activity, in analogy to our application in Section~\ref{Rainfall data}. 
So we use again the randomization approach in \eqref{noise}, this time with $\unif(0,1)$-noise due to the integer data. As a consequence, in weeks with no war-related fires, the SOPs are computed from pure uniform noise, which perfectly agrees with the ``IC-iid'' assumption considered in Section~\ref{Control Charts for Spatial Dependence} (in analogy to Section~\ref{Rainfall data}). By contrast, in weeks with a pronounced spatial pattern of fire activity (such as week 39 in 2023 in Figure \ref{fig:Ukraine_maps}), the resulting spatial dependence causes violations of the ``IC-iid'' assumption, where the effect of the noise for SOP computation now largely fizzles out since strict orders among count values are always preserved.

\begin{figure}[!htb]
     \begin{subfigure}{0.49\textwidth}
         \centering
         \includegraphics[width=\linewidth]{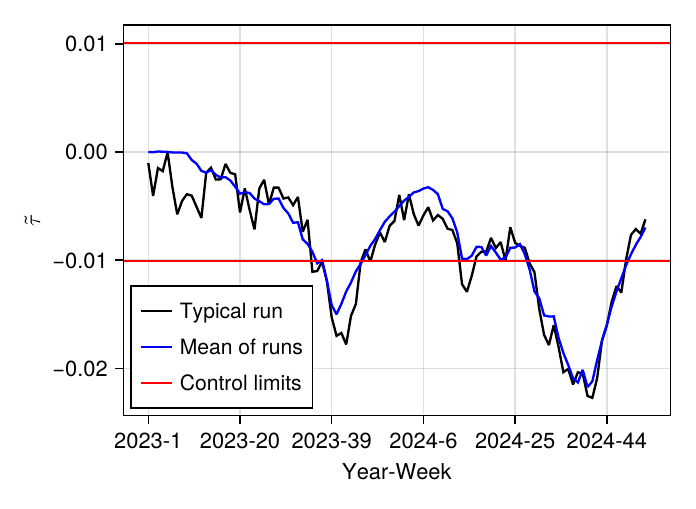}
         \caption{EWMA chart ($\lambda = 0.1$)}
         \label{fig:EWMA_Ukraine_sop}
     \end{subfigure}
     \begin{subfigure}{0.49\textwidth}
         \centering
         \includegraphics[width=\linewidth]{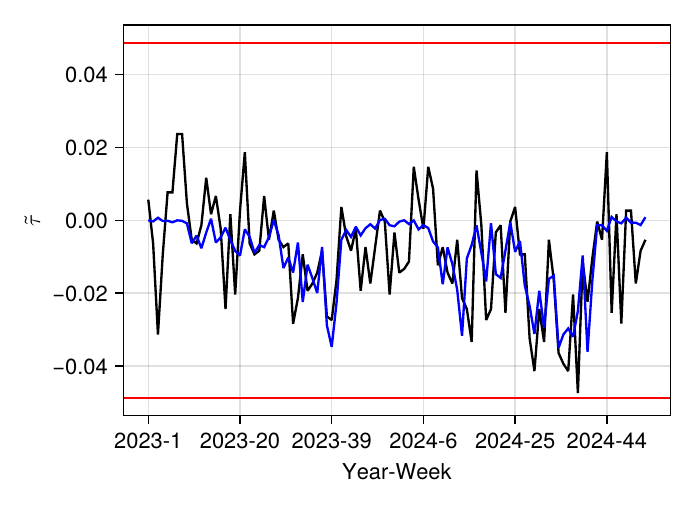}
         \caption{Shewhart chart ($\lambda = 1.0$)}
         \label{fig:Shewhart_Ukraine_sop}
     \end{subfigure}    
        \caption{
    $\widetilde{\tau}$-charts with $\lambda=0.1$ (left) and $\lambda=1.0$ (right) applied to data on war-related fires in eastern Ukraine (see Figure~\ref{fig:Ukraine_maps}) for the years 2023 and 2024 (black curve). Grid sizes are $m = 40$ and $n = 25$.  The blue curve is the average chart computed across 1000 charts.
    }
    \label{fig:SOP_charts_ukraine}
\end{figure}

Figure~\ref{fig:EWMA_Ukraine_sop} shows the resulting $\widetilde{\tau}$-EWMA chart with $\lambda=0.1$, the chart design of which is chosen according to Table~\ref{tab:ICiid_design}. While the EWMA statistics are centered around zero in the first weeks of 2023, we observe a decreasing trend starting in March, culminating in several alarms for the weeks $\geq 35$ (August/September 2023), and later again in 2024. This increasing activity in war-related fires can also be recognized from the lower panel of Figure~\ref{fig:WeeksFire_Ukraine}. 
Like before, it is instructive to look at the corresponding Shewhart chart in Figure~\ref{fig:Shewhart_Ukraine_sop}. Although it does not lead to any alarm, we recognize phases with slightly negative $\widetilde{\tau}$-values, roughly between weeks~30 and~40 in 2023, as well as in 2024. As we are concerned with only mild violations of the $\widetilde{\tau}$'s IC-value of~0 this time, by contrast to Section~\ref{Rainfall data}, the EWMA chart's inherent memory clearly turns out to be beneficial for monitoring war-related fires. Recalling the definition of $\widetilde{\tau}$ according to \eqref{StatType}, negative values of $\widetilde{\tau}$ indicate that type~3 occurs less frequently than expected under IC-assumptions, \ie the highest ranks tend to occur (slightly) more often along rows or columns rather than on a diagonal. Such ``dominant rows or columns'' can also be recognized in Figure~\ref{fig:WeeksFire_Ukraine}, which constitute themselves as horizontal or vertical stripes within the grid. 

\begin{figure}[H]
     \begin{subfigure}{0.49\textwidth}
         \centering
         \includegraphics[width=\linewidth]{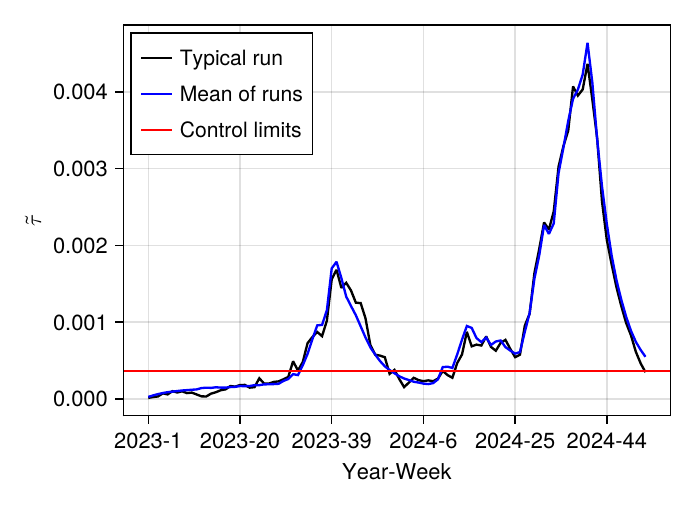}
         \caption{BP-EWMA chart ($\lambda = 0.1$)}
         \label{fig:EWMA_Ukraine_bp}
     \end{subfigure}
     \begin{subfigure}{0.49\textwidth}
         \centering
         \includegraphics[width=\linewidth]{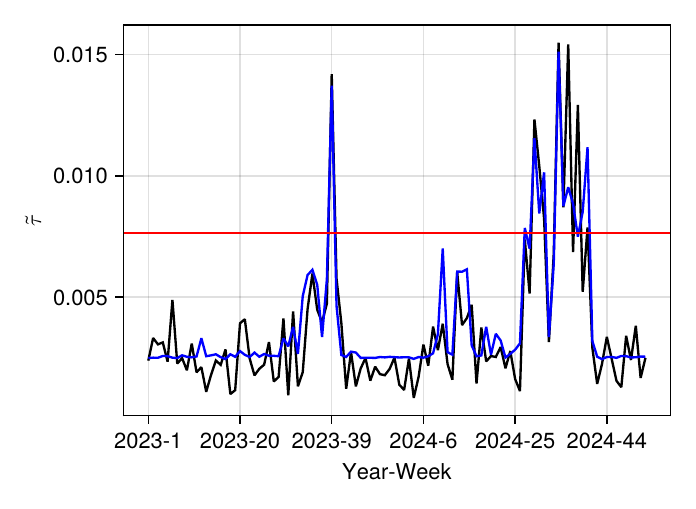}
         \caption{BP-Shewhart chart ($\lambda = 1.0$)}
         \label{fig:Shewhart_Ukraine_bp}
     \end{subfigure}    
        \caption{
    $\widetilde{\tau}^{\textup{BP}(3)}$-charts with $\lambda=0.1$ (left) and $\lambda=1.0$ (right) applied to data on war-related fires in east Ukraine (see Figure~\ref{fig:Ukraine_maps}) for the years 2023 and 2024 (black curve). Grid sizes correspond to $m = 40$ and $n = 25$.  The blue curve is the average chart computed across 1000 charts.
    }
    \label{fig:SOP_BP_charts_ukraine}
\end{figure}

Finally, we again applied the $\widetilde{\tau}^{\textup{BP}(3)}$-chart to the war-related fires data in Figure \ref{fig:SOP_BP_charts_ukraine}. While the EWMA version does not lead to novel alarms compared to the basic $\widetilde{\tau}$-EWMA chart with $\fd=\1$, we recognize an advantage of the BP approach for the Shewhart version in Figure \ref{fig:Shewhart_Ukraine_bp} this time, with several alarms in 2023 and 2024. So for the memory-less Shewhart chart, it is advantageous to aggregate the information across multiple delays~$\fd$, whereas for the memory-type EWMA chart, the information about first-order dependence is sufficient.

\subsection{Textile images}
\label{Textile images}
Our last real-data example considers the monitoring of textile images, which are provided by the R-package \href{https://cran.r-project.org/package=textile}{\nolinkurl{textile}} and further described in \citet{Bui.Apley_2018}. More precisely, 100 images with resolution $250\times 250$ (so $m=n=249$) are available, where the first 94 are without local defects according to \citet{Bui.Apley_2018} and thus constitute our Phase-I data. The last six images (Phase-II-data), by contrast, are affected by different issues such as fiber direction changes, tears, or holes. 
Strictly speaking, we are again concerned with discrete-valued data as the available images of textile material are plotted on a grey scale with levels from~0 (black) to~255 (white). However, as the range is rather large (``quasi-continuous''), ties within $2\times 2$ squares are not that frequent such that a prior randomization is omitted this time. 

\begin{figure}[!htb]
     \begin{subfigure}{0.45\textwidth}
         \centering
         \includegraphics[width=0.87\linewidth]{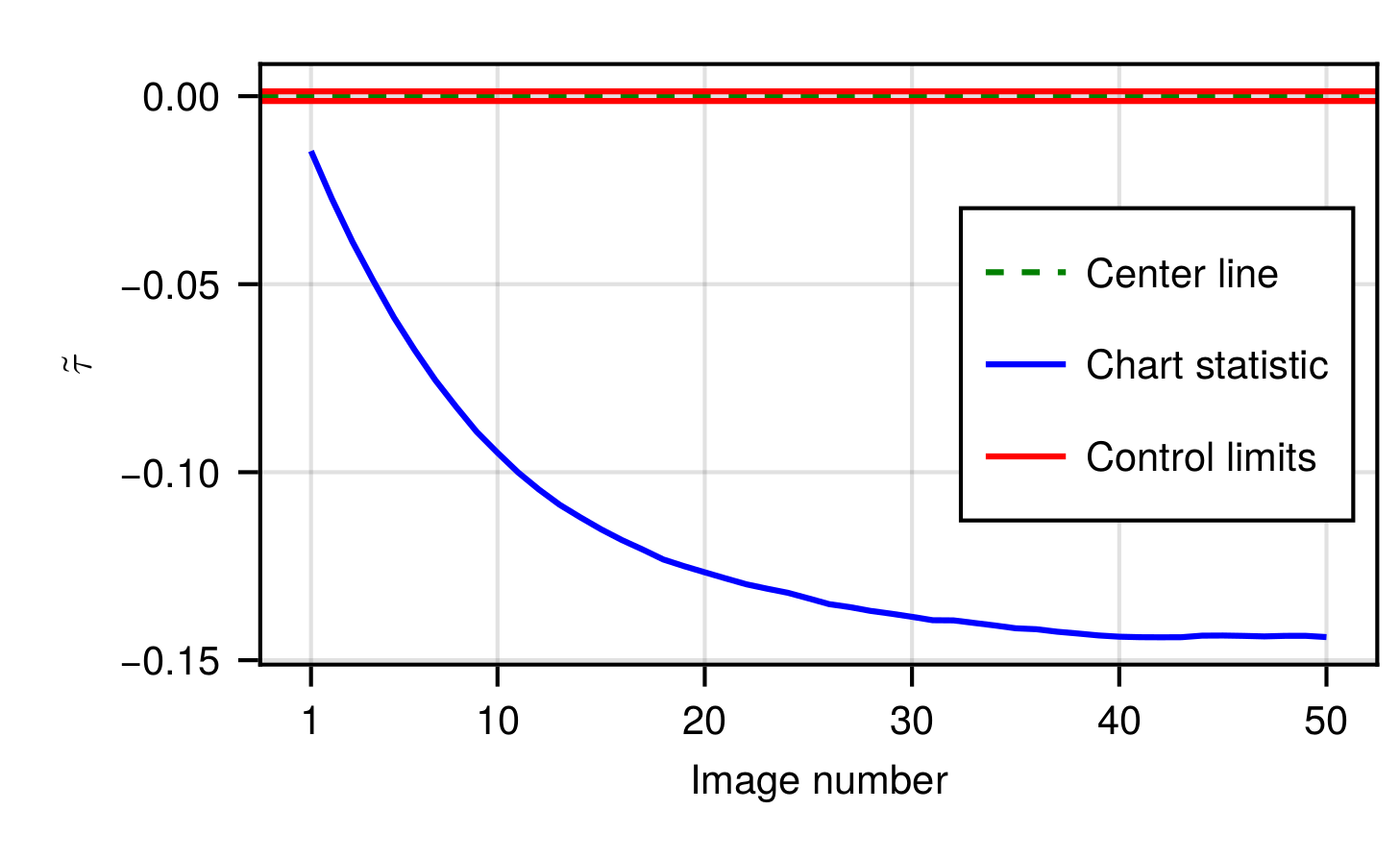}
         \caption{$\widetilde{\tau}$-chart ($\lambda = 0.1$, IC-iid)}
         \label{fig:EWMA_textile}
     \end{subfigure}    
     \begin{subfigure}{0.25\textwidth}
         \centering
         \includegraphics[width=\linewidth]{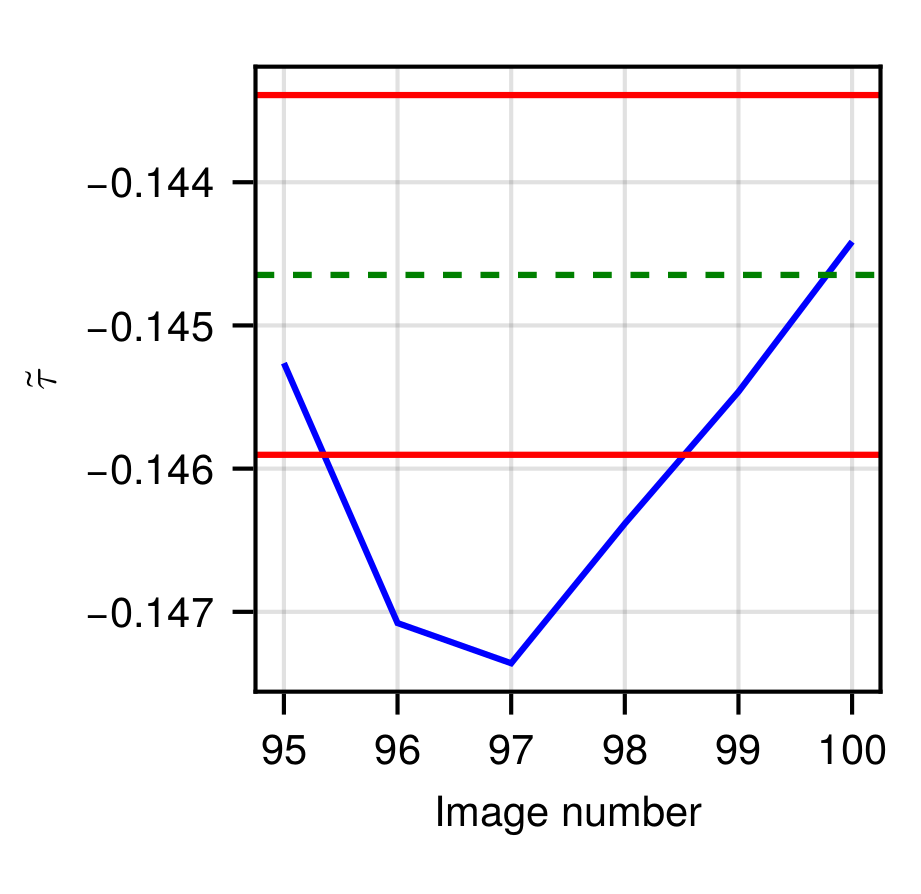}
         \caption{$\widetilde{\tau}$-chart ($\lambda = 0.1$)}
         \label{fig:EWMA_textile_short}
     \end{subfigure}  
      \begin{subfigure}{0.245\textwidth}
         \centering
         \includegraphics[width=1.04\linewidth]{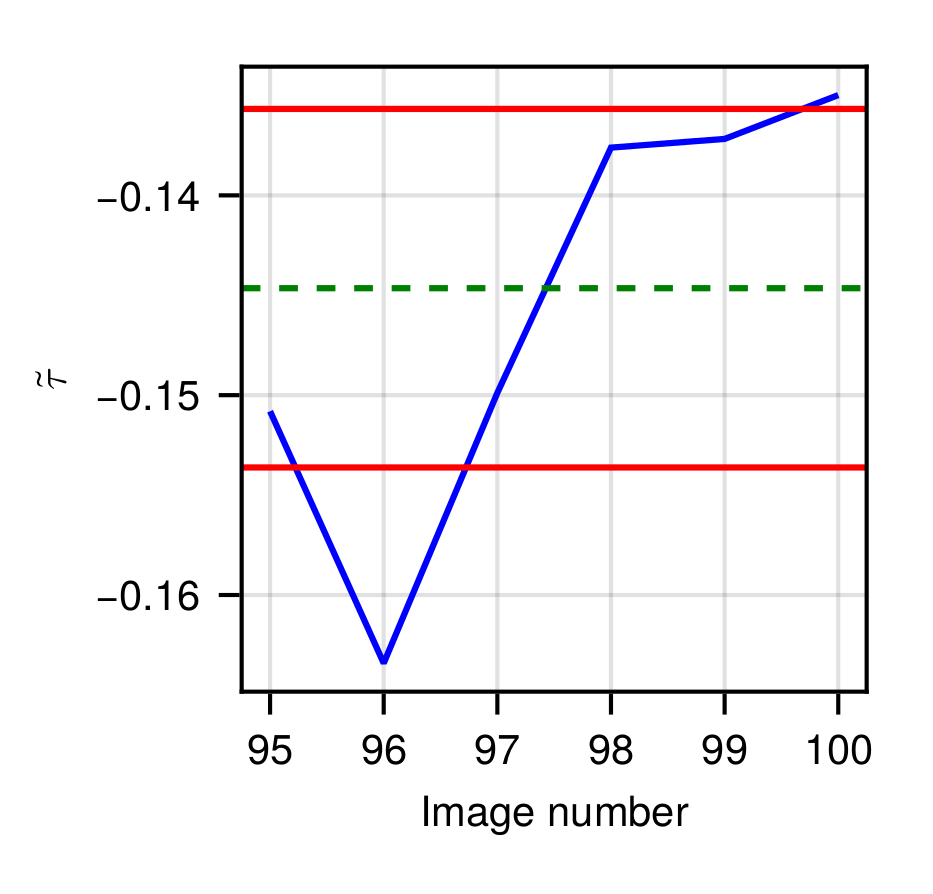}
         \caption{$\widetilde{\tau}$-chart ($\lambda = 1.0$)}
         \label{fig:Shewhart_textile_short}
     \end{subfigure}  
        \caption{$\widetilde{\tau}$-EWMA chart with $\lambda=0.1$ and ARL$_0\approx100$ under ``IC-iid'' assumption in (a). $\widetilde{\tau}$-charts for OOC data with (b) $\lambda=0.1$ and (c) $\lambda=1$, using resampling approach with ARL$_0\approx20$. Grid sizes are $m = n = 249$.}
    \label{fig:SOP_charts_textile}
\end{figure}

\begin{figure}[!htb]
    \centering
    \includegraphics[width=1\linewidth]{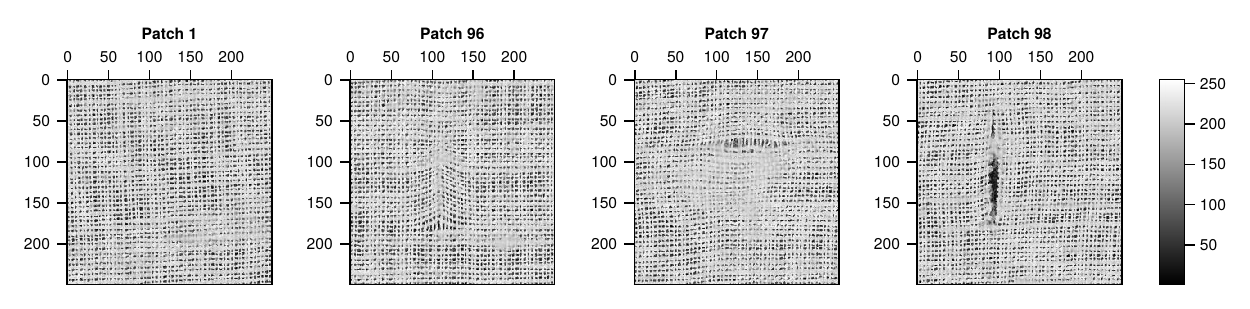}
    \caption{Textile images from IC-sample (Patch~1) and OOC-sample (Patches 96--98) with grid sizes are $m = n = 249$. 
    The local defects are a fiber direction change (96), a tear (97), and a hole (98), see  \citet[Figure~8]{Bui.Apley_2018}.}
    \label{fig:textile_patches}
\end{figure}

To monitor the sequence of textile images, we first start with the ``default IC-model'' again, namely the ``IC-iid'' assumption. Since the sample size $m=n=249$ is not covered by Table~\ref{tab:ICiid_design}, we first have to determine the chart design for the intended $\widetilde{\tau}$-EWMA chart with $\lambda=0.1$. In view of having only 100 images available in the R-package \href{https://cran.r-project.org/package=textile}{\nolinkurl{textile}}, we set the IC-target to ARL$_0=100$, leading to the control limit $l_{\widetilde{\tau}, \lambda}=0.0010174$. Figure \ref{fig:EWMA_textile} shows the resulting $\widetilde{\tau}$-EWMA chart for the first 50 Phase-I observations. It is clearly visible that the ``IC-iid'' assumption for the control chart does not hold, \ie the chosen IC-model is not adequate for the textile data. This is plausible given the regular spatial structure of the textile images; see Patch~1 in Figure~\ref{fig:textile_patches} as an example, which implies a particular form of spatial dependence. However, it is not clear with which kind of stochastic model the apparent spatial dependence can be described. 

Thus, to apply the $\widetilde{\tau}$-EWMA chart to the data, we proceed in a data-driven but still nonparametric manner and utilize a kind of Efron bootstrap approach. Here, we use the 96 IC-observations from our Phase-I data for resampling, the IC-mean of which equals $\approx -0.145$ (instead of~0 as under the ``IC-iid'' assumption). 
Note that negative $\widetilde{\tau}$-values imply that the maximal ranks mainly occur along rows and columns, which is plausible because of the vertical and horizontal fiber directions. 
According to our nonparametric bootstrap approach, we sequentially resample from the 96 IC-values of~$\widetilde{\tau}$, apply the $\widetilde{\tau}$-EWMA chart (initialized with the IC-mean) to the resulting image sequence, and determine the respective run length. ARLs are then computed based on $10^6$ replications again. Since the R-package \href{https://cran.r-project.org/package=textile}{\nolinkurl{textile}} provides only six OOC-images for Phase~II, we set the IC-target to ARL$_0=20$, leading to the control limit $l_{\widetilde{\tau}, \lambda}=0.0012559$.
The application of the $\widetilde{\tau}$-EWMA chart with $\lambda=0.1$ to the Phase-II data is shown in Figure~\ref{fig:EWMA_textile_short}. It triggers an alarm for Patches 96, 97, and 98, showing different local defects; see Figure \ref{fig:textile_patches}. 
For interpretation, we also show the raw $\widetilde{\tau}$-values in the form of a Shewhart chart; see Figure~\ref{fig:Shewhart_textile_short}. 
Its control limit is $l_{\widetilde{\tau}, \lambda}=0.0089714$, leading to $\mathrm{ARL}_0\approx 18.8$ (note that due to the limited amount of Phase-I data, the IC-target ARL can only be met approximately for the Shewhart chart).
It can be seen that Patch~96 leads to a particularly strong negative value, \ie the maximal ranks occur even rarer along the diagonals. This is probably explained by the long vertical fibers below the fiber direction change; see Figure~\ref{fig:textile_patches}. Altogether, the $\widetilde{\tau}$-EWMA chart successfully detected the disturbances of the regular textile structure, which correspond to changes in spatial dependence. 
The same conclusion applies to the more sophisticated $\widetilde{\tau}^{\textup{BP}(3)}$-chart. But as it does not lead to additional alarms, we skip the plot this time.

\section{Conclusions and Future Research}
\label{Conclusions}
There is an increasing interest in monitoring of streams of rectangular data in the SPM literature. In this study, we considered data that were collected as a regular two-dimensional grid. To monitor such streams of data for possible spatial dependence, we proposed four classes of novel nonparametric control charts based on $2\times 2$-SOPs. It is assumed that the rectangular data sets originate from continuously distributed RVs. However, for discrete RVs and limited measurement precision, a randomization approach was also proposed, as there may be ties in the rankings in the SOPs. It is further assumed that the rectangular sets are generated independently of each other for different $t$, but the RVs within the $t$th  rectangular set can follow any joint distribution. 

To construct the proposed control charts, the relative frequency vectors of the SOPs were used both solely to calculate the control statistics for the Shewhart-type control charts, which have no memory, and as an EWMA of the process of frequency vectors to develop control charts with inherent memory. Each of the proposed four classes of control charts have a different control statistic, as well as different approaches to account for higher-order spatial dependencies. All of the methods used in this study are also provided in a publicly available \texttt{Julia} package on \href{https://github.com/AdaemmerP/OrdinalPatterns.jl}{GitHub}.

Extensive performance evaluations were carried out using zero-state ARL simulations. In the OOC simulations, we used various spatially dependent DGPs---unilateral and bilateral, linear and nonlinear, contaminated and noncontaminated, discrete and continuous---to evaluate the out-of-control ARL performance of the charts.  As a natural competitor to our SOP-based charts, analogous charts based on spatial ACF were also considered.  Note that the ACF approach is parametric in nature, so the chart design requires a full specification of the in-control model. By contrast, SOP chart designs can be used immediately with any continuously distributed DGP. In particular, no prior Phase-I analysis is necessary, but these charts could even be applied within Phase-I for a retrospective data analysis. We showed that the $\widetilde{\tau}$ chart is generally superior, and in those cases where another chart is optimal, the $\widetilde{\tau}$ chart performs similarly well. Hence, the $\widetilde{\tau}$-EWMA chart constitutes a universally applicable solution to uncover spatial dependence. More precisely, the $\widetilde{\tau}^{(\onei)}$-EWMA chart with delay $\fd=\1$ appears to be most relevant for applications in practice, as long as the anticipated OOC-scenarios exhibit spatial dependence (also) at lag~1. If an OOC-situation might manifest itself solely in higher-order dependencies, then the use of a BP-modification of the $\widetilde{\tau}$ chart is recommended. Finally, three real-world cases were used to illustrate the applicability and effectiveness of the proposed method, namely heavy rainfall in Germany, war-related fires in eastern Ukraine, and manufacturing defects in textile production. 

There are various related directions for future research. For example, we used the types proposed by \citet{Bandt.Wittfeld_2023} for defining our control charts. Distinguishing three types only, however, constitutes a rather rough discretization of the original data, which may result in information loss. Therefore, we recommend developing ``refined types'' (\i.e. a finer partitioning of the set of SOPs) and the corresponding control charts, and analysing whether these lead to improved detection performance for some OOC scenarios. Another direction is inspired by the discreteness of data, often observed in real applications (recall those discussed in Section~\ref{Real-Data Application}). Our current solution is to add uniform noise to integer data to break ties, particularly useful in situations where constant rectangles might be observed under IC conditions (\eg no precipitation or no fires at all, recall Section~\ref{Real-Data Application}). In other applications, however, ties may contain valuable information about the actual spatial dependence structure, which is obviously lost after noise is added. In such cases, the concept of generalized OPs from \citet{Weiss.Schnurr_2024} might be useful. It should be analyzed whether a feasible definition of ``generalized SOPs'' is possible and how corresponding control charts work in practice. Finally, in view of our positive experience with BP-type extensions of SOP-statistics, it is recommended to consider such BP-approaches also in other fields where ordinal patterns are used, such as classical hypothesis tests for serial or spatial dependence.

\subsubsection*{Data}
The war-related fire data are available at \url{https://github.com/TheEconomist/the-economist-war-fire-model}. 
The Radolan rain data are available at \url{https://opendata.dwd.de/climate\_environment/CDC/grids\_germany/hourly/radolan/}. The images of textile material are available in the R-package \href{https://cran.r-project.org/package=textile}{\nolinkurl{textile}} \citep{Bui.Apley_2018}.

\subsubsection*{Software}
We used the programming language \texttt{Julia} \citep{Bezanson.etal_2017} for the simulations and analyses and the package \texttt{Makie} \citep{Danisch.Krumbiegel_2021} for the visualizations. We have implemented all methods used in this study in a \texttt{Julia} package that is publicly available on GitHub at \url{https://github.com/AdaemmerP/OrdinalPatterns.jl}.

\subsubsection*{Acknowledgements}
The authors thank the editor, the associate editor, and the referees for their useful comments on an earlier draft of this article. The authors are grateful to Professor Scott D.\ Grimshaw (Brigham Young University) for providing the bottle thickness data discussed in Example~\ref{bspBottles}. The authors would like to thank Dr.\ Svenja Fischer (Wageningen University) and Professor Alexander Schnurr (University of Siegen) for providing details of the Radolan rain data. The authors are also grateful to Dr. Sondre Solstad (\texttt{The Economist}) for an insightful discussion of the construction and intricacies of tracking war-related fires in Ukraine. Computational resources have been provided by the HPC clusters at the University of Greifswald and the Helmut Schmidt University.  The project hpc.bw (HPC cluster HSUper) is funded by dtec.bw -- Digitalization and Technology Research Center of the Bundeswehr. dtec.bw is funded by the European Union -- NextGenerationEU. 

\bibliographystyle{abbrvnat}

\appendix

\numberwithin{table}{section}

\section{Out-of-control ARL performance results}\label{sec:appendix_ooc_results}

\begin{table}[!htb]
    \centering
    \caption{OOC-ARLs of SOP-EWMA charts and competing $\widehat{\rho}$-chart for unilateral SAR$(1,1)$ model, for different $m, n, \alpha_1, \alpha_2, \alpha_3$ combinations, smoothing parameter $\lambda=0.1$, and target ARL$_0\approx370$; simulated with $10^5$ replications.}
    \begin{threeparttable}
    \begin{tabular}{ccccrcrcrcrcc}
    \toprule
    $m,n$ & $\alpha_1$ & $\alpha_2$ & $\alpha_3$
    &  $\widehat{\tau}$-chart
    &&$\widehat{\kappa}$-chart
    &&$\widetilde{\tau}$-chart
    &&$\widetilde{\kappa}$-chart
    & $\widehat{\rho}$-chart \\
    \midrule
    $10,10$ & $0.1$ & $0.1$ & $0.1$ & $72.01$ &  & $90.46$ &  & $52.78$ &  & $208.1$ & $\bf{9.49}$\\
    $15,15$ &  &  &  & $36.88$ &  & $45.33$ &  & $25.86$ &  & $140.42$ & $\bf{5.05}$\\
    $25,25$ &  &  &  & $15.97$ &  & $19.04$ &  & $11.53$ &  & $69.06$ & $\bf{2.80}$\\
    $40,25$ &  &  &  & $11.32$ &  & $13.18$ &  & $8.34$ &  & $46.98$ & $\bf{2.22}$\\
    \midrule
    $10,10$ & $0.2$ & $0.2$ & $0.2$ & $19.38$ &  & $28.96$ &  & $15.52$ &  & $66.94$ & $\bf{3.16}$\\
    $15,15$ &  &  &  & $10.6$ &  & $14.78$ &  & $8.63$ &  & $34.85$ & $\bf{2.06}$\\
    $25,25$ &  &  &  & $5.51$ &  & $7.25$ &  & $4.62$ &  & $15.42$ & $\bf{1.13}$\\
    $40,25$ &  &  &  & $4.22$ &  & $5.45$ &  & $3.59$ &  & $10.97$ & $\bf{1.0}$\\
    \midrule
    $10,10$ & $0.2$ & $0.2$ & $0.5$ & $25.51$ &  & $171.05$ &  & $212.29$ &  & $24.4$ & $\bf{1.57}$\\
    $15,15$ &  &  &  & $13.61$ &  & $100.11$ &  & $134.7$ &  & $13.09$ & $\bf{1.0}$\\
    $25,25$ &  &  &  & $6.89$ &  & $43.76$ &  & $64.49$ &  & $6.61$ & $\bf{1.0}$\\
    $40,25$ &  &  &  & $5.21$ &  & $29.38$ &  & $44.01$ &  & $5.0$ & $\bf{1.0}$\\
    \midrule
    $10,10$ & $0.4$ & $0.3$ & $0.1$ & $5.05$ &  & $6.90$ &  & $4.33$ &  & $13.03$ & $\bf{2.47}$\\
    $15,15$ &  &  &  & $3.27$ &  & $4.29$ &  & $2.86$ &  & $7.57$ & $\bf{1.63}$\\
    $25,25$ &  &  &  & $2.07$ &  & $2.55$ &  & $1.96$ &  & $4.17$ & $\bf{1.0}$\\
    $40,25$ &  &  &  & $1.81$ &  & $2.09$ &  & $1.56$ &  & $3.26$ & $\bf{1.0}$\\
    \bottomrule
    \end{tabular}
    \begin{tablenotes}
    NOTE: The table's largest standard error for the ARL is 0.64 
    \end{tablenotes} 
    \end{threeparttable}
    \label{tab:DGP3}
\end{table}

\newpage
 \begin{table}[!htb]
    \centering
    \caption{OOC-ARLs of SOP-EWMA charts and competing $\widehat{\rho}$-chart for unilateral SINAR$(1,1)$ model, for different $m, n, \alpha_1, \alpha_2, \alpha_3$ combinations, smoothing parameter $\lambda=0.1$, and target ARL$_0\approx370$; simulated with $10^5$ replications.}
    \begin{threeparttable}
    \begin{tabular}{ccccrcrcrcrcc}
     \toprule
    $m,n$ & $\alpha_1$ & $\alpha_2$ & $\alpha_3$
    &  $\widehat{\tau}$-chart
    &&$\widehat{\kappa}$-chart
    &&$\widetilde{\tau}$-chart
    &&$\widetilde{\kappa}$-chart
    & $\widehat{\rho}$-chart \\
    \midrule
    $10,10$ & $0.1$ & $0.1$ & $0.1$ & $74.44$ &  & $92.08$ &  & $54.62$ &  & $214.0$ & $\bf{9.49}$\\
    $15,15$ &  &  &  & $38.29$ &  & $46.23$ &  & $26.63$ &  & $146.33$ & $\bf{5.05}$\\
    $25,25$ &  &  &  & $16.6$ &  & $19.43$ &  & $11.86$ &  & $72.59$ & $\bf{2.79}$\\
    $40,25$ &  &  &  & $11.72$ & & $13.46$ &  & $8.54$ &  & $49.83$ & $\bf{2.23}$\\
    \midrule
    $10,10$ & $0.2$ & $0.2$ & $0.2$ & $19.83$ &  & $29.25$ &  & $15.77$ &  & $70.03$ & $\bf{3.16}$\\
    $15,15$ &  &  &  & $10.82$ &  & $14.93$ &  & $8.73$ &  & $36.2$ & $\bf{2.06}$\\
    $25,25$ &  &  &  & $5.62$ &  & $7.3$ &  & $4.68$ &  & $16.02$ & $\bf{1.13}$\\
    $40,25$ &  &  &  & $4.31$ &  & $5.48$ &  & $3.63$ &  & $11.37$ & $\bf{1.0}$\\
    \midrule
    $10,10$ & $0.2$ & $0.2$ & $0.5$ & $25.68$ &  & $172.17$ &  & $214.69$ &  & $24.8$ & $\bf{1.57}$\\
    $15,15$ &  &  &  & $13.85$ &  & $100.61$ &  & $137.38$ &  & $13.26$ & $\bf{1.0}$\\
    $25,25$ &  &  &  & $6.97$ &  & $43.85$ &  & $66.3$ &  & $6.68$ & $\bf{1.0}$\\
    $40,25$ &  &  &  & $5.25$ &  & $29.47$ &  & $45.33$ &  & $5.06$ & $\bf{1.0}$\\
    \midrule
    $10,10$ & $0.4$ & $0.3$ & $0.1$ & $5.12$ &  & $6.92$ &  & $4.35$ &  & $13.32$ & $\bf{2.47}$\\
    $15,15$ &  &  &  & $3.3$ &  & $4.3$ &  & $2.88$ &  & $7.71$ & $\bf{1.63}$\\
    $25,25$ &  &  &  & $2.08$ &  & $2.56$ &  & $1.97$ &  & $4.22$ & $\bf{1.0}$\\
    $40,25$ &  &  &  & $1.83$ &  & $2.09$ &  & $1.58$ &  & $3.31$ & $\bf{1.0}$ \\
    \bottomrule
    \end{tabular}
    \begin{tablenotes}
    NOTE: The table's largest standard error for the ARL is 0.65. 
    \end{tablenotes} 
    \end{threeparttable}
    \label{tab:DGP4}
\end{table}

\begin{table}[!htb]
    \centering
    \caption{OOC-ARLs of SOP-EWMA charts and competing $\widehat{\rho}$-chart for unilateral SAR$(1,1)$ model like in Table~\ref{tab:DGP3}, smoothing parameter $\lambda=0.1$, but where 10\,\% of observations contaminated by AOs; simulated with $10^5$ replications.}
    \begin{threeparttable}
    \begin{tabular}{cccccrcrcrcrcc}
     \toprule
    $m,n$ & $\alpha_1$ & $\alpha_2$ & $\alpha_3$ & AO 
    &  $\widehat{\tau}$-chart
    &&$\widehat{\kappa}$-chart
    &&$\widetilde{\tau}$-chart
    &&$\widetilde{\kappa}$-chart
    & $\widehat{\rho}$-chart \\
    \midrule
    $10,10$ & $0.1$ & $0.1$ & $0.1$ & $10$ & $149.34$ &  & $123.08$ &  & $\bf{90.48}$ &  & $323.06$ & $463.76$\\
    $15,15$ &  &  &  &  & $85.98$ &  & $64.54$ &  & $\bf{44.73}$ &  & $296.1$ & $216.56$\\
    $25,25$ &  &  &  &  & $37.45$ &  & $26.75$ &  & $\bf{18.63}$ &  & $234.09$ & $76.89$\\
    $40,25$ &  &  &  &  & $25.16$ &  & $18.21$ &  & $\bf{12.97}$ &  & $195.44$ & $47.65$\\
    \midrule
    $10,10$ & $0.2$ & $0.2$ & $0.2$ & $10$ & $44.94$ &  & $40.8$ &  & $\bf{26.04}$ &  & $201.65$ & $96.28$\\
    $15,15$ &  &  &  &  & $22.93$ &  & $20.31$ &  & $\bf{13.51}$ &  & $135.47$ & $33.41$\\
    $25,25$ &  &  &  &  & $10.55$ &  & $9.43$ &  & $\bf{6.72}$ &  & $66.61$ & $12.32$\\
    $40,25$ &  &  &  &  & $7.7$ &  & $6.94$ &  & $\bf{5.07}$ &  & $45.59$ & $8.6$\\
    \midrule
    $10,10$ & $0.2$ & $0.2$ & $0.5$ & $10$ & $52.1$ &  & $277.17$ &  & $299.35$ &  & $51.02$ & $\bf{9.22}$\\
    $15,15$ &  &  &  &  & $26.56$ &  & $197.52$ &  & $212.28$ &  & $25.84$ & $\bf{4.8}$\\
    $25,25$ &  &  &  &  & $12.03$ &  & $103.42$ &  & $112.29$ &  & $11.72$ & $\bf{2.6}$\\
    $40,25$ &  &  &  &  & $8.69$ &  & $71.33$ &  & $78.24$ &  & $8.49$ & $\bf{2.09}$\\
    \midrule
    $10,10$ & $0.4$ & $0.3$ & $0.1$ & $10$ & $9.46$ &  & $8.92$ &  & $\bf{6.25}$ &  & $49.48$ & $34.36$\\
    $15,15$ &  &  &  &  & $5.69$ &  & $5.37$ &  & $\bf{3.95}$ &  & $26.32$ & $12.95$\\
    $25,25$ &  &  &  &  & $3.25$ &  & $3.1$ &  & $\bf{2.37}$ &  & $12.15$ & $5.72$\\
    $40,25$ &  &  &  &  & $2.58$ &  & $2.46$ &  & $\bf{2.01}$ &  & $8.83$ & $4.25$\\
    \midrule 
    $10,10$ & $0.1$ & $0.1$ & $0.1$ & $\pm 10$ & $154.87$ &  & $120.35$ &  & $\bf{91.08}$ &  & $331.11$ & $550.84$\\
    $15,15$ &  &  &  &  & $90.59$ &  & $63.2$ &  & $\bf{44.44}$ &  & $307.85$ & $273.57$\\
    $25,25$ &  &  &  &  & $39.34$ &  & $26.26$ &  & $\bf{18.66}$ &  & $256.87$ & $93.36$\\
    $40,25$ &  &  &  &  & $26.34$ &  & $17.85$ &  & $\bf{13.02}$ &  & $221.77$ & $57.19$\\
    \midrule 
    $10,10$ & $0.2$ & $0.2$ & $0.2$ & $\pm 10$ & $47.7$ &  & $39.74$ &  & $\bf{26.12}$ &  & $218.07$ & $127.87$\\
    $15,15$ &  &  &  &  & $24.01$ &  & $19.86$ &  & $\bf{13.51}$ &  & $153.42$ & $39.92$\\
    $25,25$ &  &  &  &  & $10.99$ &  & $9.27$ &  & $\bf{6.73}$ &  & $78.91$ & $14.1$\\
    $40,25$ &  &  &  &  & $7.99$ &  & $6.82$ &  & $\bf{5.09}$ &  & $54.49$ & $9.73$\\
    \midrule 
    $10,10$ & $0.2$ & $0.2$ & $0.5$ & $\pm 10$ & $53.39$ &  & $282.04$ &  & $300.79$ &  & $52.76$ & $\bf{10.2}$\\
    $15,15$ &  &  &  &  & $27.13$ &  & $204.26$ &  & $213.11$ &  & $26.54$ & $\bf{5.27}$\\
    $25,25$ &  &  &  &  & $12.29$ &  & $107.91$ &  & $112.81$ &  & $11.99$ & $\bf{2.82}$\\
    $40,25$ &  &  &  &  & $8.89$ &  & $74.71$ &  & $78.47$ &  & $8.69$ & $\bf{2.23}$\\
    \midrule 
    $10,10$ & $0.4$ & $0.3$ & $0.1$ & $\pm 10$ & $9.85$ &  & $8.74$ &  & $\bf{6.27}$ &  & $57.94$ & $41.13$\\
    $15,15$ &  &  &  &  & $5.88$ &  & $5.28$ &  & $\bf{3.95}$ &  & $30.91$ & $14.73$\\
    $25,25$ &  &  &  &  & $3.35$ &  & $3.06$ &  & $\bf{2.37}$ &  & $14.05$ & $6.34$\\
    $40,25$ &  &  &  &  & $2.66$ &  & $2.43$ &  & $\bf{2.01}$ &  & $10.1$ & $4.69$\\
    \bottomrule
    \end{tabular}
    \begin{tablenotes}
    NOTE: The table's largest standard error for the ARL is 1.72. 
    \end{tablenotes} 
    \end{threeparttable}
    \label{tab:DGP5b}
\end{table}

\newpage
\begin{table}[!htb]
    \centering
    \caption{OOC-ARLs of SOP-EWMA charts and competing $\widehat{\rho}$-chart for unilateral SINAR$(1,1)$ model like in Table~\ref{tab:DGP4}, smoothing parameter $\lambda=0.1$, but where 10\,\% of observations contaminated by AOs from~$\poi(25)$; simulated with $10^5$ replications.}
    \begin{threeparttable}
    \begin{tabular}{ccccrcrcrcrcc}
     \toprule
    $m,n$ & $\alpha_1$ & $\alpha_2$ & $\alpha_3$
    &  $\widehat{\tau}$-chart
    &&$\widehat{\kappa}$-chart
    &&$\widetilde{\tau}$-chart
    &&$\widetilde{\kappa}$-chart
    & $\widehat{\rho}$-chart \\
    \midrule
    $10,10$ & $0.1$ & $0.1$ & $0.1$ & $160.16$ &  & $122.85$ &  & $\bf{92.96}$ &  & $332.93$ & $437.13$\\
    $15,15$ &  &  &  & $94.15$ &  & $64.34$ &  & $\bf{46.3}$ &  & $315.38$ & $197.22$\\
    $25,25$ &  &  &  & $41.06$ &  & $26.8$ &  & $\bf{19.17}$ &  & $267.66$ & $69.41$\\
    $40,25$ &  &  &  & $27.65$ &  & $18.19$ &  & $\bf{13.33}$ &  & $233.61$ & $43.22$\\
    \midrule
    $10,10$ & $0.2$ & $0.2$ & $0.2$ & $48.43$ &  & $40.53$ &  & $\bf{26.6}$ &  & $224.04$ & $37.57$\\
    $15,15$ &  &  &  & $24.59$ &  & $20.09$ &  & $\bf{13.73}$ &  & $158.31$ & $15.03$\\
    $25,25$ &  &  &  & $11.21$ &  & $9.35$ &  & $6.8$ &  & $82.51$ & $\bf{6.64}$\\
    $40,25$ &  &  &  & $8.16$ &  & $6.89$ &  & $5.14$ &  & $56.99$ & $\bf{4.92}$\\
    \midrule
    $10,10$ & $0.2$ & $0.2$ & $0.5$ & $52.44$ &  & $281.82$ &  & $295.92$ &  & $51.45$ & $\bf{2.68}$\\
    $15,15$ &  &  &  & $26.51$ &  & $202.89$ &  & $207.22$ &  & $26.11$ & $\bf{1.79}$\\
    $25,25$ &  &  &  & $12.09$ &  & $108.6$ &  & $109.4$ &  & $11.83$ & $\bf{1.02}$\\
    $40,25$ &  &  &  & $8.73$ &  & $75.15$ &  & $75.99$ &  & $8.58$ & $\bf{1.0}$\\
    \midrule
    $10,10$ & $0.4$ & $0.3$ & $0.1$ & $9.86$ &  & $8.85$ &  & $\bf{6.3}$ &  & $57.27$ & $8.71$\\
    $15,15$ &  &  &  & $5.88$ &  & $5.32$ &  & $\bf{3.96}$ &  & $30.35$ & $4.49$\\
    $25,25$ &  &  &  & $3.35$ &  & $3.09$ &  & $\bf{2.38}$ &  & $13.84$ & $2.47$\\
    $40,25$ &  &  &  & $2.65$ &  & $2.45$ &  & $\bf{2.02}$ &  & $9.95$ & $\bf{2.02}$\\
    \bottomrule
    \end{tabular}
    \begin{tablenotes}
    NOTE: The table's largest standard error for the ARL is 1.36. 
    \end{tablenotes} 
    \end{threeparttable}
    \label{tab:DGP6_mu5}
\end{table}

\begin{table}[!htb]
    \centering
    \caption{OOC-ARLs of SOP-EWMA charts and competing $\widehat{\rho}$-chart for unilateral SINAR$(1,1)$ model like in Table~\ref{tab:DGP3}, smoothing parameter $\lambda=0.1$, but with ZIP$(0.9, 5)$ innovations; simulated with $10^5$ replications.}
    \begin{threeparttable}
    \begin{tabular}{ccccrcrcrcrcc}
     \toprule
    $m,n$ & $\alpha_1$ & $\alpha_2$ & $\alpha_3$
    &  $\widehat{\tau}$-chart
    &&$\widehat{\kappa}$-chart
    &&$\widetilde{\tau}$-chart
    &&$\widetilde{\kappa}$-chart
    & $\widehat{\rho}$-chart \\
    \midrule
    $10,10$ & $0.1$ & $0.1$ & $0.1$ & $6.21$ &  & $5.85$ &  & $\bf{4.26}$ &  & $29.92$ & $9.63$\\
    $15,15$ &  &  &  & $3.93$ &  & $3.71$ &  & $\bf{2.82}$ &  & $15.89$ & $5.08$\\
    $25,25$ &  &  &  & $2.36$ &  & $2.25$ &  & $\bf{1.96}$ &  & $7.84$  & $2.8$\\
    $40,25$ &  &  &  & $2.01$ &  & $1.97$ &  & $\bf{1.52}$ &  & $5.9$   & $2.22$\\
    \midrule
    $10,10$ & $0.2$ & $0.2$ & $0.2$ & $3.7$ &  & $4.77$ &  & $3.17$ &  & $9.07$ & $\bf{3.15}$\\
    $15,15$ &  &  &  & $2.48$ &  & $3.11$ &  & $2.13$ &  & $5.48$ & $\bf{2.05}$\\
    $25,25$ &  &  &  & $1.72$ &  & $2.02$ &  & $1.38$ &  & $3.16$ & $\bf{1.13}$\\
    $40,25$ &  &  &  & $1.2$  &  & $1.78$ &  & $1.01$ &  & $2.52$ & $\bf{1.0}$\\
    \midrule
    $10,10$ & $0.2$ & $0.2$ & $0.5$ & $5.11$ &  & $298.41$ &  & $15.2$ &  & $5.72$ & $\bf{1.57}$\\
    $15,15$ &  &  &  & $3.32$ &  & $214.87$ &  & $8.59$ &  & $3.66$ & $\bf{1.0}$\\
    $25,25$ &  &  &  & $2.09$ &  & $113.44$ &  & $4.64$ &  & $2.25$ & $\bf{1.0}$\\
    $40,25$ &  &  &  & $1.78$ &  & $78.24$  &  & $3.6$  &  & $1.91$ & $\bf{1.0}$\\
    \midrule
    $10,10$ & $0.4$ & $0.3$ & $0.1$ & $2.37$ &  & $3.29$ &  & $\bf{2.11}$ &  & $4.71$ & $2.46$\\
    $15,15$ &  &  &  & $1.8$ &  & $2.2$ &  & $1.75$ &  & $3.08$ & $\bf{1.64}$\\
    $25,25$ &  &  &  & $\bf{1.0}$ &  & $1.51$ &  & $\bf{1.0}$ &  & $2.0$ & $\bf{1.0}$\\
    $40,25$ &  &  &  & $\bf{1.0}$ &  & $1.02$ &  & $\bf{1.0}$ &  & $1.68$ & $\bf{1.0}$\\
    \bottomrule
    \end{tabular}
    \begin{tablenotes}
    NOTE: The table's largest standard error for the ARL is 0.92. 
    \end{tablenotes} 
    \end{threeparttable}
    \label{tab:DGP7}
\end{table}

\newpage

\begin{table}[!htb]
    \centering
    \caption{OOC-ARLs of SOP-EWMA charts and competing $\widehat{\rho}$-chart for unilateral SQMA$(1,1)$-model for $\beta_1=\beta_2=\beta_3=0.8$, different $m, n$ combinations, smoothing parameter $\lambda=0.1$, and target ARL$_0\approx370$; simulated with $10^5$ replications.}
    \begin{threeparttable}
    \begin{tabular}{ccrcrcrcrcrc}
     \toprule
    $m,n$ & Model &  $\widehat{\tau}$-chart
    &&$\widehat{\kappa}$-chart
    &&$\widetilde{\tau}$-chart
    &&$\widetilde{\kappa}$-chart
    && $\widehat{\rho}$-chart \\
    \midrule
    $10,10$ & ``$1^2\;2^2\;3^2$'' & $13.64$ &  & $8.87$ &  & $\bf{6.9}$ &  & $165.91$ &  & $79.13$\\
    $15,15$ &  & $7.78$ &  & $5.33$ &  & $\bf{4.29}$ &  & $108.8$ &  & $85.5$\\
    $25,25$ &  & $4.25$ &  & $3.09$ &  & $\bf{2.56}$ &  & $52.0$  &  & $90.03$\\
    $40,25$ &  & $3.32$ &  & $2.45$ &  & $\bf{2.09}$ &  & $35.57$ &  & $91.58$\\
    \midrule
    $10,10$ & ``$1^2\;2^1\;3^2$'' & $7.86$ &  & $4.67$ &  & $\bf{3.92}$ &  & $168.38$ &  & $87.33$\\
    $15,15$ &  & $4.83$ &  & $3.06$ &  & $\bf{2.61}$ &  & $130.93$ &  & $84.24$\\
    $25,25$ &  & $2.83$ &  & $2.0$  &  & $\bf{1.89}$ &  & $78.02$ &  & $80.73$\\
    $40,25$ &  & $2.27$ &  & $1.73$ &  & $\bf{1.3}$  &  & $57.22$ &  & $79.96$\\
    \midrule
    $10,10$ & ``$1^1\;2^1\;3^2$'' & $7.37$ &  & $7.45$ &  & $\bf{5.16}$ &  & $32.62$ &  & $79.01$\\
    $15,15$ &  & $4.56$ &  & $4.58$ &  & $\bf{3.34}$ &  & $17.25$ &  & $84.12$\\
    $25,25$ &  & $2.69$ &  & $2.7$  &  & $\bf{2.08}$ &  & $8.42$ &  & $87.19$\\
    $40,25$ &  & $2.18$ &  & $2.17$ &  & $\bf{1.89}$ &  & $6.27$ &  & $87.81$\\
    \midrule
    $10,10$ & ``$1^2\;2^1\;3^1$'' & $12.58$ &  & $10.79$ &  & $7.68$ &  & $82.82$ &  & $\bf{4.13}$\\
    $15,15$ &  & $7.3$ &  & $6.35$ &  & $4.72$ &  & $44.64$ &  & $\bf{2.59}$\\
    $25,25$ &  & $4.02$ &  & $3.58$ &  & $2.78$ &  & $19.6$ &  & $\bf{1.76}$\\
    $40,25$ &  & $3.15$ &  & $2.83$ &  & $2.22$ &  & $13.77$ &  & $\bf{1.2}$\\
    \bottomrule
    \end{tabular}
    \begin{tablenotes}
    NOTE: The table's largest standard error for the ARL is 0.51. Label ``$1^a\;2^b\;3^c$'' indicates which MA-terms are squared. 
    \end{tablenotes} 
    \end{threeparttable}
    \label{tab:DGP8}
\end{table}

\newpage

\begin{table}[!htb]
    \centering
    \caption{OOC-ARLs of SOP-EWMA charts and competing $\widehat{\rho}$-chart for unilateral SQINMA$(1,1)$-model with \iid\ $\poi(5)$-innovations~$\epsilon_{t_{1}, t_{2}}$ and $\beta_1=\beta_2=\beta_3=0.8$, different $m, n$ combinations, smoothing parameter $\lambda=0.1$, and target ARL$_0\approx370$; simulated with $10^5$ replications.}
    \begin{threeparttable}
    \begin{tabular}{ccrcrcrcrcrc}
     \toprule
    $m,n$ & Model &  $\widehat{\tau}$-chart
    &&$\widehat{\kappa}$-chart
    &&$\widetilde{\tau}$-chart
    &&$\widetilde{\kappa}$-chart
    && $\widehat{\rho}$-chart \\
    \midrule
    $10,10$ & ``$1^2\;2^2\;3^2$'' & $4.92$ &  & $6.3$ &  & $\bf{4.09}$ &  & $13.45$ &  & $138.17$\\
    $15,15$ &  & $3.2$ &  & $3.96$ &  & $\bf{2.72}$ &  & $7.78$ &  & $45.14$\\
    $25,25$ &  & $2.04$ &  & $2.38$ &  & $\bf{1.92}$ &  & $4.26$ &  & $14.55$\\
    $40,25$ &  & $1.78$ &  & $2.02$ &  & $\bf{1.42}$ &  & $3.33$ &  & $9.78$\\
    \midrule
    $10,10$ & ``$1^2\;2^1\;3^2$'' & $5.06$ &  & $4.58$ &  & $\bf{3.45}$ &  & $24.12$ &  & $53.25$\\
    $15,15$ &  & $3.28$ &  & $3.0$ &  & $\bf{2.3}$ &  & $13.34$ &  & $18.72$\\
    $25,25$ &  & $2.07$ &  & $1.99$ &  & $\bf{1.65}$ &  & $6.8$ &  & $7.77$\\
    $40,25$ &  & $1.81$ &  & $1.7$  &  & $\bf{1.05}$ &  & $5.14$ &  & $5.65$\\
    \midrule
    $10,10$ & ``$1^1\;2^1\;3^2$'' & $45.04$ &  & $78.89$ &  & $39.23$ &  & $128.63$ &  & $\bf{12.36}$\\
    $15,15$ &  & $22.96$ &  & $39.15$ &  & $19.56$ &  & $72.88$ &  & $\bf{6.34}$\\
    $25,25$ &  & $10.63$ &  & $16.69$ &  & $9.18$ &  & $31.51$ &  & $\bf{3.38}$\\
    $40,25$ &  & $7.75$  &  & $11.71$ &  & $6.77$ &  & $21.53$ &  & $\bf{2.65}$\\
    \midrule
    $10,10$ & ``$1^2\;2^1\;3^1$'' & $23.85$ &  & $47.63$ &  & $\bf{22.0}$ &  & $67.97$ &  & $484.57$\\
    $15,15$ &  & $12.75$ &  & $23.37$ &  & $\bf{11.67}$ &  & $35.35$ &  & $351.6$\\
    $25,25$ &  & $6.45$ &  & $10.61$ &  & $\bf{5.96}$ &  & $15.66$ &  & $144.19$\\
    $40,25$ &  & $4.88$ &  & $7.73$ &  & $\bf{4.54}$ &  & $11.12$ &  & $89.31$\\
    \bottomrule
    \end{tabular}
    \begin{tablenotes}
    NOTE: The table's largest standard error for the ARL is 1.5. Label ``$1^a\;2^b\;3^c$'' indicates which MA-terms are squared.  
    \end{tablenotes} 
    \end{threeparttable}
    \label{tab:DGP9}
\end{table}

\newpage

\begin{table}[!htb]
    \centering
    \caption{OOC-ARLs of SOP-EWMA charts and competing $\widehat{\rho}$-chart for bilateral SAR$(1)$-model for different $m, n, a_1, a_2, a_3,a_4$ combinations, smoothing parameter $\lambda=0.1$, and target ARL$_0\approx370$; simulated with $10^5$ replications.}
    \begin{threeparttable}
    \begin{tabular}{cccccrcrcrcrcc}
     \toprule
    $m,n$ & $a_1$ & $a_2$ & $a_3$ & $a_4$ 
    &  $\widehat{\tau}$-chart
    &&$\widehat{\kappa}$-chart
    &&$\widetilde{\tau}$-chart
    &&$\widetilde{\kappa}$-chart
    & $\widehat{\rho}$-chart \\
    \midrule
    $10,10$ & $0.1$ & $0.1$ & $0.1$ & $0.1$ & $8.77$ &  & $10.92$ &  & $\bf{6.81}$ &  & $30.42$ & $36.47$\\
    $15,15$ &  &  &  &  & $5.31$ &  & $6.37$ &  & $\bf{4.25}$ &  & $16.23$ & $12.98$\\
    $25,25$ &  &  &  &  & $3.07$ &  & $3.58$ &  & $\bf{2.53}$ &  & $7.93$ & $5.7$\\
    $40,25$ &  &  &  &  & $2.45$ &  & $2.83$ &  & $\bf{2.08}$ &  & $5.95$ & $4.26$\\
    \midrule
    $10,10$ & $0.05$ & $0.05$ & $0.15$ & $0.15$ & $8.86$ &  & $10.98$ &  & $\bf{6.85}$ &  & $30.74$ & $30.84$\\
    $15,15$ &  &  &  &  & $5.36$ &  & $6.42$ &  & $\bf{4.27}$ &  & $16.34$ & $11.53$\\
    $25,25$ &  &  &  &  & $3.09$ &  & $3.6$ &  & $\bf{2.54}$ &  & $8.01$ & $5.23$\\
    $40,25$ &  &  &  &  & $2.46$ &  & $2.85$ &  & $\bf{2.09}$ &  & $6.0$ & $3.93$\\
    \midrule
    $10,10$ & $0.05$ & $0.15$ & $0.05$ & $0.15$ & $8.85$ &  & $11.01$ &  & $\bf{6.87}$ &  & $30.75$ & $46.61$\\
    $15,15$ &  &  &  &  & $5.35$ &  & $6.41$ &  & $\bf{4.27}$ &  & $16.36$ & $15.31$\\
    $25,25$ &  &  &  &  & $3.09$ &  & $3.61$ &  & $\bf{2.55}$ &  & $8.03$ & $6.43$\\
    $40,25$ &  &  &  &  & $2.46$ &  & $2.85$ &  & $\bf{2.08}$ &  & $6.0$ & $4.73$\\
    \bottomrule
    \end{tabular}
    \begin{tablenotes}
    NOTE: The table's largest standard error for the ARL is 0.12. 
    \end{tablenotes} 
    \end{threeparttable}
    \label{tab:DGP10}
    \vspace{3ex}
    \caption{OOC-ARLs of SOP-EWMA charts and competing $\widehat{\rho}$-chart for bilateral SAR$(1)$-model like in Table~\ref{tab:DGP10}, smoothing parameter $\lambda=0.1$, but where 10\,\% of observations are contaminated by AOs; simulated with $10^5$ replications.
    }
    \begin{threeparttable}
    \begin{tabular}{cccccrcrcrcrcc}
     \toprule
    $m,n$ & $a_1$ & $a_2$ & $a_3$ & $a_4$ 
    &  $\widehat{\tau}$-chart
    &&$\widehat{\kappa}$-chart
    &&$\widetilde{\tau}$-chart
    &&$\widetilde{\kappa}$-chart
    & $\widehat{\rho}$-chart \\
    \midrule
    $10,10$ & $0.1$ & $0.1$ & $0.1$ & $0.1$ & $19.28$ &  & $14.52$ &  & $\bf{10.37}$ &  & $146.32$ & $371.24$\\
    $15,15$ &  &  &  &  & $10.55$ &  & $8.13$ &  & $\bf{6.12}$ &  & $90.48$ & $115.7$\\
    $25,25$ &  &  &  &  & $5.5$ &  & $4.41$ &  & $\bf{3.47}$ &  & $41.88$ & $33.06$\\
    $40,25$ &  &  &  &  & $4.22$ &  & $3.43$ &  & $\bf{2.74}$ &  & $28.51$ & $20.72$\\
    \midrule
    $10,10$ &  $0.05$ & $0.05$ & $0.15$ & $0.15$ & $19.45$ &  & $14.63$ &  & $\bf{10.46}$ &  & $147.33$ & $336.06$\\
    $15,15$ &  &  &  &  & $10.65$ &  & $8.2$ &  & $\bf{6.17}$ &  & $91.41$ & $101.0$\\
    $25,25$ &  &  &  &  & $5.55$ &  & $4.43$ &  & $\bf{3.49}$ &  & $42.33$ & $29.18$\\
    $40,25$ &  &  &  &  & $4.25$ &  & $3.45$ &  & $\bf{2.76}$ &  & $28.93$ & $18.62$\\
    \midrule
    $10,10$ & $0.05$ & $0.15$ & $0.05$ & $0.15$ & $19.5$ &  & $14.58$ &  & $\bf{10.47}$ &  & $147.77$ & $426.56$\\
    $15,15$ &  &  &  &  & $10.68$ &  & $8.19$ &  & $\bf{6.19}$ &  & $91.35$ & $141.25$\\
    $25,25$ &  &  &  &  & $5.55$ &  & $4.43$ &  & $\bf{3.49}$ &  & $42.13$ & $40.18$\\
    $40,25$ &  &  &  &  & $4.25$ &  & $3.46$ &  & $\bf{2.76}$ &  & $28.94$ & $24.73$\\
    \bottomrule
    \end{tabular}
    \begin{tablenotes}
    NOTE: The table's largest standard error for the ARL is 1.32. 
    \end{tablenotes} 
    \end{threeparttable}
    \label{tab:DGP11}
\end{table}

\newpage

\begin{table}[!htb]
    \centering
    \caption{OOC-ARLs of SOP-EWMA charts and competing $\widehat{\rho}$-chart for bilateral SQMA$(1)$-model 
    for $b_1=\ldots=b_4=0.8$, for different $m, n$ combinations, smoothing parameter $\lambda=0.1$, and target ARL$_0\approx370$; simulated with $10^5$ replications.}
    \begin{threeparttable}
    \begin{tabular}{ccrcrcrcrcrc}
     \toprule
    $m,n$ & Model &  $\widehat{\tau}$-chart
    &&$\widehat{\kappa}$-chart
    &&$\widetilde{\tau}$-chart
    &&$\widetilde{\kappa}$-chart
    && $\widehat{\rho}$-chart\\
    \midrule
    $10,10$ & ``$1^2\;2^2\;3^2\;4^2$'' & $43.3$ &  & $\bf{28.49}$ &  & $34.5$ &  & $32.68$ &  & $37.11$\\
    $15,15$ &  & $30.67$ &  & $\bf{17.63}$ &  & $26.68$ &  & $18.27$ &  & $33.68$\\
    $25,25$ &  & $17.66$ &  & $9.45$ &  & $17.16$ &  & $\bf{9.1}$ &  & $31.13$\\
    $40,25$ &  & $13.31$ &  & $7.16$ &  & $13.42$ &  & $\bf{6.78}$ &  & $30.34$\\
    \midrule
    $10,10$ & ``$1^2\;2^1\;3^2\;4^1$'' & $12.83$ &  & $5.68$ &  & $\bf{5.05}$ &  & $405.43$ &  & $48.05$\\
    $15,15$ &  & $7.4$ &  & $3.65$ &  & $\bf{3.28}$ &  & $296.66$ &  & $41.95$\\
    $25,25$ &  & $4.07$ &  & $2.24$ &  & $\bf{2.07}$ &  & $156.89$ &  & $38.11$\\
    $40,25$ &  & $3.19$ &  & $1.91$ &  & $\bf{1.77}$ &  & $108.16$ &  & $36.94$\\
    \midrule
    $10,10$ & ``$1^2\;2^2\;3^1\;4^1$'' & $11.82$ &  & $6.05$ &  & $\bf{5.23}$ &  & $421.68$ &  & $6.14$\\
    $15,15$ &  & $6.98$ &  & $3.86$ &  & $\bf{3.39}$ &  & $414.0$ &  & $3.6$\\
    $25,25$ &  & $3.89$ &  & $2.35$ &  & $\bf{2.13}$ &  & $409.19$ &  & $2.15$\\
    $40,25$ &  & $3.07$ &  & $1.97$ &  & $\bf{1.79}$ &  & $403.95$ &  & $1.82$\\
    \bottomrule
    \end{tabular}
    \begin{tablenotes}
    NOTE: The table's largest standard error for the ARL is 1.31. Label ``$1^a\;2^b\;3^c\;4^d$'' indicates which MA-terms are squared. 
    \end{tablenotes} 
    \end{threeparttable}
    \label{tab:DGP12}
\end{table}

\newpage

\begin{table}[!htb]
\setlength{\tabcolsep}{5.2pt}
    \centering
    \footnotesize
    \caption{IC-ARLs of EWMA $\widehat{\rho}^{(\fhi)}$ and $\widehat{\rho}^{\textup{BP}(w)}$-charts with $\lambda=0.1$ and target ARL$_0\approx370$, for different $m, n$ combinations and different marginal distributions, but using the CLs corresponding to $\norm(0,1)$-distributed data; simulated with $10^5$ replications.}
    \begin{threeparttable}
\begin{tabular}{crrrrrrrrrrrr}
\toprule
& \multicolumn{6}{c}{$\widehat{\rho}^{(\fhi)}$} & \multicolumn{6}{c}{$\widehat{\rho}^{\textup{BP}(w)}$} \\
\cmidrule(lr){2-7} \cmidrule(lr){8-13}
& \multicolumn{3}{c}{t(2)} & \multicolumn{3}{c}{Exp(1)} & \multicolumn{3}{c}{t(2)} & \multicolumn{3}{c}{Exp(1)}
\\ \cmidrule(lr){2-4} \cmidrule(lr){5-7} \cmidrule(lr){8-10} \cmidrule(lr){11-13}
$m,n$ & $\fh = \1$ & $\fh = \2$ & $\fh = \3$ & $\fh = \1$ &$ \fh = \2 $& $\fh = \3 $&$ w = 1$ &$ w = 2$ & $w = 3$ & $w = $1 & $w = 2$ &$ w = 3$ \\ 
  \cmidrule(lr){1-7} \cmidrule(lr){8-13}
$10, 10$ & 586.55 & 567.75 & 539.21 & 464.35 & 464.42 & 462.61 & 764.78 & 1140.39 & 1644.67 & 520.71 & 551.48 & 569.07 \\ 
  $15,15$ & 524.62 & 512.91 & 494.88 & 418.38 & 416.37 & 417.25 & 608.26 & 839.62 & 1174.09 & 444.27 & 464.33 & 477.33 \\ 
  $25,25$ & 453.77 & 448.56 & 442.99 & 388.94 & 388.11 & 389.20 & 435.86 & 497.39 & 627.80 & 400.31 & 407.00 & 414.64 \\ 
  $40,25$ & 435.65 & 428.80 & 425.16 & 382.73 & 382.73 & 380.58 & 381.57 & 392.60 & 455.47 & 388.81 & 392.90 & 397.32 \\ 
   \bottomrule
    \end{tabular}
    \begin{tablenotes}
    NOTE: The table's largest standard error for the ARL is 5.12.
    \end{tablenotes} 
    \end{threeparttable}
    \label{tab:rename11}
\end{table}

\begin{table}[!htb]
\setlength{\tabcolsep}{4.2pt}
    \centering
    \footnotesize
    \caption{OOC-ARLs of EWMA $\widetilde{\tau}^{(\fdi)}$-charts and competing $\widehat{\rho}^{(\fhi)}$-chart for unilateral SAR$(1,1)$ model like in Tables~\ref{tab:DGP3} and~\ref{tab:DGP5b}, for different $m, n$ combinations and  $\alpha_1=0.4, \alpha_2=0.3, \alpha_3=0.1$, smoothing parameter $\lambda=0.1$, with and without 10\,\% of observations contaminated by AOs; simulated with $10^5$ replications.}
    \begin{threeparttable}
\begin{tabular}{crrrrrrrrrrrr}
\toprule
& \multicolumn{6}{c}{No outlier} & \multicolumn{6}{c}{With outlier} \\
\cmidrule(lr){2-7} \cmidrule(lr){8-13}
& \multicolumn{3}{c}{$\widetilde{\tau}^{(\fdi)}$} & \multicolumn{3}{c}{$\widehat{\rho}^{(\fhi)}$} & \multicolumn{3}{c}{$\widetilde{\tau}^{(\fdi)}$} & \multicolumn{3}{c}{$\widehat{\rho}^{(\fhi)}$}  
\\ \cmidrule(lr){2-4} \cmidrule(lr){5-7} \cmidrule(lr){8-10} \cmidrule(lr){11-13}
$m,n$ & $\fd = \1$ & $\fd = \2$ & $ \fd = \3$ & $\fh = \1$ & $\fh = \2$ & $ \fh = \3$ & $\fd = \1$ & $\fd = \2$ & $ \fd = \3$ & $\fh = \1$ & $\fh = \2$ & $ \fh = \3$ \\ 
  \cmidrule(lr){1-7} \cmidrule(lr){8-13}
$10, 10$ & 4.33 & 16.32 & 70.17 & \textbf{2.47} & 7.46 & 40.58 & \textbf{6.26} & 27.45 & 118.54 & 34.33 & 206.26 & 484.79 \\ 
  $15,15$ & 2.86 & 8.77 & 37.62 & \textbf{1.63} & 3.36 & 8.74 & \textbf{3.94} & 13.68 & 65.70 & 12.88 & 52.09 & 197.69 \\ 
  $25,25$ & 1.96 & 4.59 & 17.07 & \textbf{1.00} & 1.81 & 3.29 & \textbf{2.37} & 6.64 & 28.41 & 5.73 & 14.77 & 44.97 \\ 
  $40,25$ & 1.57 & 3.55 & 12.16 & \textbf{1.00} & 1.39 & 2.44 & \textbf{2.02} & 4.99 & 19.43 & 4.25 & 9.73 & 25.54 \\ 
   \bottomrule
    \end{tabular}
    \begin{tablenotes}
    NOTE: The table's largest standard error for the ARL is 1.51. 
    \end{tablenotes} 
    \end{threeparttable}
    \label{tab:rename12}
\end{table}

\begin{table}[!htb]
\setlength{\tabcolsep}{4.3pt}
    \centering
    \footnotesize
        \caption{OOC-ARLs of EWMA $\widetilde{\tau}^{\textup{BP}(w)}$-charts and competing $\widehat{\rho}^{\textup{BP}(w)}$-chart for unilateral SAR$(1,1)$ model like in Tables~\ref{tab:DGP3} and~\ref{tab:DGP5b}, for different $m, n$ combinations and  $\alpha_1=0.4, \alpha_2=0.3, \alpha_3=0.1$, smoothing parameter $\lambda=0.1$, with and without 10\,\% of observations contaminated by AOs; simulated with $10^5$ replications.}
    \begin{threeparttable}
\begin{tabular}{crrrrrrrrrrrr}
\toprule
& \multicolumn{6}{c}{No outlier} & \multicolumn{6}{c}{With outlier} \\
\cmidrule(lr){2-7} \cmidrule(lr){8-13}
& \multicolumn{3}{c}{$\widetilde{\tau}^{\textup{BP}(w)}$} & \multicolumn{3}{c}{$\widehat{\rho}^{\textup{BP}(w)}$} & \multicolumn{3}{c}{$\widetilde{\tau}^{\textup{BP}(w)}$} & \multicolumn{3}{c}{$\widehat{\rho}^{\textup{BP}(w)}$}  
\\ \cmidrule(lr){2-4} \cmidrule(lr){5-7} \cmidrule(lr){8-10} \cmidrule(lr){11-13}
$m,n$ & $w=1$ & $w=2$ & $w=3$ & $w=1$ & $w=2$ & $w=3$ & $w=1$ & $w=2$ & $w=3$ & $w=1$ & $w=2$ & $w=3$ \\ 
  \cmidrule(lr){1-7} \cmidrule(lr){8-13}
$10, 10$ & 4.32 & 3.88 & 3.87 & \textbf{1.83} & 2.01 & 2.21 & 6.26 & 5.45 & \textbf{5.40} & 16.26 & 19.46 & 22.71 \\ 
  $15,15$ & 2.86 & 2.57 & 2.56 & \textbf{1.08} & 1.28 & 1.53 & 3.95 & 3.48 & \textbf{3.43} & 7.30 & 8.15 & 9.39 \\ 
  $25,25$ & 1.96 & 1.84 & 1.84 & \textbf{1.00} & \textbf{1.00} & \textbf{1.00} & 2.37 & 2.13 & \textbf{2.10} & 3.71 & 4.07 & 4.63 \\ 
  $40,25$ & 1.57 & 1.26 & 1.23 & \textbf{1.00} & \textbf{1.00} & \textbf{1.00} & 2.01 & \textbf{1.93} & \textbf{1.93} & 2.89 & 3.16 & 3.57 \\ 
   \bottomrule
    \end{tabular}
    \begin{tablenotes}
    NOTE: The table's largest standard error for the ARL is 0.04. 
    \end{tablenotes} 
    \end{threeparttable}
    \label{tab:rename13}
\end{table}

\begin{table}[!htb]
\setlength{\tabcolsep}{4.7pt}
    \centering
    \footnotesize
     \caption{OOC-ARLs of EWMA $\widetilde{\tau}^{(\fdi)}$-charts and competing $\widehat{\rho}^{(\fhi)}$-chart for unilateral SAR$(2,2)$ model in analogy to Table~\ref{tab:rename12}, for different $m, n$ combinations and  $\alpha_1=0.4, \alpha_2=0.3, \alpha_3=0.1$, smoothing parameter $\lambda=0.1$, with and without 10\,\% of observations contaminated by AOs; simulated with $10^5$ replications.}
    \begin{threeparttable}
\begin{tabular}{ccccccccccccc}
\toprule
& \multicolumn{6}{c}{No outlier} & \multicolumn{6}{c}{With outlier} \\
\cmidrule(lr){2-7} \cmidrule(lr){8-13}
& \multicolumn{3}{c}{$\widetilde{\tau}^{(\fdi)}$} & \multicolumn{3}{c}{$\widehat{\rho}^{(\fhi)}$} & \multicolumn{3}{c}{$\widetilde{\tau}^{(\fdi)}$} & \multicolumn{3}{c}{$\widehat{\rho}^{(\fhi)}$}  
\\ \cmidrule(lr){2-4} \cmidrule(lr){5-7} \cmidrule(lr){8-10} \cmidrule(lr){11-13}
$m,n$ & $\fd = \1$ & $\fd = \2$ & $ \fd = \3$ & $\fh = \1$ & $\fh = \2$ & $ \fh = \3$ & $\fd = \1$ & $\fd = \2$ & $ \fd = \3$ & $\fh = \1$ & $\fh = \2$ & $ \fh = \3$ \\ 
  \cmidrule(lr){1-7} \cmidrule(lr){8-13}
$10, 10$ & 20.51 & 4.81 & 44.66 & 16.85 & \textbf{2.57} & 23.14 & 33.77 & \textbf{7.03} & 77.53 & 187.93 & 34.89 & 281.40 \\ 
  $15,15$ & 17.68 & 3.04 & 33.30 & 14.87 & \textbf{1.71} & 19.21 & 29.25 & \textbf{4.22} & 59.17 & 176.01 & 13.69 & 238.74 \\ 
  $25,25$ & 15.65 & 1.99 & 27.05 & 13.32 & \textbf{1.00} & 16.95 & 26.07 & \textbf{2.45} & 48.28 & 178.14 & 5.94 & 229.12 \\ 
  $40,25$ & 15.06 & 1.66 & 25.35 & 12.93 & \textbf{1.00} & 16.39 & 25.00 & \textbf{2.04} & 45.65 & 179.41 & 4.39 & 232.12 \\ 
   \bottomrule
    \end{tabular}
    \begin{tablenotes}
    NOTE: The table's largest standard error for the ARL is 0.86. 
    \end{tablenotes} 
    \end{threeparttable}
    \label{tab:rename14}
\end{table}

\begin{table}[!htb]
\setlength{\tabcolsep}{4.3pt}
    \centering
    \footnotesize
        \caption{OOC-ARLs of EWMA $\widetilde{\tau}^{\textup{BP}(w)}$-charts and competing $\widehat{\rho}^{\textup{BP}(w)}$-chart for unilateral SAR$(2,2)$ model in analogy to Table~\ref{tab:rename13}, for different $m, n$ combinations and  $\alpha_1=0.4, \alpha_2=0.3, \alpha_3=0.1$, smoothing parameter $\lambda=0.1$, with and without 10\,\% of observations contaminated by AOs; simulated with $10^5$ replications.}
    \begin{threeparttable}
\begin{tabular}{crrrrrrrrrrrr}
\toprule
& \multicolumn{6}{c}{No outlier} & \multicolumn{6}{c}{With outlier} \\
\cmidrule(lr){2-7} \cmidrule(lr){8-13}
& \multicolumn{3}{c}{$\widetilde{\tau}^{\textup{BP}(w)}$} & \multicolumn{3}{c}{$\widehat{\rho}^{\textup{BP}(w)}$} & \multicolumn{3}{c}{$\widetilde{\tau}^{\textup{BP}(w)}$} & \multicolumn{3}{c}{$\widehat{\rho}^{\textup{BP}(w)}$}  
\\ \cmidrule(lr){2-4} \cmidrule(lr){5-7} \cmidrule(lr){8-10} \cmidrule(lr){11-13}
$m,n$ & $w=1$ & $w=2$ & $w=3$ & $w=1$ & $w=2$ & $w=3$ & $w=1$ & $w=2$ & $w=3$ & $w=1$ & $w=2$ & $w=3$ \\ 
  \cmidrule(lr){1-7} \cmidrule(lr){8-13}
$10, 10$ & 20.53 & 3.39 & 3.34 & 8.04 & \textbf{2.12} & 2.35 & 33.76 & 4.72 & \textbf{4.62} & 115.87 & 19.57 & 21.13 \\ 
  $15,15$ & 17.75 & 2.29 & 2.25 & 7.12 & \textbf{1.48} & 1.73 & 29.32 & 3.11 & \textbf{3.04} & 109.96 & 9.55 & 10.96 \\ 
  $25,25$ & 15.69 & 1.62 & 1.58 & 6.34 & \textbf{1.00} & \textbf{1.00} & 25.95 & 2.01 & \textbf{1.98} & 114.35 & 4.86 & 5.71 \\ 
  $40,25$ & 15.03 & 1.07 & 1.06 & 6.10 & \textbf{1.00} & \textbf{1.00} & 25.08 & 1.80 & \textbf{1.77} & 117.73 & 3.75 & 4.43 \\ 
   \bottomrule
    \end{tabular}
    \begin{tablenotes}
    NOTE: The table's largest standard error for the ARL is 0.34. 
    \end{tablenotes} 
    \end{threeparttable}
    \label{tab:rename15}
\end{table}

\begin{table}[!htb]
\setlength{\tabcolsep}{4.3pt}
    \centering
    \footnotesize
    \caption{OOC-ARLs of EWMA $\widetilde{\tau}^{(\fdi)}$ and $\widetilde{\tau}^{\textup{BP}(w)}$-charts and competing $\widehat{\rho}^{(\fhi)}$ and $\widehat{\rho}^{\textup{BP}(w)}$-chart for unilateral SQMA$(1,1)$-model ($1^2\;2^2\;3^2$) like in Table~\ref{tab:DGP8}, for $\beta_1=\beta_2=\beta_3=0.8$, different $m, n$ combinations, smoothing parameter $\lambda=0.1$, and target ARL$_0\approx370$; simulated with $10^5$ replications.}
    \begin{threeparttable}
\begin{tabular}{crrrrrrrrrrrr}
\toprule
& \multicolumn{3}{c}{$\widetilde{\tau}^{(\fdi)}$} & \multicolumn{3}{c}{$\widehat{\rho}^{(\fhi)}$} & \multicolumn{3}{c}{$\widetilde{\tau}^{\textup{BP}(w)}$} & \multicolumn{3}{c}{$\widehat{\rho}^{\textup{BP}(w)}$}  
\\ \cmidrule(lr){2-4} \cmidrule(lr){5-7} \cmidrule(lr){8-10} \cmidrule(lr){11-13}
$m,n$ & $\fd = \1$ & $\fd = \2$ & $ \fd = \3$ & $\fh = \1$ & $\fh = \2$ & $ \fh = \3$ & $w = 1$ & $w = 2$ & $ w = 3$ & $w = 1$ & $w = 2$ & $ w = 3$ \\ 
  \cmidrule(lr){1-7} \cmidrule(lr){8-13}
$10, 10$ & \textbf{6.91} & 242.73 & 214.43 & 79.19 & 117.15 & 128.74 & 6.90 & 6.27 & 6.36 & \textbf{2.69} & 3.28 & 3.73 \\ 
  $15,15$ & \textbf{4.30} & 235.00 & 206.44 & 85.92 & 123.60 & 128.33 & 4.30 & 3.94 & 3.97 & \textbf{1.93} & 2.16 & 2.58 \\ 
  $25,25$ & \textbf{2.56} & 230.47 & 202.08 & 90.59 & 127.91 & 129.83 & 2.56 & 2.37 & 2.38 & \textbf{1.01} & 1.51 & 1.95 \\ 
  $40,25$ & \textbf{2.09} & 229.64 & 200.75 & 91.66 & 129.44 & 130.71 & 2.09 & 2.01 & 2.01 & \textbf{1.00} & 1.01 & 1.39 \\ 
    \bottomrule
    \end{tabular}
    \begin{tablenotes}
    NOTE: The table's largest standard error for the ARL is 0.74. 
    \end{tablenotes} 
    \end{threeparttable}
    \label{tab:rename16}
\end{table}

\begin{table}[!htb]
\setlength{\tabcolsep}{4.8pt}
    \centering
    \footnotesize
    \caption{OOC-ARLs of EWMA $\widetilde{\tau}^{(\fdi)}$ and $\widetilde{\tau}^{\textup{BP}(w)}$-charts and competing $\widehat{\rho}^{(\fhi)}$ adn $\widehat{\rho}^{\textup{BP}(w)}$-chart for unilateral SQMA$(2,2)$-model in analogy to Table~\ref{tab:rename16}, for $\beta_1=\beta_2=\beta_3=0.8$, different $m, n$ combinations, smoothing parameter $\lambda=0.1$, and target ARL$_0\approx370$; simulated with $10^5$ replications.}
    \begin{threeparttable}
\begin{tabular}{crrrrrrrrrrrr}
\toprule
& \multicolumn{3}{c}{$\widetilde{\tau}^{(\fdi)}$} & \multicolumn{3}{c}{$\widehat{\rho}^{(\fhi)}$} & \multicolumn{3}{c}{$\widetilde{\tau}^{\textup{BP}(w)}$} & \multicolumn{3}{c}{$\widehat{\rho}^{\textup{BP}(w)}$}   
\\ \cmidrule(lr){2-4} \cmidrule(lr){5-7} \cmidrule(lr){8-10} \cmidrule(lr){11-13}
$m,n$ & $\fd = \1$ & $\fd = \2$ & $ \fd = \3$ & $\fh = \1$ & $\fh = \2$ & $ \fh = \3$ & $w = 1$ & $w = 2$ & $ w = 3$ & $w = 1$ & $w = 2$ & $ w = 3$ \\ 
  \cmidrule(lr){1-7} \cmidrule(lr){8-13}
$10, 10$ & 72.21 & \textbf{7.78} & 230.81 & 109.14 & 93.60 & 144.88 & 72.08 & 6.09 & 6.20 & 36.74 & \textbf{3.66} & 4.18 \\ 
  $15,15$ & 67.84 & \textbf{4.60} & 217.03 & 110.07 & 94.08 & 137.91 & 67.82 & 3.86 & 3.91 & 34.64 & \textbf{2.34} & 2.77 \\ 
  $25,25$ & 65.07 & \textbf{2.66} & 206.74 & 109.43 & 94.82 & 135.43 & 65.06 & 2.33 & 2.35 & 32.66 & \textbf{1.64} & 1.97 \\ 
  $40,25$ & 64.41 & \textbf{2.14} & 203.71 & 108.78 & 94.75 & 134.72 & 64.41 & 2.01 & 2.01 & 32.02 & \textbf{1.03} & 1.53 \\ 
    \bottomrule
    \end{tabular}
    \begin{tablenotes}
    NOTE: The table's largest standard error for the ARL is 0.71. 
    \end{tablenotes} 
    \end{threeparttable}
    \label{tab:rename17}
\end{table}

\begin{table}[!htb]
\setlength{\tabcolsep}{4.3pt}
    \centering
    \footnotesize
          \caption{OOC-ARLs of EWMA $\widetilde{\tau}^{(\fdi)}$-charts and competing $\widehat{\rho}^{(\fhi)}$-chart for bilateral SAR$(1)$-model like in Tables~\ref{tab:DGP10} and~\ref{tab:DGP11}, smoothing parameter $\lambda=0.1$, with and without 10\%\, of observations contaminated by AOs $\pm 5$ for different $m, n$ combinations and $a_1, a_2, a_3, a_4 = 0.1$, and target ARL$_0\approx370$; simulated with $10^5$ replications.}
    \begin{threeparttable}
\begin{tabular}{crrrrrrrrrrrr}
  \toprule
  & \multicolumn{6}{c}{No outlier} & \multicolumn{6}{c}{With outlier} \\
  \cmidrule(lr){2-7} \cmidrule(lr){8-13}
  & \multicolumn{3}{c}{$\widetilde{\tau}^{(\fdi)}$} & \multicolumn{3}{c}{$\widehat{\rho}^{(\fhi)}$} &\multicolumn{3}{c}{$\widetilde{\tau}^{(\fdi)}$} & \multicolumn{3}{c}{$\widehat{\rho}^{(\fhi)}$}\\
  \cmidrule(lr){2-4} \cmidrule(lr){5-7} \cmidrule(lr){8-10} \cmidrule(lr){11-13}
$m,n$ & $\fd = \1$ & $\fd = \2$ & $ \fd = \3$ & $\fh = \1$ & $\fh = \2$ & $ \fh = \3$ & $\fd = \1$ & $\fd = \2$ & $ \fd = \3$ & $\fh = \1$ & $\fh = \2$ & $ \fh = \3$ \\ 
  \cmidrule(lr){1-7} \cmidrule(lr){8-13}
$10, 10$ & \textbf{6.80} & 116.42 & 289.13 & 36.39 & 167.15 & 152.05 & \textbf{10.39} & 177.38 & 325.15 & 371.17 & 327.50 & 314.87 \\ 
  $15,15$ & \textbf{4.25}& 58.19 & 271.45 & 12.95 & 207.08 & 167.61 & \textbf{6.12} & 97.98 & 313.34 & 115.23 & 348.24 & 324.54 \\ 
  $25,25$ & \textbf{2.53} & 23.63 & 228.14 & 5.72 & 225.64 & 186.98 & \textbf{3.47} & 40.39 & 280.61 & 32.92 & 361.79 & 335.18 \\ 
  $40,25$ & \textbf{2.08} & 16.04 & 198.91 & 4.25 & 208.31 & 193.29 & \textbf{2.75} & 26.68 & 255.96 & 20.79 & 364.48 & 339.33 \\ 
    \bottomrule
    \end{tabular}
    \begin{tablenotes}
    NOTE: The table's largest standard error for the ARL is 1.14. 
    \end{tablenotes} 
    \end{threeparttable}
    \label{tab:rename18}
\end{table}

\begin{table}[!htb]
\setlength{\tabcolsep}{5pt}
    \centering
    \footnotesize
       \caption{OOC-ARLs of EWMA $\widetilde{\tau}^{\textup{BP}(w)}$-charts and competing $\widehat{\rho}^{\textup{BP}(w)}$-chart for bilateral SAR$(1)$-model like in Tables~\ref{tab:DGP10} and~\ref{tab:DGP11}, smoothing parameter $\lambda=0.1$, with and without 10\%\, of observations contaminated by AOs $\pm 5$ for different $m, n$ combinations and $a_1, a_2, a_3, a_4 = 0.1$, and target ARL$_0\approx370$; simulated with $10^5$ replications.}
    \begin{threeparttable}
\begin{tabular}{crrrrrrrrrrrr}
  \toprule
  & \multicolumn{6}{c}{No outlier} & \multicolumn{6}{c}{With outlier} \\
  \cmidrule(lr){2-7} \cmidrule(lr){8-13}
  & \multicolumn{3}{c}{$\widetilde{\tau}^{\textup{BP}(w)}$} & \multicolumn{3}{c}{$\widehat{\rho}^{\textup{BP}(w)}$} &\multicolumn{3}{c}{$\widetilde{\tau}^{\textup{BP}(w)}$} & \multicolumn{3}{c}{$\widehat{\rho}^{\textup{BP}(w)}$}\\
  \cmidrule(lr){2-4} \cmidrule(lr){5-7} \cmidrule(lr){8-10} \cmidrule(lr){11-13}
$m,n$ & $w=1$ & $w=2$ & $w=3$ & $w=1$ & $w=2$ & $w=3$ & $w=1$ & $w=2$ & $w=3$ & $w=1$ & $w=2$ & $w=3$ \\ 
  \cmidrule(lr){1-7} \cmidrule(lr){8-13}
$10, 10$ & 6.80 & 7.02 & 7.64 & \textbf{3.85} & 4.74 & 5.40   & \textbf{10.38} & 10.62 & 11.55 & 19.43 & 26.36 & 30.55 \\ 
  $15,15$ & 4.25 & 4.35 & 4.64 & \textbf{2.46} & 3.07 & 3.55  & \textbf{6.12} & 6.19 & 6.59 & 8.96 & 11.64 & 13.82 \\ 
  $25,25$ & 2.53 & 2.60 & 2.78 & \textbf{1.71} & 2.00 & 2.18  & \textbf{3.47} & 3.52 & 3.71 & 4.57 & 5.82 & 6.92 \\ 
  $40,25$ & 2.08 & 2.09 & 2.17 & \textbf{1.13} & 1.83 & 1.99  & \textbf{2.75} & 2.80 & 2.97 & 3.54 & 4.49 & 5.34 \\ 
    \bottomrule
    \end{tabular}
    \begin{tablenotes}
    NOTE: The table's largest standard error for the ARL is 0.06. 
    \end{tablenotes} 
    \end{threeparttable}
    \label{tab:rename19}
\end{table}

\end{document}